\documentclass[final,3p,times]{elsarticle}

\usepackage[USenglish]{babel}
\usepackage[T1]{fontenc}
\usepackage[ansinew]{inputenc}
\usepackage{lmodern} 
\usepackage{graphicx} 
\usepackage{amsmath}
\usepackage{amsthm}
\usepackage{amsfonts}
\usepackage{mathrsfs}
\usepackage{amssymb}
\usepackage{gensymb}
\usepackage{commath}
\usepackage{listings}
\usepackage{fancyvrb}
\usepackage{textcomp}
\usepackage[labelfont=bf]{caption}
\usepackage{subcaption}
\usepackage{chngcntr}
\counterwithin{figure}{section}
\usepackage{tabularx}
\usepackage{multirow}
\usepackage{courier}
\usepackage{array}
\usepackage{enumitem}
\usepackage{float}
\usepackage{keyval}
\usepackage{epstopdf}
\usepackage{xcolor}

\def \c {\sc Cosmos++}
\def \cdg {\sc CosmosDG}
\def \nn {\nonumber}
\def \alf {Alfv{\'e}n }

\journal{Journal of Computational Physics}
\begin{document}

\begin{frontmatter}
	\title{Divergence-Free Magnetohydrodynamics on Conformally Moving, Adaptive Meshes Using a Vector Potential Method}
	\author[CofC]{P. Chris Fragile\corref{cor1}}
	\ead{fragilep@cofc.edu}
 
	\author[CofC]{Daniel Nemergut}

	\author[CofC]{Payden L. Shaw}

	\author[LLNL]{Peter Anninos}
	\ead{anninos1@llnl.gov}

	\cortext[cor1]{Principal Corresponding Author}

	\address[CofC]{Department of Physics and Astronomy, College of Charleston, Charleston, SC 29424, USA}
	\address[LLNL]{Lawrence Livermore National Laboratory, P.O. Box 808, Livermore, CA, 94550, USA} 

	\begin{abstract}
	
We present a new method for evolving the equations of magnetohydrodynamics (both Newtonian and relativistic) that is capable of maintaining a divergence-free magnetic field ($\nabla \cdot \mathbf{B} = 0$) on adaptively refined, conformally moving meshes. The method relies on evolving the magnetic vector potential and then using it to reconstruct the magnetic fields. The advantage of this approach is that the vector potential is not subject to a constraint equation in the same way the magnetic field is, and so can be refined and moved in a straightforward way. We test this new method against a wide array of problems from simple \alf waves on a uniform grid to general relativistic MHD simulations of black hole accretion on a nested, spherical-polar grid. We find that the code produces accurate results and in all cases maintains a divergence-free magnetic field to machine precision. 

	\end{abstract}
	

\end{frontmatter}

\section{Introduction}
In magnetohydrodynamics (MHD), the time component of Maxwell's equations, known independently as Gauss' law of magnetism, provides a constraint condition. Mathematically, the constraint is $\nabla \cdotp \mathbf{B} = 0$, where $\mathbf{B}$ is the magnetic field vector. Simply put, this says that a naturally occurring magnetic field should be divergence free. Physically, this is equivalent to saying that there are no magnetic monopoles.

However, most numerical implementations of MHD are not guaranteed to abide by the $\nabla \cdotp \mathbf{B} = 0$ constraint, even if the initial field is divergence free. Thus, the magnetic field is allowed to build up unphysical divergence (monopoles) during the evolution. This can potentially lead to multiple numerical problems, including nonlinear instabilities, potentially driving pressures and densities to unphysical values \citep{Rossmanith13}. More critically, for conservative schemes, the build up of $\nabla \cdot \mathbf{B}$ can generate anomalous forces parallel to the field \citep{Brackbill80} or cause an extra compressive component to arise within the magnetic field itself \citep{Balsara99}.

Several approaches for controlling the build up of $\nabla \cdotp \mathbf{B}$ have been proposed \citep[e.g.][]{Brackbill80,Evans88,Dai98,Ryu98,Balsara99,Toth00}. Some aim only to minimize it by adding supplemental fields that dampen or otherwise decay the accumulation of monopoles \citep{Powell99,Dedner02}. Although straightforward to implement in some numerical schemes, these approaches formally break the conservative nature of the MHD equations \citep{Mocz14}, which may be undesirable for certain schemes. 

Other approaches \cite{Evans88,Dai98,Ryu98,Balsara99} aim to carefully evolve the magnetic field in such a way that after each update cycle, it is again divergence free, at least for one chosen numerical representation of the divergence \citep[see][for an exploration of some of the different implementations for Godunov codes]{Toth00}. One such approach is known as Constrained Transport (CT), so named because of its aim to constrain the evolution of the magnetic fields. While many of these approaches work well for uniform, static meshes, they can be challenging to extend to unstructured or refined meshes \citep{Mocz14}. However, Balsara \& Dumser \cite{Balsara15} have recently discovered a projection operator that can be coupled with a cell-centered WENO algorithm to preserve a divergence-free magnetic field on unstructured meshes. Additional work \cite{Balsara10,Balsara12} has focused on finding suitable multi-dimensional Riemann solvers to obtain edge-centered electric fields, which are then used to update face-centered representations of the magnetic field.

In this paper we take a different approach, explored in some previous codes \citep[e.g.][]{Brandenburg02,Etienne10}, to evolve the magnetic vector potential rather than the magnetic field itself. Since the magnetic field can then be reconstructed as needed by taking the curl of the vector potential, and since the divergence of a curl is always zero, this procedure is guaranteed to maintain a divergence-free field. The benefit of this approach for our purposes is that, since there are no constraints on the vector potential, it can be refined, de-refined, and relocated without introducing divergence errors. 

There has been some concern, though, expressed in the literature about the use of the vector potential (VP) approach, mainly that it may produce spurious forces or current reversals near discontinuities \cite{Evans88}. The trouble comes from the calculation of the Lorentz force, $\mathbf{J} \times \mathbf{B} = (\nabla \times \mathbf{B}) \times \mathbf{B}/4\pi$. The thinking is that, because the Lorentz force requires taking a derivative of the magnetic field (to get the electric current), that equates to a second derivative of the vector potential. While Choptuik \cite{Choptuik86} showed that the order of accuracy of derivatives is the same as the order of the underlying reconstruction of the field for smooth functions, so that the order of accuracy of the method can be preserved, the concern is in how the VP method behaves when the field is not smooth, i.e. near discontinuities. The expectation of Evans \& Hawley \cite{Evans88} was that these methods would produce spurious current reversals in such locations. However, as we show, we do not see any evidence of spurious currents nor forces in our tests. 

This paper will focus on the development of our magnetic VP method, specifically as implemented in the {\c} computational astrophysics code \citep{Anninos05}. The paper is structured as follows: Section \ref{sec:Cosmos++} provides an overview of the capabilities and general structure of {\c}; Section \ref{sec:vectpot} details the implementation of the new VP method within {\c}; Section \ref{sec:results} presents the results of well known MHD and astrophysical test problems generated using the updated code; and conclusions are drawn in Section \ref{sec:conclusions}.

\section{\c}
\label{sec:Cosmos++}
\begin{figure}
\includegraphics[width=0.49\textwidth]{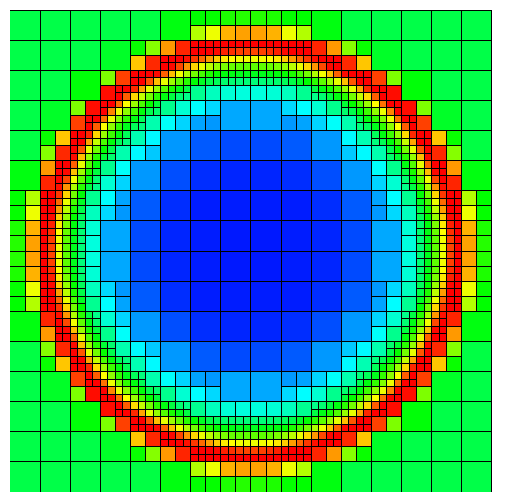}
\caption{Pseudocolor plot of the rest-mass density of gas for a cylindrical blast wave expanding into an ambient medium. This example uses adaptive mesh refinement to concentrate zones (resolution) in the vicinity of the blast wave (colored red).
}
\label{fig:sedov}
\end{figure}

{\c} is a parallel, multi-dimensional, fully covariant, modern, object-oriented (C++), radiation magnetohydrodynamics (RMHD) code for both Newtonian and general relativistic astrophysical and cosmological applications. {\c} utilizes unstructured meshes with adaptive ($h$-) refinement \citep{Anninos05}, moving-mesh ($r$-refinement) \citep{Anninos12}, and adaptive order ($p$-refinement) \citep{Anninos17} capabilities, enabling it to evolve fluid systems over a wide range of spatial scales with targeted precision. {\c} can, in principle, support triangular/tetrahedral meshes, though it currently only offers quadrilateral/hexahedral options. Nevertheless, they are unstructured in the sense that {\c} supports arbitrarily reduced and enhanced nodal connectivity. The mesh refinement is done on a local, cell-by-cell basis as illustrated in Fig. \ref{fig:sedov} for a simple example of a 3-level mesh tracking the dense gas front created by a cylindrical blast wave plowing into an ambient medium. A typical block-refinement algorithm would be much less efficient on problems such as this.  
{\c} can also use adaptive and moving mesh refinement \citep{Anninos18} or adaptive mesh and adaptive order \citep{Anninos17} in conjunction for even greater precision. 

\subsection{Equations of Magnetohydrodynamics}

Since {\c} is designed to handle problems in both Newtonian and general relativistic gravity, we present the MHD equations for both, though they are intentionally written to look quite similar. 

\subsubsection{Newtonian}

Introducing an effective Newtonian MHD stress energy tensor
\begin{equation}
{\cal T}^{\alpha\beta} =
         \rho v^\alpha v^\beta + (P+P_B) g^{\alpha\beta}
         - b^\alpha b^\beta ~,
\label{eqn:newt_tmn}
\end{equation}
where Greek indices run over the four spacetime coordinates $\{0, 1, 2, 3\}$, we can derive the four fluid equations (energy and the three components of momentum) from the conservation of stress energy: $\nabla_\mu {\cal T}^\mu_{\ \nu} = \partial_\mu {\cal T}^\mu_{\ \nu} +
 \Gamma^\mu_{\alpha\mu} {\cal T}^\alpha_{\ \nu} - \Gamma^\alpha_{\mu\nu} {\cal T}^\mu_{\ \alpha} = S_\nu$,
where $\rho$ is the fluid mass density, $v^i$ is the fluid velocity ($v^0 = 1$), $P$ is the fluid pressure (for an ideal gas $P=(\Gamma-1)e$, $e=\rho\epsilon$ is the fluid internal energy density, $\Gamma$ is the adiabatic index), $P_B = b^i b_i/2$ is the magnetic pressure, $b^i$ is the primitive Newtonian field ($b^0 = 0$), $g_{\alpha\beta}$ is the curvature metric used here to handle curvature associated with non-Cartesian meshes, $\Gamma^\alpha_{\mu\nu}$ are the Christoffel symbols representing the metric derivatives, and $S_\nu$ includes any source terms. Note that we define our magnetic field such that the factor of $\sqrt{4\pi}$ is absorbed.
In addition to energy and momentum, we also require equations
for the conservation of mass
$\nabla_\mu(\rho v^\mu) = 0$
and magnetic induction $\nabla_\mu(v^\mu b^\nu - b^\mu v^\nu) = 0$.
Note that here and throughout, we use the Einstein summation convention, where repeated indices imply summation, e.g.
\begin{equation}
\partial_i v^i = \frac{\partial v^1}{\partial x^1} + \frac{\partial v^2}{\partial x^2} + \frac{\partial v^3}{\partial x^3} ~.
\end{equation}

Expanding out the space and time coordinates and including a grid velocity $V_g^i$ for when moving mesh is active, the Newtonian conservation equations (of mass, energy, and momentum) and the magnetic induction equation can be written as
\begin{equation}
 \partial_t (\sqrt{g} \rho) +
  \partial_i \left[\sqrt{g} \rho (v^i - V_g^i)\right] +
 \sqrt{g} \rho \partial_i V_g^i = 0 ~,
 \label{eqn:newt_den}
\end{equation}
\begin{equation}
 \partial_t (\sqrt{g} E) +
  \partial_i \left\{\sqrt{g} E (v^i - V_g^i) + \sqrt{g} \left[(P+P_B) v^i - b^i b_j v^j\right] \right\} +
 \sqrt{g} E \partial_i V_g^i =
      - \sqrt{g} \rho v^i \partial_i \phi ~,
      \label{eqn:newt_ene}
\end{equation}
\begin{eqnarray}
 \partial_t( \sqrt{g} s_j) +
  \partial_i \left\{\sqrt{g} s_j (v^i - V_g^i) + \sqrt{g} \left[(P+P_B) g^{ij} - b^i b^j\right] \right\} +
 \sqrt{g} s_j \partial_i V_g^i =
        \sqrt{g} {\cal T}^{i k} \Gamma_{i k j}
       -\sqrt{g} \rho \partial_j \phi ~,
    \label{eqn:newt_mom_dn}
\end{eqnarray}
\begin{equation}
 \partial_t B^j +
  \partial_i \left[B^j(v^i - V_g^i) - B^i v^j\right] +
 B^j \partial_i V_g^i =
       0 ~,
      \label{eqn:newt_mag}
\end{equation}
where $s_i$ are the covariant momentum components, $B^i = \sqrt{g} b^i$ is the conserved Newtonian magnetic field, $\sqrt{g}$ is the determinant of the curvature metric, $\phi$ is the gravitational potential, and $E$ is the total energy density, which includes internal, magnetic, and kinetic energy contributions:
$E=e + b^i b_i/2 + \rho v^i v_i/2$.
By treating the gravitational energy as a separate source term, the Newtonian and relativistic flux and source terms are easily interchangeable within the code framework. See \citep{Anninos03b} for a more complete description of the Newtonian capabilities of {\c}.

\subsubsection{General Relativity}

For general relativistic MHD, we follow a very similar procedure to arrive at the appropriate conservation equations. We again start from the contravariant stress energy tensor for an ideal MHD fluid:
\begin{equation}
T^{\alpha\beta} =
         (\rho h + 2 P_B/c^2) u^\alpha u^\beta
         + (P+P_B) g^{\alpha\beta}
         - b^\alpha b^\beta ~.
\label{eqn:gr_tmn}
\end{equation}
Here $h=1+\epsilon/c^2 + P/(\rho c^2)$ is the specific enthalpy,
$c$ is the speed of light,
$u^\alpha = u^0 V^\alpha$ is the contravariant four-velocity,
$V^\alpha$ is the transport velocity,
$b^\alpha$ is the magnetic field measured by an observer co-moving with the fluid, 
and $P_B = b^\alpha b_\alpha/2$.

By defining the boosted mass density, total energy density, momentum density, and magnetic fields as
$D=\sqrt{-g} u^0 \rho = W\rho$, ${\cal E} = - \sqrt{-g} {T}^0_0$, ${\cal S}_j = \sqrt{-g} {T}^0_j$, and ${\cal B}^j = \sqrt{-g}(u^0 b^j - u^j b^0)$, respectively,
the equations take on a traditional conservation law formulation
\begin{equation}
\partial_t D + \partial_i \left[D (V^i - V_g^i)\right] + D \partial_i V_g^i = 0 ~,
      \label{eqn:gr_den}
\end{equation}
for mass density, 
\begin{equation}
\partial_t {\cal E} + \partial_i\left[{\cal E} (V^i - V_g^i)\right]
    + \partial_i \left[ \sqrt{-g} (P + P_B) V^i\right] + {\cal E} \partial_i V_g^i
    = - \sqrt{-g} T^\mu_\sigma \ \Gamma^\sigma_{\mu 0} ~,
      \label{eqn:gr_en}
\end{equation}
for energy,
\begin{equation}
\partial_t {\cal S}_j + \partial_i\left[{\cal S}_j (V^i - V_g^i)\right]
    + \partial_i \left[ \sqrt{-g} (P + P_B)~g^0_j~V^i\right] + {\cal S}_j \partial_i V_g^i
    = \sqrt{-g} T^\mu_\sigma \ \Gamma^\sigma_{\mu j} ~.
      \label{eqn:gr_mom}
\end{equation}
for momentum, and
\begin{equation}
 \partial_t {\cal B}^j + \partial_i\left[{\cal B}^j (V^i - V_g^i) - {\cal B}^i V^j\right] + {\cal B}^j \partial_i V_g^i
      = 0 ~,
      \label{eqn:gr_mag}
\end{equation}
for magnetic induction. These fields $\{D, {\cal E}, {\cal S}_j, {\cal B}^j\}$ serve as the conserved fields for our relativistic MHD scheme, with $\{\rho, \rho \epsilon, V^i, {\cal B}^j/\sqrt{-g}\}$ being the corresponding primitives.

\subsection{Evolution}

For this work we utilize the fully conservative High Resolution Shock Capturing (HRSC) Godunov scheme \cite{Font03} to solve the Newtonian or relativistic MHD equations.  To do so, we note that all of the equations have already been cast in conservation form
\begin{equation}
\partial_t \mathbf{U}(\mathbf{P}) + \partial_i \mathbf{F}^i(\mathbf{P}) = \mathbf{S}(\mathbf{P}) 
\label{eqn:cons}
\end{equation}
where $\mathbf{U}(\mathbf{P})$, $\mathbf{F}^i(\mathbf{P})$, and $\mathbf{S}(\mathbf{P})$ are the arrays representing the conserved quantities, fluxes, and source terms, respectively. 
By integrating both sides with respect to volume and applying Gauss' Theorem, we can rewrite the conservation equation in the form
\begin{equation}
\int_V \partial_t \mathbf{U} dV = -\oint_S \mathbf{F}^i dA_i + \int_V \mathbf{S} dV ~.
\end{equation} 
We can then discretize the equations using a finite volume representation as
\begin{equation}
\mathbf{U}^{n+1} = \mathbf{U}^n - \frac{\Delta t}{V} \sum\limits_{faces}\left(\mathbf{F}^i A_i\right) + \Delta t \mathbf{S} ~,
\label{eqn:consstep}
\end{equation}
where $V$ is the cell volume and $A_i$ is the area of face $i$. The $\mathbf{F}^i$ are determined using the two-speed HLL approximate Riemann solver
\begin{equation}
\mathbf{F} = \frac{c_{min} \mathbf{F}_R + c_{max} \mathbf{F}_L - c_{max} c_{min} ( \mathbf{U}_R - \mathbf{U}_L )}{c_{max} + c_{min}} ~,
\label{eqn:flux}
\end{equation}
where $\mathbf{F}_R = \mathbf{F}(\mathbf{P}_R)$ and $\mathbf{F}_L = \mathbf{F}(\mathbf{P}_L)$ are the fluxes at the right- and left-hand side of each zone interface and $\mathbf{U}_R = \mathbf{U}_R(\mathbf{P}_R)$ and $\mathbf{U}_L = \mathbf{U}_L(\mathbf{P}_L)$ are the corresponding conserved quantities, each calculated from the extrapolations of the primitive variables to the zone interface, $\mathbf{P}_R$ and $\mathbf{P}_L$. {\c} has options for piecewise-linear and piecewise-parabolic extrapolations, both limited by a one-dimensional MUSCL-type limiter with $1 \le \theta \le 2$, and a multi-dimensional limiter designed to work on unstructured meshes by constructing upwind and downwind tetrahedral elements \citep{Anninos05}. In this paper, all tests use the piecewise-parabolic option with the 1D MUSCL-type limiter and $\theta = 1.5$.
The bounding wave speeds are $c_\mathrm{max} \equiv \mathrm{max}(0, ~c_{+,R}, ~c_{+,L})$ and $c_\mathrm{min} \equiv -\mathrm{min}(0, ~c_{-,R}, ~c_{-,L})$, where $c_{\pm , R}$ and $c_{\pm , L}$ are the maximum right- and left-going waves speeds.
Time integration is handled by a five-stage, strong-stability-preserving Runge-Kutta (SSPRK) method \citep{Spiteri02,Anninos17}.

Table \ref{tab:evolution} outlines the steps taken during a typical compute cycle. Usually the first physics package that is called is the primitive solver. In this step, a series of coupled equations are solved to extract primitive fields (mass density, internal energy, velocity, and primitive magnetic field) from evolved conserved fields (boost density, total energy, momentum, and conserved magnetic field or vector potential), after which the equation of state is applied to compute thermodynamic quantities, pressure, sound speed, and temperature. For Newtonian systems this procedure is straightforward, but relativity introduces a nonlinear inter-dependency of primitives so their extraction from conserved fields requires special numerical treatments. We have implemented several procedures for doing this, solving one, two, or five dimensional inversion schemes for MHD \citep{Noble06}.

\begin{table}
  \centering
  \begin{tabular}{p{\textwidth}}
    \hline
    Evolution algorithm \\
    \hline
    \begin{enumerate}[leftmargin=*, topsep=-10pt,itemsep=0ex,partopsep=0ex,parsep=0ex]
      \item calculate timestep
		\begin{enumerate}[leftmargin=*, topsep=-10pt,itemsep=0ex,partopsep=0ex,parsep=0ex]
        		\item Each physics package is polled for its smallest timestep
        		\item All packages, in all zones, on all mesh levels, are advanced using the smallest overall timestep
      		\end{enumerate}
      \item move mesh
      \item refine mesh
      \item second-order time-stepping loop
      		\begin{enumerate}[leftmargin=*, topsep=-10pt,itemsep=0ex,partopsep=0ex,parsep=0ex]
		\item invert conserved fields to primitives
 			\begin{enumerate}[leftmargin=*, topsep=-10pt,itemsep=0ex,partopsep=0ex,parsep=0ex]
        			\item Use current $\mathbf{A}$ to recover $\mathbf{B}$
      			\end{enumerate}     
      		\item HRSC update
        			\begin{enumerate}[leftmargin=*, topsep=-10pt,itemsep=0ex,partopsep=0ex,parsep=0ex]
        			\item construct face-centered fluxes $\mathbf{F}_R$ and $\mathbf{F}_L$
				\begin{enumerate}[leftmargin=*, topsep=-10pt,itemsep=0ex,partopsep=0ex,parsep=0ex]
				\item Use magnetic fluxes to construct emfs used to update $\mathbf{A}$
				\end{enumerate}
			\item update conserved fields from fluxes
			\item update conserved fields from source terms
			\end{enumerate}
    		\end{enumerate}
	\item final reconstruction of updated conserved fields
	\end{enumerate} \\
    \hline
  \end{tabular}
  \caption{Steps taken during a single evolution step of {\c}.}
  \label{tab:evolution}
\end{table}

\subsection{Refinement}
\label{sec:amr}

\subsubsection{Adaptive Mesh Refinement}

\begin{figure}
\includegraphics[width=0.49\textwidth]{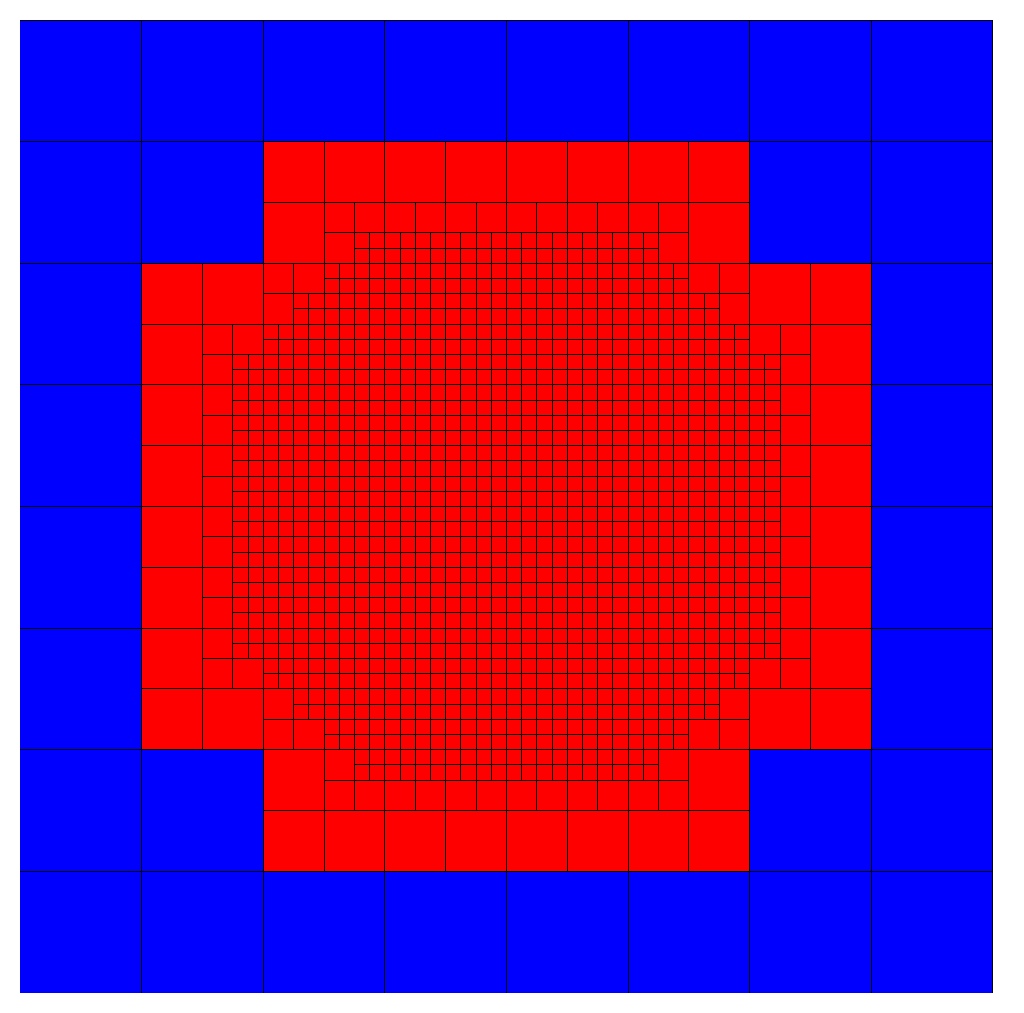}
\includegraphics[width=0.49\textwidth]{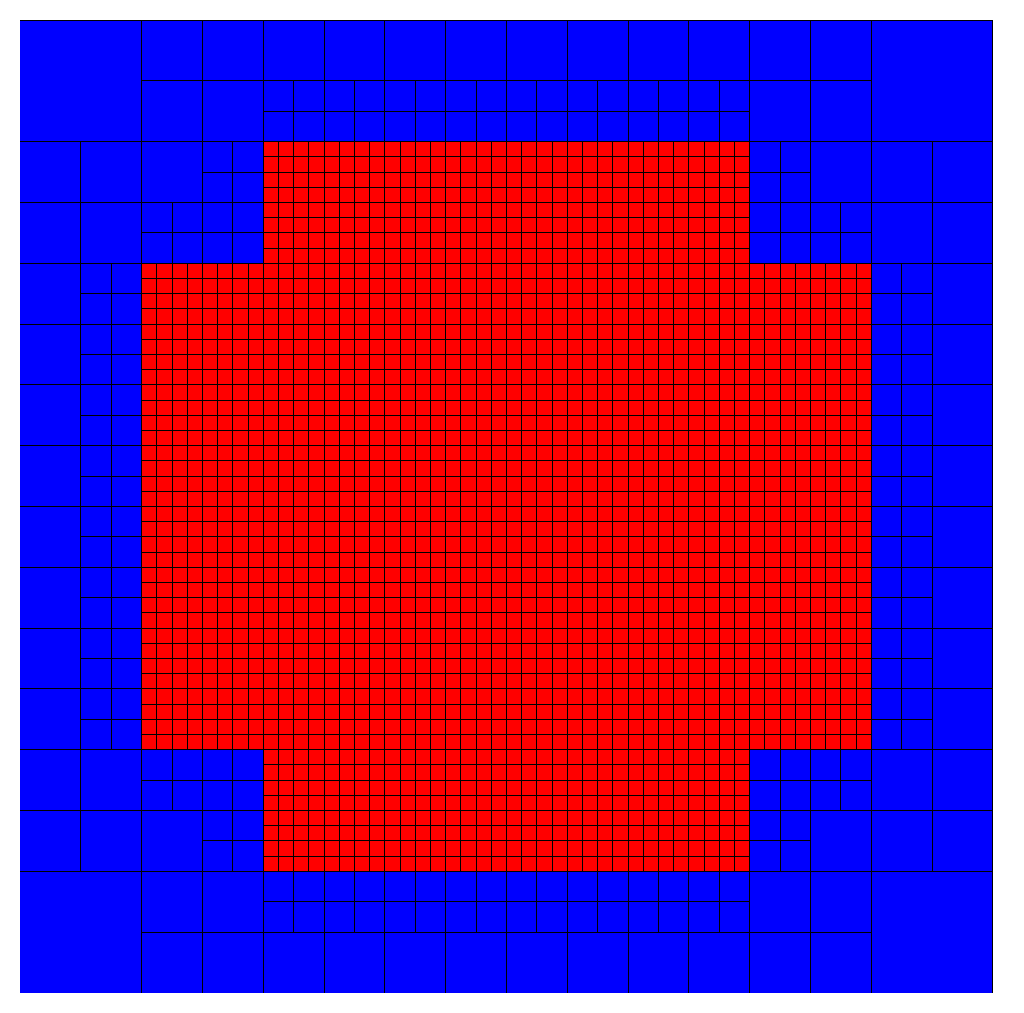}
\caption{Comparison of the four-level mesh resulting from application of the fast (left) and strict (right) refinement methods three times to an $8 \times 8$ base mesh. In both cases, all red zones are tagged for refinement.}
\label{fig:fast_strict}
\end{figure}

Adaptive mesh refinement (AMR) in {\c} uses a set of refinement criteria based on the local field, slope, or curvature of any user-specified code variable to actively adapt the mesh in each zone to best match the problem. A simple example of a field refinement criteria might be to refine a zone whenever the fluid density exceeds a given value and to de-refine if the density drops below a different value.  A firm requirement in {\c} is that neighboring zones never differ by more than one level of refinement. We have two options in the code for how to enforce this requirement. The first, referred to as ``strict,'' guarantees that any zone that meets a refinement criteria will indeed refine. In order to ensure that it will not then differ from any neighbor by more than one level of refinement, the refining zone queries the refinement level of all its neighbors and then forces them to refine as necessary to preserve the one level separation. The ``fast'' option takes the opposite approach. Whenever a zone is tagged for refinement, similar to ``strict,'' it polls the refinement level of its neighbors. If the refinement of the tagged zone will cause it to differ from any of its neighbors by more than one level of refinement (i.e. if any neighbors are currently at a lower level of refinement), then that zone is no longer tagged for refinement. Fig. \ref{fig:fast_strict} illustrates the practical difference. In this example, we start with a coarse $8\times8$ base mesh and define a ``circular'' region of interest (red zones). These zones are then marked for refinement three separate times, for a total of four mesh levels. Note how the fast method refines ``inward'' from the edges of the targeted refinement region, whereas the strict method refines ``outward.'' Incidentally, this example also demonstrates why it is often useful to start a simulation fully refined and then allow it to de-refine as it can. Otherwise, the coarse initial resolution will imprint itself upon the problem as it has here.

De-refinement, on the other hand, solely follows the ``fast'' approach, i.e. zones only de-refine if after doing so they will not differ by more than one level from any direct neighbor. In this way, we err on the side of keeping greater resolution than we may need. The basic algorithmic steps for each option are presented in Table \ref{table:refinement_algorithm}.

\begin{table}
  \centering
  \begin{tabular}{|p{0.5\linewidth}|p{0.5\linewidth}|}
    \hline
    Strict refinement algorithm & Fast refinement algorithm \\
    \hline
      tag zones based on refinement criteria & tag zones based on refinement criteria \\
      \textbf{for each} level $i$ of refinement, from highest to lowest & \textbf{for each} zone $j$ tagged for refinement \\      
      $\quad$ \textbf{for each} zone $j$ tagged for refinement & $\quad$ doRefine = true \\
      $\quad$ $\quad$  \textbf{for each} neighbor zone $k$ & $\quad$ \textbf{for each} neighbor zone $k$ \\
      $\quad$ $\quad$ $\quad$ \textbf{if} $j$ is already more refined than $k$ & $\quad$ $\quad$ \textbf{if} $j$ is already more refined than $k$ \\
      $\quad$ $\quad$ $\quad$ refine($k$) & $\quad$ $\quad$ $\quad$ doRefine = false \\
      $\quad$ $\quad$ refine($j$) & $\quad$ $\quad$ $\quad$ break \\
      & $\quad$ \textbf{if} doRefine == true \\
      & $\quad$ $\quad$ refine($j$) \\
    \hline
  \end{tabular}
  \caption{Refinement algorithms in {\c}.}
  \label{table:refinement_algorithm}
\end{table}

Whenever a zone is refined, the conserved fields are interpolated to the child zones in such a way as to maintain the same volume-integrated value as the parent. For de-refinement, the conserved fields in parent zones are set from the volume integral of all child zones. Thus, the conservative nature of our scheme is preserved for both refinement and de-refinement. New primitive fields are assigned to each zone after refinement/de-refinement is complete.

\subsubsection{Moving Mesh Refinement}
\label{sec:movingmesh}

Mesh motion is implemented by a straight-forward
replacement of the generic advective terms
\begin{equation}
\partial_t (\sqrt{-g} T^0_\alpha) + \partial_i(\sqrt{-g} T^0_\alpha V^i)
\end{equation}
with
\begin{equation}
\partial_t (\sqrt{-g} T^0_\alpha) + \partial_i(\sqrt{-g} T^0_\alpha (V^i - V_g^i)) + \sqrt{-g} T^0_\alpha \partial_i V_g^i ~.
\end{equation}
Here $T^0_\alpha$ is used here to represent any evolved field.

The specification of the grid velocity $\mathbf{V}_g$ is a highly problem-dependent choice, and we do not in this paper advocate one method over another. Rather than developing and locking into specific algorithms, we have instead taken the tact of designing the code in such a way that users can easily implement their favorite options. However, whatever choices are made for mesh motion, they will almost certainly require rezoning strategies for mesh optimization in order to prevent tangling and highly distorted cells from forming and degrading the numerical solution. This is, in fact, the more crucial element needed for robust mesh motion. To this end, Cosmos++ has numerous options for mesh relaxation and optimization, including softening procedures generally adapted to Lagrangian flows, potential-based optimal equidistribution methods \citep{Budd09}, velocity-based geometric conservation law methods \citep{Cao02}, parabolic gradient flow relaxation methods \citep{Huang01}, and variational equidistribution and mass optimization methods \citep{Chacon11,Sulman11}.

{\c} supports grid velocity specification at either the cell centers or nodes, although when specified at the centers, it must be projected onto the nodes in order to guarantee cell separation does not occur. This is done with simple mass weighting. After the node positions are physically relocated according to the nodal velocity and timestep, zone attributes (including face area vectors, cell volumes, and the spacetime metric) are updated for each cell \cite{Anninos17,Anninos18}.

\section{Divergence-Free Evolution of Magnetic Fields}
\label{sec:vectpot}

Along with the evolution equations, (\ref{eqn:newt_den})-(\ref{eqn:newt_mag}) for Newtonian or (\ref{eqn:gr_den})-(\ref{eqn:gr_mag}) for general relativity, the time component of Maxwell's equations provides a constraint condition:  $\nabla \cdotp \mathbf{B} = 0$ in Newtonian MHD or $\partial_j \mathcal{B}^j = 0$ in general relativity. In both cases, the meaning is the same: there can never be more magnetic field entering a region of space than leaving it, or that the divergence of the magnetic field through a closed surface must always be zero. However, most numerical implementations of the induction equation [(\ref{eqn:newt_mag}) or (\ref{eqn:gr_mag})] will not be guaranteed to preserve the $\nabla \cdotp \mathbf{B} = 0$ constraint, even if the initial field is divergence free. 

\begin{figure}
\includegraphics[width=\textwidth]{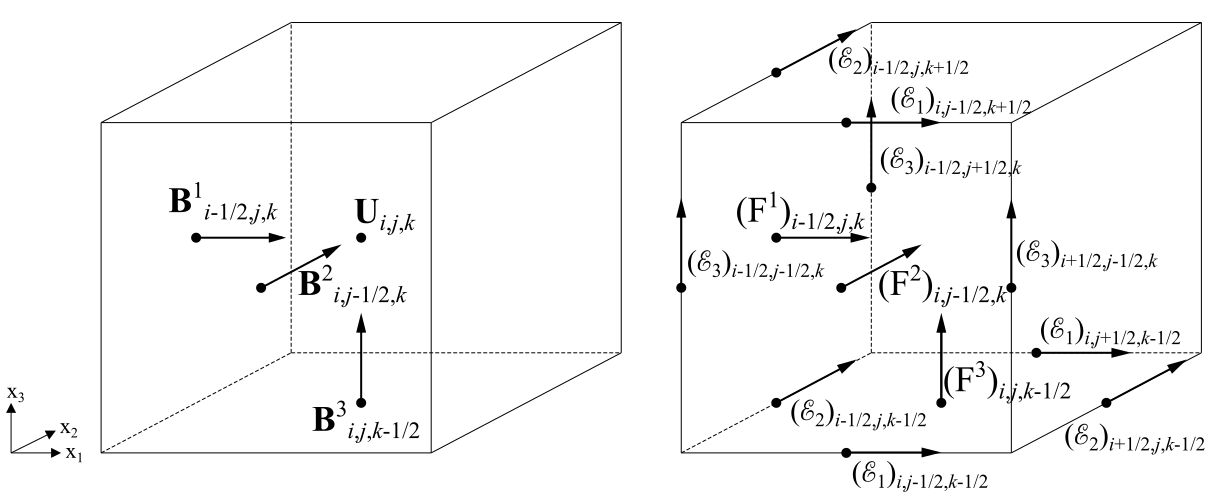}
\caption{{\it Left:} Representation of the field locations within a typical computational cell.  Most conserved fields, $\mathbf{U}$, are located at the zone centers, while the primary representation of the magnetic fields are located at the left-hand zone faces.  {\it Right:} Locations of the fluxes, $\mathbf{F}$, that come from the Riemann step of HRSC and are used to calculate the electric field components, $\mathscr{E}$, along zone edges.}
\label{fig:cell_emf}
\end{figure}

The original version of {\c} \cite{Anninos05} included a simple parabolic divergence cleanser option \cite{Dedner02}, intended to dampen any accumulated divergence error. This option was extended in later work \cite[e.g.][]{Fragile05a,Fragile07} to include elliptic and hyperbolic elements to filter and propagate away the divergence error. Such methods are generally successful at maintaining $\nabla \cdot \mathbf{B} = 0$ to about the level of truncation errors. Once the fully conservative HRSC MHD method was implemented in {\c} \cite{Fragile12}, we introduced a staggered representation of the magnetic field, with the primary (conserved) fields ($B^i$ for Newtonian or ${\cal B}^i$ for GR) being treated as area-weighted variables located at face centers (see Fig. \ref{fig:cell_emf}) and implemented a form of CT. The new VP method shares many elements with our staggered CT scheme.

\subsection{Vector Potential Evolution}

Instead of evolving the magnetic field directly using the induction equation [(\ref{eqn:newt_mag}) or (\ref{eqn:gr_mag})], another option is to evolve the magnetic vector potential \citep{DelZanna03,Etienne10}. In doing so, we replace the evolution equations (\ref{eqn:newt_mag}) and (\ref{eqn:gr_mag}) with the following
\begin{align}
\frac{\partial \vec{\mathcal{A}}}{\partial t} - (\mathbf{V}_g \cdotp \nabla) \vec{\mathcal{A}} & = \mathbf{v} \times \mathbf{B} - \nabla \Phi & \mathrm{(Newtonian)} \label{eq:vect_Newt} \\
\partial_{t} \mathcal{A}_i  - V_g^j \partial_j \mathcal{A}_i & = \epsilon_{ijk} V^{j} \mathcal{B}^{k} + \partial_{i}\left[\frac{1}{g^{00}}(\Phi - g^{0j} \mathcal{A}_{j})\right] ~ & \mathrm{(GR)} \label{eq:vect_GR} \\
 & = -\mathscr{E}_i + \partial_{i}\left[\frac{1}{g^{00}}(\Phi - g^{0j} \mathcal{A}_{j})\right] ~, & \nonumber
\end{align}
where $\mathscr{E}_i = -\epsilon_{ijk} V^j \mathcal{B}^k$, is the $i^\mathrm{th}$-component of the electric field (or emf), centered on the appropriate cell edge (Fig. \ref{fig:cell_emf}) and $\Phi$ is the electrostatic potential. 

The most robust method we have found so far for constructing the edge-centered electric fields is to simply average the surrounding face-centered fluxes recovered from the Riemann solver (being careful of signs), as described in \citep{Balsara99}. For example,
\begin{equation}
\left(\mathscr{E}_3\right)_{i-1/2,j-1/2,k} = \frac{1}{4}\left[\left(F_{B^1}^2\right)_{i,j-1/2,k} - \left(F_{B^2}^1\right)_{i-1/2,j,k} + \left(F_{B^1}^2\right)_{i-1,j-1/2,k} - \left(F_{B^2}^1\right)_{i-1/2,j-1,k}\right] ~,
\label{eqn:edgeEMF}
\end{equation}
where $F_{B^1}^2 = B^1 V^2 - B^2 V^1$ and $F_{B^2}^1 = B^2 V^1 - B^1 V^2$ are the properly upwinded Godunov fluxes resulting from equation (\ref{eqn:flux}). Similar procedures are used for the other components of $\mathscr{E}_i$. Note here that we always use the fluid velocity, $V^j$, in defining the electric field and not the relative velocity with the grid motion subtracted out, $(V^j - V_g^j)$. This reconstruction is second-order accurate in space and retains the temporal accuracy of whatever time integration is used \citep{Balsara99}. This is consistent with the order of all source terms in Eqs. (\ref{eqn:newt_ene}), (\ref{eqn:newt_mom_dn}), (\ref{eqn:gr_en}), and (\ref{eqn:gr_mom}). We will explore higher-order interpolations once we port this method over to our discontinuous Galerkin (DG) version of {\sc Cosmos} \citep{Anninos17}. Another option would be to implement a multi-step or multi-dimensional Riemann solver \citep{DelZanna03,Balsara10,Balsara12}.

The electrostatic potential in Eq. (\ref{eq:vect_Newt})-(\ref{eq:vect_GR}) is dependent on a choice of gauge. For our work, we have adopted the simplest, algebraic EM gauge condition, in which 
\begin{equation}
\Phi = g^{0j} \mathcal{A}_j
\end{equation}
and the evolution simplifies to 
\begin{equation}
\partial_{t} \mathcal{A}_i  - V_g^j \partial_j \mathcal{A}_i = -\mathscr{E}_i ~.
\end{equation}
Another choice would be the Lorentz gauge \cite{Etienne12}, which requires an additional evolution equation for $\Phi$:
\begin{equation}
\partial_t(\sqrt{-g} \Phi) - V_g^i \partial_i (\sqrt{-g} \Phi) + \partial_i \left(\sqrt{-g} \mathcal{A}^i + \frac{g^{0i}}{g^{00}} \sqrt{-g} \Phi \right) = 0 ~.
\end{equation}
This could easily be added to {\c}, but has, so far, proven unnecessary.

\begin{figure}
\centering
\includegraphics[width=0.5\textwidth]{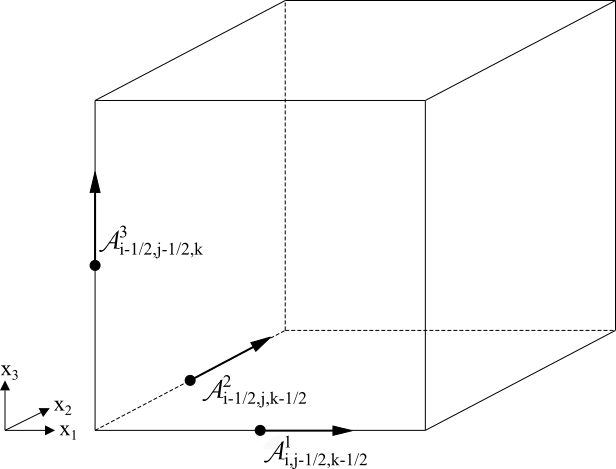}
\caption{Location of the edge-centered vector potentials as used in {\c}.}
  \label{figure:vectpot_center_to_edge_unit3vec}
\end{figure}

In {\c}, we choose to have each zone store its own bottom-left edge-centered vector potential components in two dimensions or its back-bottom-left edge-centered vector potential components in three dimensions, as illustrated in Fig. \ref{figure:vectpot_center_to_edge_unit3vec}. This allows each zone to only store a single set of edge-centered vector potentials, while still facilitating easy reconstruction of the magnetic field. This is done during the primitive inversion step by taking the curl of the vector potential:
\begin{align}
\mathbf{B} & = \nabla \times \vec{\mathcal{A}} & \mathrm{(Newtonian)} \label{eq:B_from_A} \\
\mathcal{B}^i & = \epsilon^{ijk} \partial_j \mathcal{A}_k ~, & \mathrm{(GR)}
\end{align}
or in component form, explicitly, as
\begin{align}
 & \mathrm{(Newtonian)} & & \mathrm{(GR) } \nn \\
B^1 &= \frac{\partial \mathcal{A}^3}{\partial x_2} - \frac{\partial \mathcal{A}^2}{\partial x_3} & \mathcal{B}^1 & = \partial_2 \mathcal{A}_3 - \partial_3 \mathcal{A}_2 ~, \label{equation:b_field_components_x} \\
B^2 &= \frac{\partial \mathcal{A}^1}{\partial x_3} - \frac{\partial \mathcal{A}^3}{\partial x_1} & \mathcal{B}^2 & = \partial_3 \mathcal{A}_1 - \partial_1 \mathcal{A}_3 ~, \label{equation:b_field_components_y} \\
B^3 &= \frac{\partial \mathcal{A}^2}{\partial x_1} - \frac{\partial \mathcal{A}^1}{\partial x_2} & \mathcal{B}^3 & = \partial_1 \mathcal{A}_2 - \partial_2 \mathcal{A}_1 ~. \label{equation:b_field_components_z}
\end{align}
Since the divergence of a curl is always zero,
\begin{align}
\nabla \cdotp \mathbf{B} = \nabla \cdotp (\nabla \times \mathbf{A}) = 0 ~,
\label{eq:divB}
\end{align}
this procedure is mathematically guaranteed to maintain a divergence-free field. In practice, it preserves
the following finite volume representation of divergence:
\begin{eqnarray}
\partial_i \mathcal{B}^i & = & \frac{1}{V} \left[ (\mathcal{B}^1 A_1)_{i+1/2,j,k} - (\mathcal{B}^1 A_1)_{i-1/2,j,k} + (\mathcal{B}^2 A_2)_{i,j+1/2,k} \right. \nonumber \\
 & & \left. - (\mathcal{B}^2 A_2)_{i,j-1/2,k} + (\mathcal{B}^3 A_3)_{i,j,k+1/2} - (\mathcal{B}^3 A_3)_{i,j,k-1/2} \right] ~,
\label{eqn:divfreepres}
\end{eqnarray}
where again $V$ is the cell volume and $A_i$ is the area of face $i$. When normalized by $\sqrt{P_B}/\Delta x$, where $\Delta x$ is a characteristic zone length, this is preserved to round-off error in each zone.
 the same representation of $\nabla  \cdotp \mathbf{B} = 0$ as the CT scheme (\ref{eqn:divfreepres}). To see this explicitly, we can expand Eq. (\ref{equation:b_field_components_x})-(\ref{eq:divB}).  First, the magnetic field components at the faces are
\begin{align}
B^1_{i-\frac{1}{2},j,k} &= \frac{\mathcal{A}^3_{i-\frac{1}{2}, j+\frac{1}{2}, k} - \mathcal{A}^3_{i-\frac{1}{2}, j-\frac{1}{2}, k}}{\Delta x_2} - \frac{\mathcal{A}^2_{i-\frac{1}{2}, j, k+\frac{1}{2}} - \mathcal{A}^2_{i-\frac{1}{2}, j, k-\frac{1}{2}}}{\Delta x_3} ~, \label{eq:bx} \\
B^2_{i,j-\frac{1}{2},k} &= \frac{\mathcal{A}^1_{i, j-\frac{1}{2}, k+\frac{1}{2}} - \mathcal{A}^1_{i, j-\frac{1}{2}, k-\frac{1}{2}}}{\Delta x_3} - \frac{\mathcal{A}^3_{i+\frac{1}{2}, j-\frac{1}{2}, k} - \mathcal{A}^3_{i-\frac{1}{2}, j-\frac{1}{2}, k}}{\Delta x_1} ~, \label{eq:by} \\
B^3_{i,j,k-\frac{1}{2}} &= \frac{\mathcal{A}^2_{i+\frac{1}{2}, j, k-\frac{1}{2}} - \mathcal{A}^2_{i-\frac{1}{2}, j, k-\frac{1}{2}}}{\Delta x_1} - \frac{\mathcal{A}^1_{i, j+\frac{1}{2}, k-\frac{1}{2}} - \mathcal{A}^1_{i, j-\frac{1}{2}, k-\frac{1}{2}}}{\Delta x_2} ~. \label{eq:bz}
\end{align}
The magnetic field divergence through the surface of the cell at position ($i$, $j$, $k$) is then
\begin{align}
\nabla \cdotp \mathbf{B} = &\frac{B^1_{i+\frac{1}{2},j,k} - B^1_{i-\frac{1}{2},j,k}}{\Delta x_1} + \frac{B^2_{i,j+\frac{1}{2},k} - B^2_{i,j-\frac{1}{2},k}}{\Delta x_2} + \frac{B^3_{i,j,k+\frac{1}{2}} - B^3_{i,j,k-\frac{1}{2}}}{\Delta x_3} \nn \\
 = & \frac{\mathcal{A}^3_{i+\frac{1}{2},j+\frac{1}{2},k} - \mathcal{A}^3_{i+\frac{1}{2},j-\frac{1}{2},k}}{\Delta x_1 \Delta x_2} - \frac{\mathcal{A}^2_{i+\frac{1}{2},j,k+\frac{1}{2}} - \mathcal{A}^2_{i+\frac{1}{2},j,k-\frac{1}{2}}}{\Delta x_1 \Delta x_3} - \nn \\
& \frac{\mathcal{A}^3_{i-\frac{1}{2},j+\frac{1}{2},k} - \mathcal{A}^3_{i-\frac{1}{2},j-\frac{1}{2},k}}{\Delta x_1 \Delta x_2} + \frac{\mathcal{A}^2_{i-\frac{1}{2},j,k+\frac{1}{2}} - \mathcal{A}^2_{i-\frac{1}{2},j,k-\frac{1}{2}}}{\Delta x_1 \Delta x_3} + \nn \\
& \frac{\mathcal{A}^1_{i,j+\frac{1}{2},k+\frac{1}{2}} - \mathcal{A}^1_{i,j+\frac{1}{2},k-\frac{1}{2}}}{\Delta x_2 \Delta x_3} - \frac{\mathcal{A}^3_{i+\frac{1}{2},j+\frac{1}{2},k} - \mathcal{A}^3_{i-\frac{1}{2},j+\frac{1}{2},k}}{\Delta x_1 \Delta x_2} - \nn \\
& \frac{\mathcal{A}^1_{i,j-\frac{1}{2},k+\frac{1}{2}} - \mathcal{A}^1_{i,j-\frac{1}{2},k-\frac{1}{2}}}{\Delta x_2 \Delta x_3} + \frac{\mathcal{A}^3_{i+\frac{1}{2},j-\frac{1}{2},k} - \mathcal{A}^3_{i-\frac{1}{2},j-\frac{1}{2},k}}{\Delta x_1 \Delta x_2} + \nn \\
& \frac{\mathcal{A}^2_{i+\frac{1}{2},j,k+\frac{1}{2}} - \mathcal{A}^2_{i-\frac{1}{2},j,k+\frac{1}{2}}}{\Delta x_1 \Delta x_3} - \frac{\mathcal{A}^1_{i,j+\frac{1}{2},k+\frac{1}{2}} - \mathcal{A}^1_{i,j-\frac{1}{2},k+\frac{1}{2}}}{\Delta x_2 \Delta x_3} - \nn \\
& \frac{\mathcal{A}^2_{i+\frac{1}{2},j,k-\frac{1}{2}} - \mathcal{A}^2_{i-\frac{1}{2},j,k-\frac{1}{2}}}{\Delta x_1 \Delta x_3} + \frac{\mathcal{A}^1_{i,j+\frac{1}{2},k-\frac{1}{2}} - \mathcal{A}^1_{i,j-\frac{1}{2},k-\frac{1}{2}}}{\Delta x_2 \Delta x_3} ~.
\label{equation:div_b_expanded}
\end{align}
It is easy to see that all of the terms in Eq. (\ref{equation:div_b_expanded}) cancel out pairwise.

The staggered magnetic field recovered in this way is used for two purposes. The first is to overwrite the appropriate component of the extrapolated primitive field used in the flux reconstruction [Eq. (\ref{eqn:flux})] at each face. The other is to calculate cell-centered values of the primitive magnetic fields as required, for instance, for calculating the magnetic pressure and recovering the internal energy of the gas. In this case we use the volume-averaged fields
\begin{eqnarray}
(B^1)_{i,j,k} & = & \frac{1}{2} \left[ (B^1)_{i-1/2,j,k} + (B^1)_{i+1/2,j,k} \right] \label{eq:b1} \\
(B^2)_{i,j,k} & = & \frac{1}{2} \left[ (B^2)_{i,j-1/2,k} + (B^2)_{i,j+1/2,k} \right] \label{eq:b2} \\
(B^3)_{i,j,k} & = & \frac{1}{2} \left[ (B^3)_{i,j,k-1/2} + (B^3)_{i,j,k+1/2} \right] \label{eq:b3} ~.
\end{eqnarray}
We store both the zone-centered and staggered magnetic fields along with the edge-centered vector potential components. 

Occasionally, we may want to solve problems containing symmetries that may allow us to reduce the dimensionality. In two dimensions, the $x_3$-component of the magnetic field is divergence free by definition, as all gradients in that direction are assumed to be zero.  Looking at the expressions for $B^1$ and $B^2$ in Eq. (\ref{eq:bx}) and (\ref{eq:by}), it is clear that there is no reason to evolve $\mathcal{A}^1$ nor $\mathcal{A}^2$ in two dimensions, as they are not needed. Thus, when considering two-dimensional problems, we only evolve $\mathcal{A}^3$ and use it to recover $B^1$ and $B^2$. $B^3$ is updated through the standard induction equation (\ref{eqn:newt_mag}). In one dimension, there is no need to evolve any of the vector potential components, as there is no chance of building up divergence in this case. Here, the induction equation (\ref{eqn:newt_mag}) is entirely sufficient.

Before moving on, we mention one slight drawback of the VP method: it can make the initialization of problems slightly more difficult. This is because, in cases where one has a starting magnetic field configuration in mind, it may not be an easy matter to find the corresponding vector potential, as there is no straightforward way to invert Eq. (\ref{eq:B_from_A}). For simple magnetic field configurations, as in all the test problems in this paper, this is not a major hurdle, but for complex configurations, this could be challenging.

\subsubsection{Mesh Refinement Considerations}

For a uniform mesh, the VP method is functionally equivalent to a conventional staggered CT scheme. Effectively, it just changes the order of the spatial and temporal operations in updating the magnetic field. However, because the vector potential itself is not subject to any constraint, we can refine and de-refine the mesh and evolved fields without introducing magnetic monopoles. This is the principle advantage of the VP method. Still, care must be taken in updating the vector potential and magnetic field components near refinement boundaries, where neighbors may be at different refinement levels. We discuss these considerations in this section.

\begin{figure}
\centering
\includegraphics[width=0.5\linewidth]{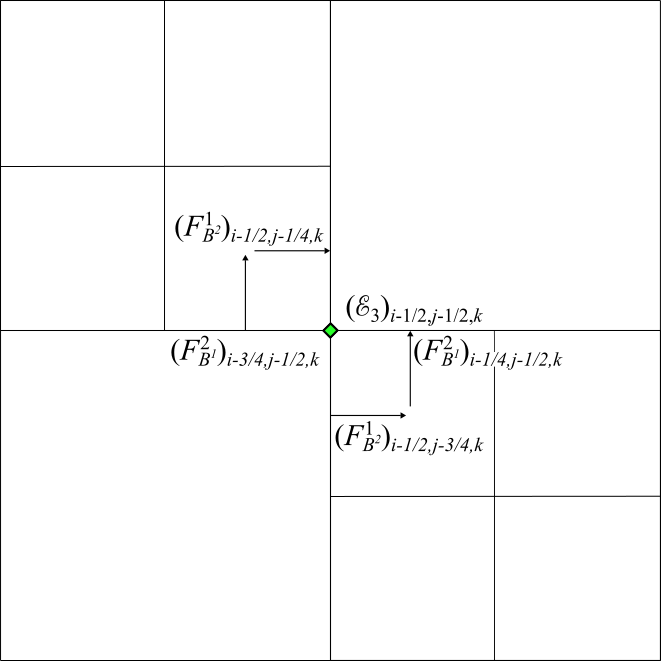}
 \caption{Example of a case where the zone owning the edge-centered electric field, $(\mathscr{E}_3)_{i-1/2,j-1/2,k}$, would only use fluxes from refined neighbors in its calculation.}
 \label{fig:refined_emf}
\end{figure}

Let us look first at the VP update. The source term for the VP update equation [(\ref{eq:vect_Newt}) or (\ref{eq:vect_GR})] is just the edge-centered electric field. We see from Eq. (\ref{eqn:edgeEMF}) that this field comes from an average of the face-centered magnetic field fluxes on the four surrounding faces. However, near refinement boundaries, a given edge could have as many as eight (four) surrounding faces in three (two) dimensions or as few as three, with the faces lying at potentially different distances from the edge center. For this reason, we must replace Eq. (\ref{eqn:edgeEMF}) with one that weights the fluxes appropriately. As a basic rule, we always use the closest set of fluxes in each direction. This can sometimes mean that a given zone does not even use its own fluxes when calculating one (or more) of its edge-centered vector potentials. Fig. \ref{fig:refined_emf} shows a two-dimensional example where the zone in question only uses fluxes from its more refined neighbors. The corresponding expression for the edge-centered electric field is 
\begin{eqnarray}
\left(\mathscr{E}_3\right)_{i-1/2,j-1/2,k} & = & \frac{1}{w_T}\left[\frac{\left(F_{B^1}^2\right)_{i-1/4,j-1/2,k}}{\vert \mathbf{x}_{i-1/4,j-1/2,k} - \mathbf{x}_{i-1/2,j-1/2,k} \vert} - \frac{\left(F_{B^2}^1\right)_{i-1/2,j-1/4,k}}{\vert \mathbf{x}_{i-1/2,j-1/4,k} - \mathbf{x}_{i-1/2,j-1/2,k} \vert} + \right. \\ \nonumber
& & \left. \frac{\left(F_{B^1}^2\right)_{i-3/4,j-1/2,k}}{\vert \mathbf{x}_{i-3/4,j-1/2,k} - \mathbf{x}_{i-1/2,j-1/2,k} \vert} - \frac{\left(F_{B^2}^1\right)_{i-1/2,j-3/4,k}}{\vert \mathbf{x}_{i-1/2,j-3/4,k} - \mathbf{x}_{i-1/2,j-1/2,k} \vert} \right] ~,
\end{eqnarray}
where $w_T$ is the sum of all the individual weights
\begin{equation}
w_T = \sum_\mathrm{faces} \frac{1}{\vert \mathbf{x}_\mathrm{face} - \mathbf{x}_\mathrm{edge} \vert} ~,
\end{equation}
and $\mathbf{x}$ is the appropriate position vector. For a uniform mesh, the result is identical to Eq. (\ref{eqn:edgeEMF}).

\begin{figure}
\centering
\includegraphics[width=0.5\linewidth]{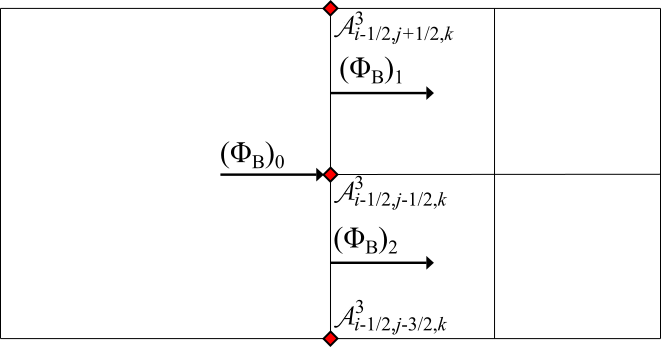}
 \caption{Example of a case where the zone owning the edge-centered vector potential, $\mathcal{A}^3_{i-1/2,j-1/2,k}$, has fewer than four surrounding faces.}
 \label{fig:hanging_emf}
\end{figure}

As mentioned, there is a possibility for ``hanging'' edges to have fewer than four surrounding faces. Fig. \ref{fig:hanging_emf} illustrates such a case. Here, we simply replace the edge-vector potential in question with the average of the already calculated edge-vector potentials along the neighboring edges
\begin{equation}
\mathcal{A}^3_{i-1/2,j-1/2,k} = \frac{1}{2} \left(\mathcal{A}^3_{i-1/2,j-3/2,k} + \mathcal{A}^3_{i-1/2,j+1/2,k} \right) ~.
\label{eqn:hanging}
\end{equation}
The physical motivation for doing so is that this procedure guarantees that the sum of the magnetic fluxes, $\Phi_B$, through the refined faces equals the magnetic flux through the neighbor's unrefined face as it should, i.e. $(\Phi_B)_0 = (\Phi_B)_1 + (\Phi_B)_2$, where 
\begin{equation}
\Phi_B = \iint_S \mathbf{B} \cdotp d\mathrm{S} = \oint_{\partial S} \vec{\mathcal{A}} \cdotp d\ell ~.
\end{equation}
Following the same principles, we can extend these procedures for updating the vector potential components at refinement boundaries to three dimensions.

The next step is to obtain the face-centered magnetic fields from the edge vector potentials by computing their curl. This is accomplished computationally by taking gradients of the edge vector potentials surrounding each face as in Eq. (\ref{eq:bx})-(\ref{eq:bz}). Again, though, there are special considerations at refinement boundaries. For example, in the case of hanging edges, there may be no corresponding vector potential at that location. For these cases, we follow the same procedure as for calculating vector potentials for edges that have fewer than four surrounding faces, namely we average the already calculated edge-vector potentials from the neighboring edges as in Eq. (\ref{eqn:hanging}) before taking any gradients.

\subsubsection{Mesh Motion Considerations}

As currently implemented, the new VP method requires that magnetic field components lie normal to their respective faces as in Fig. \ref{fig:cell_emf}. For this reason, we limit our considerations in this paper to conformal motion, such as uniform translation of the grid, which preserves the orthogonality, though conformal expansion and contraction could be treated as well. However, since {\c} supports full mesh motion, including mesh distortion, as described in Sec. \ref{sec:movingmesh}, we plan to expand the mesh motion capabilities of the VP method in the future. One way to go about that would be to perform the magnetic field updates in an orthogonal reference zone. To construct the reference zone, we could use the properties of the distorted zone (edge lengths and inner products of face areas) to construct a local metric that encapsulates the distortion. We would then use that metric to transform the vector potential and magnetic field components into the orthogonal reference frame, where the magnetic fields would be normal to the zone faces, allowing the VP method to be applied. This would result in a magnetic field that is divergence free in the reference frame. We already follow a similar procedure for solving the hydrodynamic equations in {\cdg} \citep{Anninos17}, so this would be a natural thing to do when we port the VP method over to that version of the code.

\section{Code Tests}
\label{sec:results}

In this section, we verify that our new MHD method is able to handle a large suite of standard code tests. We only consider multi-dimensional tests, as the VP method plays no role in one dimension, and one dimensional MHD tests using {\c} were already presented in \cite{Anninos05}. We pay particular attention to any accumulation of $\nabla \cdotp \mathbf{B}$ and report the $L$-2 norm error of this quantity 
\begin{equation}
\vert \vert \nabla \cdotp \mathbf{B} \vert \vert_2 =\frac{1}{N_iN_jN_k} \sqrt{\sum_{i,j,k} \left(\frac{\vert (\nabla \cdotp \mathbf{B})_{i,j,k} \vert}{\vert (\mathbf{B}/\delta \mathbf{x})_{i,j,k} \vert} \right)^2}~,
\end{equation}
for each test. We also compare to known analytic and published solutions whenever possible to assess how our method compares and calculate its convergence order. All tests are done with the time integration set to third order. This ensures that our errors will reveal the true spatial accuracy of the method.

\subsection{Code Tests without Mesh Refinement}

Before considering test problems that involve refinement (adaptive and moving mesh), we consider a few test problems on uniform, static meshes, just to confirm things are working correctly.

\subsubsection{Circularly Polarized \alf Wave}

Since a circularly polarized \alf wave in a periodic box has a known analytic solution \cite{Toth00}, it provides an excellent starting test and way to confirm the accuracy and convergence order of our code. To engage all components of the VP method, we make this a three-dimensional test following the approach of \cite{Gardiner08} by initializing the problem in the rotated coordinate system, $(x_1,x_2,x_3)$, where the orientation with respect to the standard $(x,y,z)$ coordinates is given by
\begin{eqnarray}
x_1 & = & x\cos(\alpha)\cos(\beta)+y\cos(\alpha)\sin(\beta)+z\sin(\alpha)\\
x_2 & = & -x\sin(\beta)+y\cos(\beta)\\
x_3 & = & -x\sin(\alpha)\cos(\beta)-y\sin(\alpha)\sin(\beta)+z\cos(\alpha)~,
\label{eqn:alfven_rot}
\end{eqnarray}
where $\sin \alpha = 2/3$ and $\sin \beta = 2/\sqrt{5}$. The mesh extends over the range $0 \le x \le 3$, $0 \le y \le 1.5$, and $0 \le z \le 1.5$, resolved by a $2N \times N \times N$ grid, and is periodic in all directions (except the vector potential $\vec{\mathcal{A}}$ is {\em not} periodic). The initial solution has mass density $\rho=1$, gas pressure $P=0.1$, adiabatic index $\Gamma=5/3$, and magnetic field $\mathbf{B}=\{1,0.1\sin(k x_1), 0.1\cos(k x_1)\}$ in $(x_1,x_2,x_3)$ coordinates, with wave vector $k=2 \pi/\lambda$ and wavelength $\lambda = 1$. Consistent with our new approach, we actually initialize the magnetic vector potential $\vec{\mathcal{A}} = \{0, [0.1 \sin(k x_1)]/k-0.5 x_3, [0.1 \cos(k x_1)]/k+0.5 x_2\}$, which ensures $\nabla \cdot \mathbf{B} = 0$ initially. However, because of truncation errors associated with recovering the magnetic field from the vector potential, the initial conditions are subject to grid-level magnetic pressure perturbations, which drive secondary compressive waves. 

\begin{figure}
  \centering
  \includegraphics[width=0.5\textwidth]{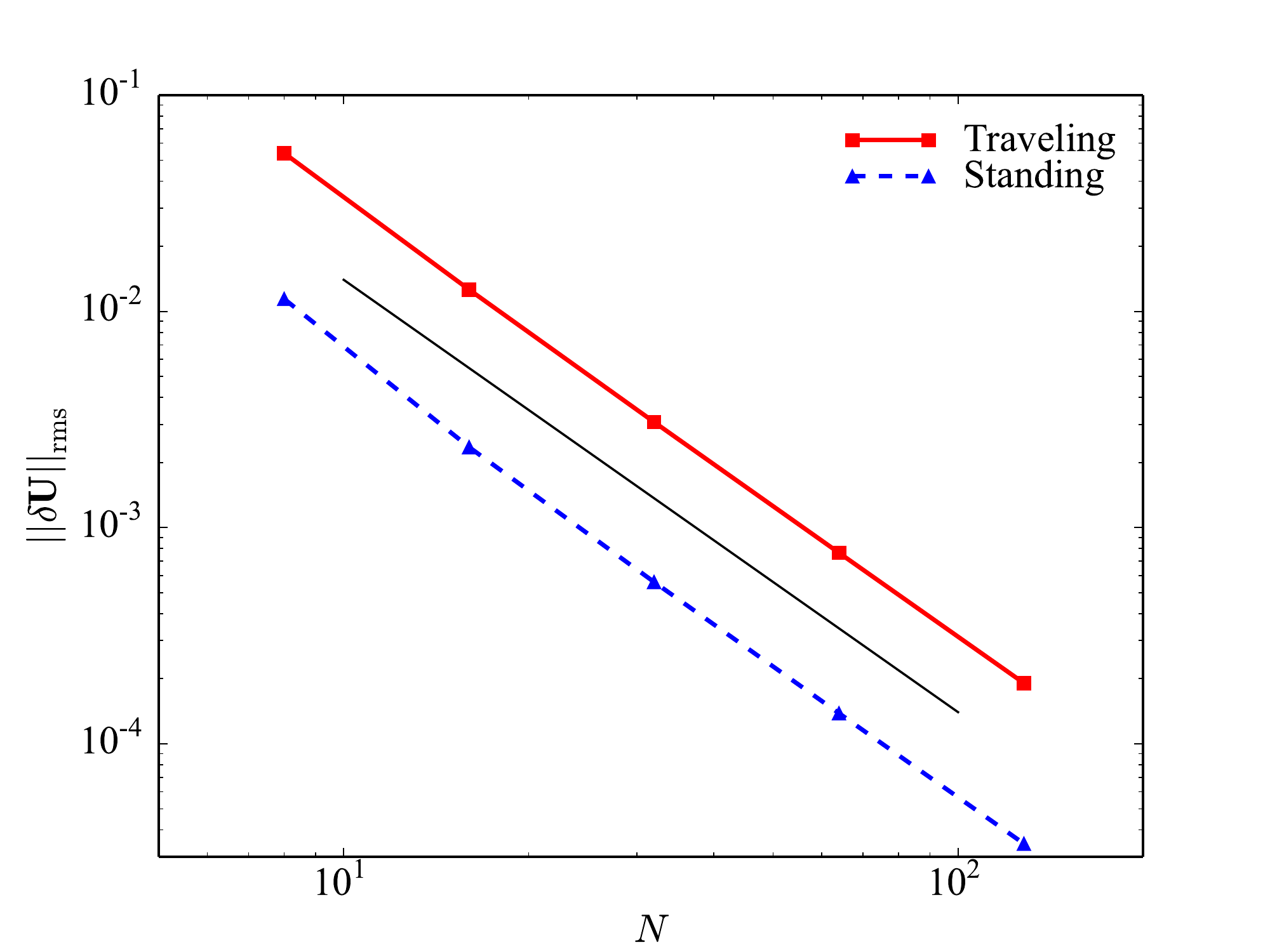}
  \caption{RMS of $L$-1 norm errors for the circularly polarized \alf wave test, both the traveling (red) and standing (blue) wave cases after $t=1$ and 0.25, respectively. The slopes of both lines are close to second-order convergence (black line).}
  \label{fig:alfvenwave}
\end{figure}

The wave can either be set up to propagate parallel to the $x_1$-axis, in which case the corresponding fluid velocity component is $v_1 = 0$ (traveling case), or remain stationary, in which case the fluid velocity component is $v_1=1$ (standing case). The other velocity components are $v_2=0.1\sin(2\pi x_1/\lambda)$ and $v_3=0.1\cos(2\pi x_1/\lambda)$. Since $v_2$, $v_3$, and the boundaries are periodic, then for these parameter choices, the traveling wave should return to its initial state after a period of $t=1$.  This makes it easy to test the convergence of our method by comparing the final state against the initial one. We first calculate the $L$-1 norm errors of each of our conserved fields, $U_s$, as
\begin{equation}
\vert \vert \delta U_s \vert \vert_1 =\frac{1}{2N^3} \sum_{i,j,k} \vert (U_s)^n_{i,j,k} - (U_s)^0_{i,j,k} \vert~,
\end{equation} 
where $(U_s)^n_{i,j,k}$ is the final numerical solution and $(U_s)^0_{i,j,k}$ is the initial (analytical) solution for field $s$. Most of these errors are reported in Table \ref{tab:alfven}. In Fig. \ref{fig:alfvenwave}, we present a plot of the root mean square of the $L$-1 norm errors 
\begin{equation}
\vert \vert \delta \mathbf{U} \vert \vert_\mathrm{rms} =\sqrt{\sum_s\left(\delta U_s\right)^2 } ~
\end{equation}
as a function of the grid resolution, $N$. Our errors compare favorably to those reported \cite{Gardiner08} and show almost exactly second order convergence, as expected. Note, though, that for the standing wave case, we could only run the test to a time of $t=0.25$ (a quarter of a traveling wave period). For longer run times, the solution became unstable, possibly due to a parametric instability \cite{Goldstein78,DelZanna01}, especially at higher resolutions. The $L$-2 norm error, $\vert \vert \nabla \cdotp \mathbf{B} \vert \vert_2$, never exceeds $1.05 \times 10^{-17}$ for the $N=8$ traveling or standing test.

\begin{table}
\caption{$L$-1 Norm Errors for Circularly Polarized \alf Wave \label{tab:alfven}}
\centering
\begin{tabular}{cccccc}
\hline\hline
$N$ & $\vert \vert E(D) \vert \vert_1$ & $\vert \vert E({\cal E}) \vert \vert_1$ & $\vert \vert E({\cal B}^x) \vert \vert_1$ & $\vert \vert E({\cal B}^y) \vert \vert_1$ & $\vert \vert E({\cal B}^z) \vert \vert_1$ \\
\hline
\multicolumn{6}{c}{Traveling} \\
\hline
8 & $9.07 \times 10^{-3}$ & $2.57 \times 10^{-3}$ & $1.77 \times 10^{-2}$ & $2.33 \times 10^{-2}$ & $2.41 \times 10^{-2}$  \\
16 & $1.84 \times 10^{-3}$ & $5.30 \times 10^{-4}$ & $3.98 \times 10^{-3}$ & $5.57 \times 10^{-3}$ & $5.58 \times 10^{-3}$  \\
32 & $4.31 \times 10^{-4}$ & $1.29 \times 10^{-4}$ & $9.64 \times 10^{-4}$ & $1.36 \times 10^{-3}$ & $1.36 \times 10^{-3}$ \\
64 & $1.06 \times 10^{-4}$ & $3.21 \times 10^{-5}$ & $2.39 \times 10^{-4}$ & $3.39 \times 10^{-4}$ & $3.38 \times 10^{-4}$ \\
128 & $2.64 \times 10^{-5}$ & $8.00 \times 10^{-6}$ & $5.97 \times 10^{-5}$ & $8.46 \times 10^{-5}$ & $8.43 \times 10^{-5}$ \\
\hline
\multicolumn{6}{c}{Standing} \\
\hline
8 & $4.40 \times 10^{-3}$ & $2.37 \times 10^{-3}$ & $2.28 \times 10^{-3}$ & $5.92 \times 10^{-3}$ & $5.96 \times 10^{-3}$  \\
16 & $8.29 \times 10^{-4}$ & $4.74 \times 10^{-4}$ & $4.42 \times 10^{-4}$ & $1.09 \times 10^{-3}$ & $1.09 \times 10^{-3}$  \\
32 & $1.92 \times 10^{-4}$ & $1.11 \times 10^{-4}$ & $1.01 \times 10^{-4}$ & $2.41 \times 10^{-4}$ & $2.41 \times 10^{-4}$ \\
64 & $4.69 \times 10^{-5}$ & $2.72 \times 10^{-5}$ & $2.46 \times 10^{-5}$ & $5.85 \times 10^{-5}$ & $5.83 \times 10^{-5}$ \\
128 & $1.17 \times 10^{-5}$ & $6.76 \times 10^{-6}$ & $6.12 \times 10^{-6}$ & $1.45 \times 10^{-5}$ & $1.45 \times 10^{-5}$ \\
\hline
\end{tabular}
\end{table}

\subsubsection{MHD Vortex}
\label{sec:vortex}

The MHD vortex problem \cite{Balsara04} is another test that can be used for convergence studies. In fact, as it is an exact solution of the MHD equations in two dimensions, it perhaps gives a better measure of the true error than the circularly polarized \alf wave, which is really a one-dimensional problem. The vortex is set up in an $N \times N$ computational domain covering $-5 \le x \le 5$, $-5 \le y \le 5$, with periodic boundaries in both directions. The unperturbed background has properties $\{\rho, P, v^x, v^y, B^x, B^y\} = \{1, 1, 1, 1, 0, 0\}$, with $\Gamma = 5/3$. The vortex is initialized via the following perturbations
\begin{equation}
\{\delta v^x, \delta v^y\} = \frac{1}{2\pi} e^{0.5(1-r^2)} \{-y, x\} ~,
\end{equation}
where $r = \sqrt{x^2 + y^2}$. The perturbations for $\{\delta B^x, \delta B^y\}$ have the exact same form, though for the VP method, we initialize the perturbation
\begin{equation}
\delta {\mathcal A}_z = \frac{1}{2\pi} e^{0.5(1-r^2)} ~.
\end{equation}
The centrifugal force of the vortex motion, the centripetal force of the magnetic tension, and the magnetic pressure all contribute to the dynamical force balance, along with gas pressure. To maintain the balance, the pressure must have a perturbation of the form
\begin{equation}
\delta P = -\left(\frac{1}{2\pi}\right)^2 \frac{r^2}{2} e^{(1-r^2)} ~.
\end{equation}
The problem is run until $t = 10$, by which time the vortex returns to its initial location.

We again test convergence by comparing the final state against the initial one, with $L$-1 errors reported in Table \ref{tab:vortex}. Our errors are about an order of magnitude lower than those reported \cite{Balsara04} and show almost exactly second-order convergence. Furthermore, the $L$-2 norm error, $\vert \vert \nabla \cdotp \mathbf{B} \vert \vert_2$, never exceeds $6.98 \times 10^{-18}$ for the $N=50$ test.

\begin{table}
\caption{$L$-1 Norm Errors for MHD Vortex \label{tab:vortex}}
\centering
\begin{tabular}{ccccccc}
\hline\hline
$N$ & $\vert \vert E(D) \vert \vert_1$ & $\vert \vert E(E) \vert \vert_1$ & $\vert \vert E(s_x) \vert \vert_1$ & $\vert \vert E(s_y) \vert \vert_1$ & $\vert \vert E(B^x) \vert \vert_1$ & $\vert \vert E(B^y) \vert \vert_1$ \\
\hline
50 & $1.54 \times 10^{-4}$ & $7.43 \times 10^{-4}$ & $5.39 \times 10^{-4}$ & $5.55 \times 10^{-4}$ & $2.16 \times 10^{-3}$ & $1.58 \times 10^{-3}$  \\
100 & $4.01 \times 10^{-5}$ & $1.79 \times 10^{-4}$ & $1.34 \times 10^{-4}$ & $1.32 \times 10^{-4}$ & $5.37 \times 10^{-4}$ & $4.00 \times 10^{-4}$  \\
200 & $1.01 \times 10^{-5}$ & $4.47 \times 10^{-5}$ & $3.36 \times 10^{-5}$ & $3.24 \times 10^{-5}$ & $1.34 \times 10^{-4}$ & $1.00 \times 10^{-4}$ \\
400 & $2.54 \times 10^{-6}$ & $1.12 \times 10^{-5}$ & $8.41 \times 10^{-6}$ & $8.08 \times 10^{-6}$ & $3.35 \times 10^{-5}$ & $2.52 \times 10^{-5}$ \\
\hline
\end{tabular}
\end{table}

\subsubsection{Current Sheet}

The current sheet test \cite{Hawley95a} is another useful multi-dimensional MHD test. It particularly challenges the robustness of codes in highly magnetized regions. In it, oppositely directed magnetic fields abut one another at multiple current sheets. An oscillating sheer flow is then set up, which forces some level of reconnection within the current sheets. Some MHD schemes go unstable and ultimately fail this test problem. 

This relativistic test is set up on a $200\times200$ grid with $0\leq{x}\leq2$, $0\leq{y}\leq2$, uniform mass density $\rho=1$ and gas pressure $P=\beta/2$. We test values in the range $10^{-3}\leq\beta\leq10^{-1}$. The velocity is initialized as $V^x=V_0\cos(\pi y)$ and $V^y=V^z=0$ with amplitude  $V_0=0.2$.  The magnetic field is initialized as $\mathcal{B}^x=\mathcal{B}^z=0$, $\mathcal{A}_x=\mathcal{A}_y=0$, $\mathcal{B}^y=-1$ and $\mathcal{A}_z = \mathcal{B}^y x$ for $0.5\leq{x}\leq1.5$, $\mathcal{B}^y=1$ and $\mathcal{A}_z = \mathcal{B}^y (x-1)$ for $x < 0.5$, and $\mathcal{B}^y=1$ and $\mathcal{A}_z = \mathcal{B}^y (x-3)$ for $x > 1.5$.

\begin{figure}
\includegraphics[width=0.33\linewidth]{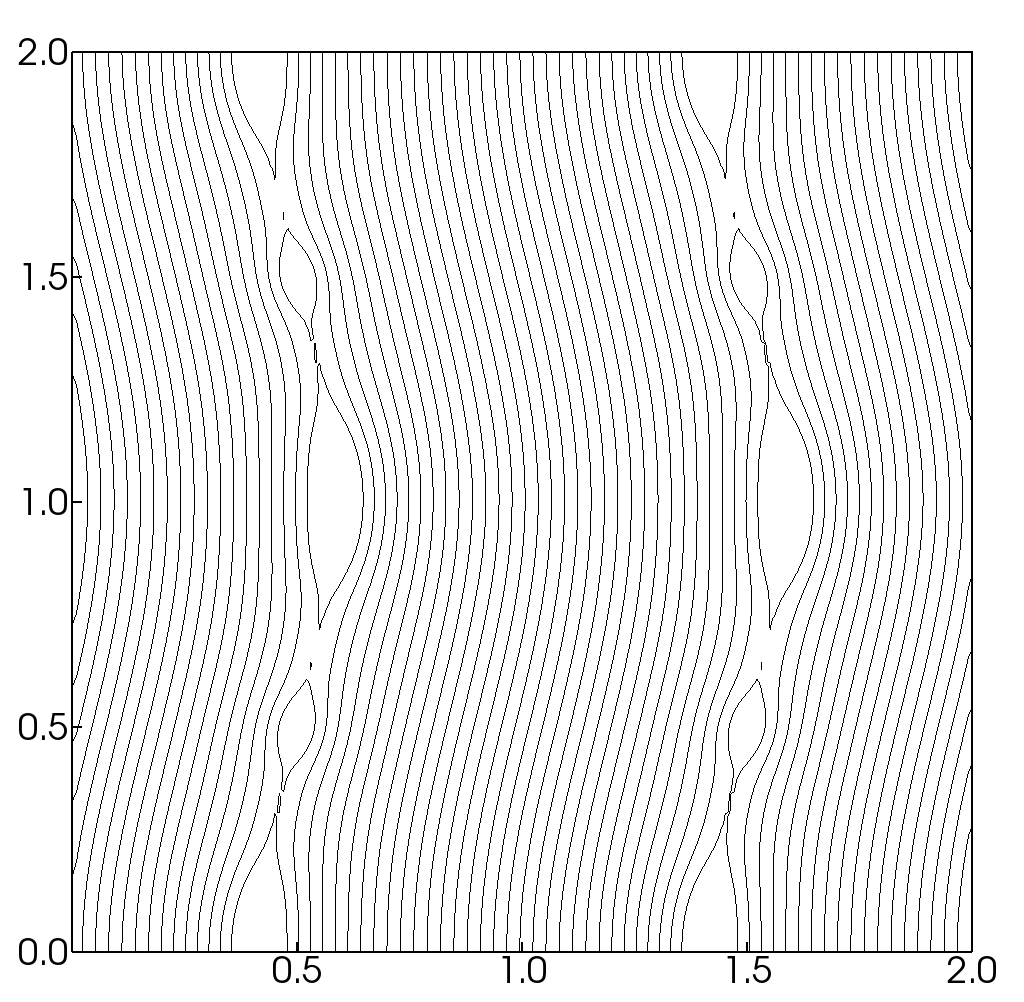}
\includegraphics[width=0.33\linewidth]{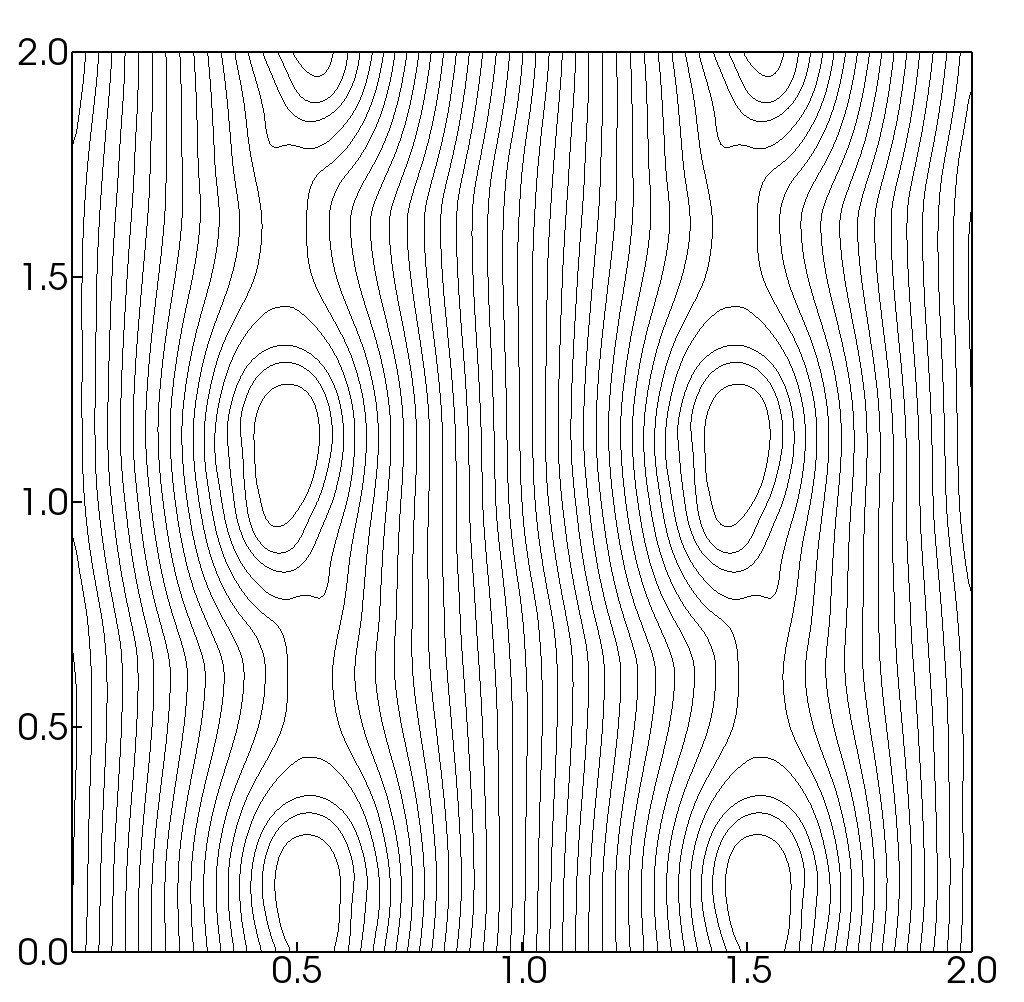}
\includegraphics[width=0.33\linewidth]{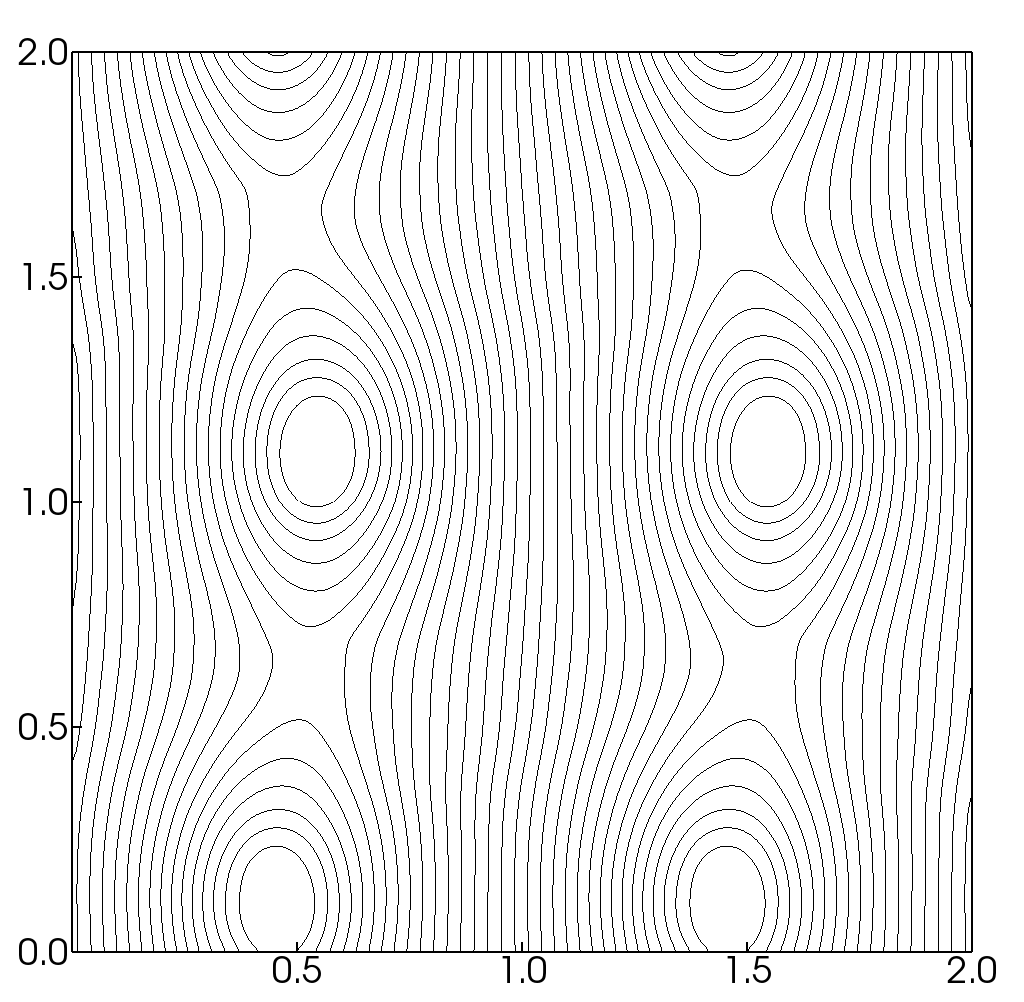}
  \caption{Contours of $A^z$ (equivalent to magnetic field lines) for a $\beta = 0.01$ current sheet at times $t = 2.0$ (left), 5.0 (center), and 10.0 (right).}
  \label{fig:current_sheet}
\end{figure}

The evolution of this test for $\beta=0.01$ is shown in Fig. \ref{fig:current_sheet}. Along the current sheets, magnetic islands begin to form as grid-scale reconnection occurs. As the simulation proceeds, smaller magnetic islands merge into larger structures until a stable configuration of four islands is reached. At no point does the $L$-2 norm error, $\vert \vert \nabla \cdotp \mathbf{B} \vert \vert_2$, exceed $5.26 \times 10^{-19}$ for this test. Only for very strong magnetic fields ($\beta < 10^{-3}$ for $V_0=0.2$) or high amplitudes ($V_0>0.5$ for $\beta < 10^{-1}$) are no stable solutions reached. This range of parameter space is very similar to what is achieved with the HLLE Riemann solver in Athena \cite{Beckwith11}.  

\subsubsection{Field-loop Advection}
\label{sec:field_loop}

The field-loop advection test \cite{Beckwith11} is another multi-dimensional MHD test that has been used to demonstrate failure modes of some MHD codes \cite{Gardiner05}. In this case, some schemes designed to preserve a divergence-free field yield unphysical results owing to an incorrect treatment of the Lorentz force. The failure manifests itself as an anomalous build up of fluid velocity (acceleration) in the direction normal to the field loop propagation. 

We initialize the relativistic version of this problem \cite{Beckwith11}, setting the mass density $\rho=1$ and the gas pressure $P=3$. The problem is conducted on a $2N\times{N}$ grid with $N=128$, $-1 \le x \le 1$, and $-0.5 \le y \le 0.5$. The velocity in the $x$-direction is set to $V^x=0.2/\sqrt{6}$,  while the velocity in the $y$-direction is set to $V^y=0.1/\sqrt{6}$. The velocity in the $z$-direction is also set to $V^z=0.1/\sqrt{6}$ and should remain constant.  The magnetic field is initialized as 
\begin{equation}
\mathcal{A}_z=
\begin{cases}
\mathcal{A}_0(R-r), & r\leq{R}~, \\
0, & r>R ~,
\end{cases}
\label{eq:Az}
\end{equation}
with $\mathcal{A}_0=10^{-3}$, $R=0.3$, and $r=\sqrt{x^2+y^2}$.

\begin{figure}
\includegraphics[width=0.33\linewidth]{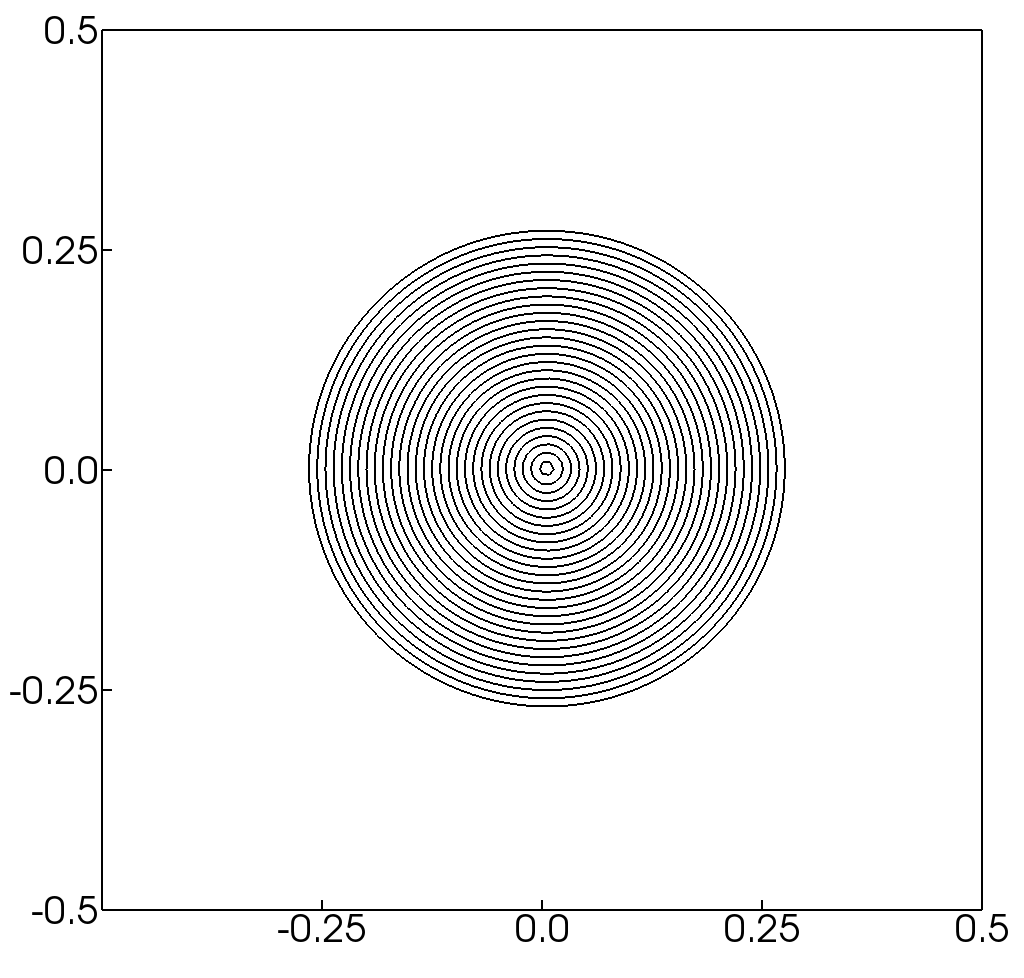}
\includegraphics[width=0.33\linewidth]{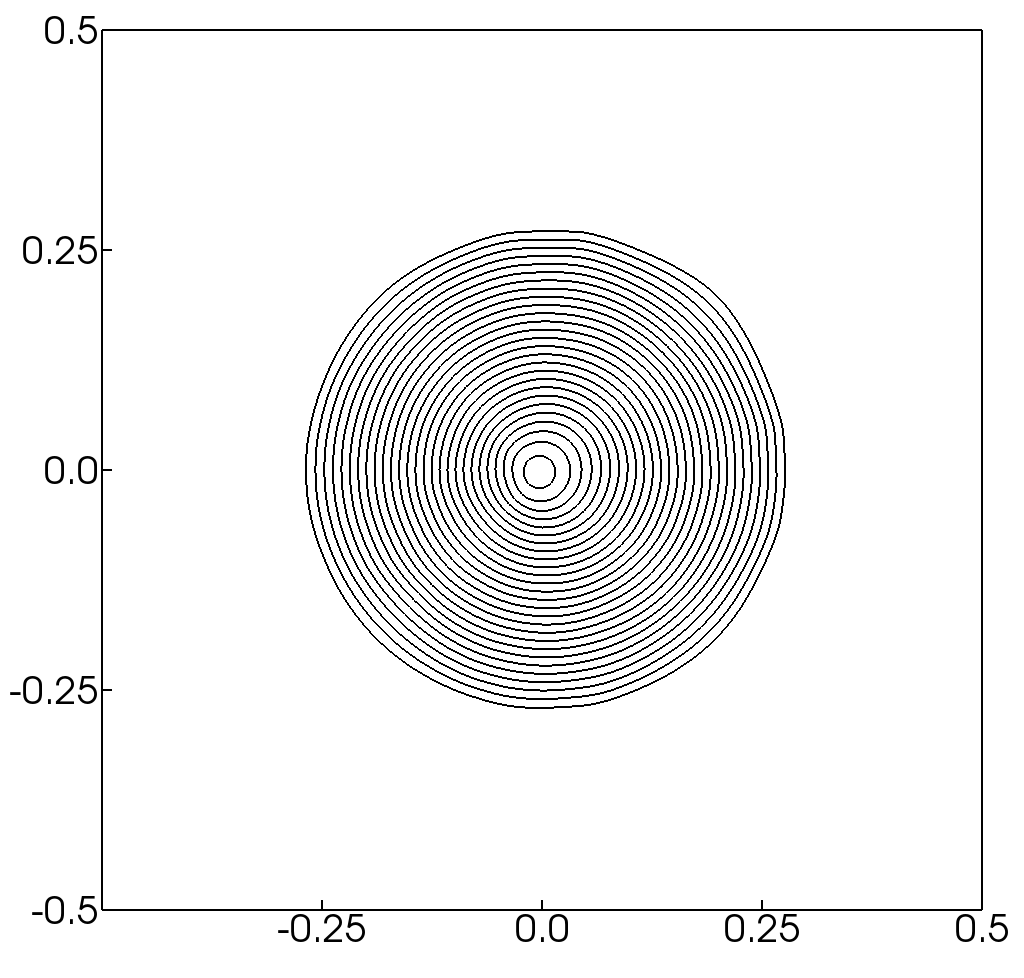}
\includegraphics[width=0.33\linewidth]{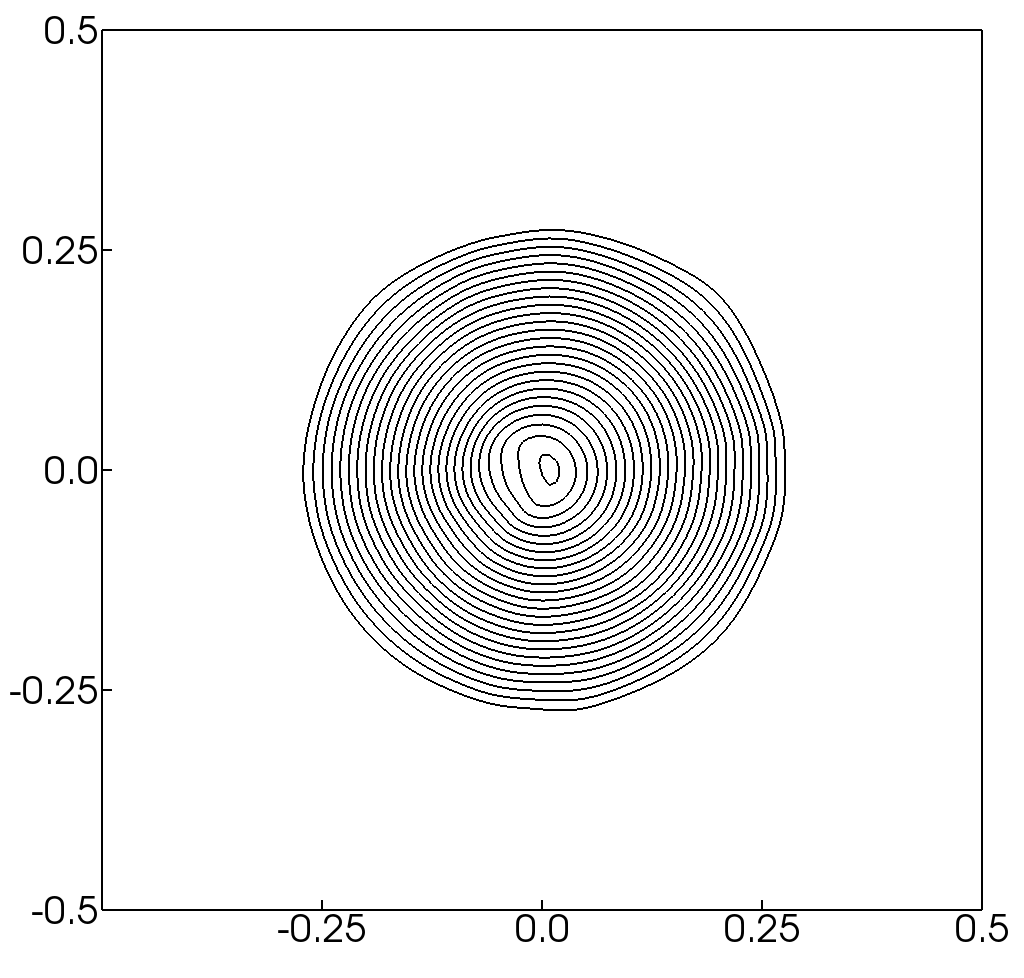}
\caption{Contours of $\mathcal{A}_z$ for the field loop advection test on a $256\times 128$ grid after zero (left), one (center), and two (right) grid crossings. $A^z$ is plotted using 30 linearly spaced contours between $3.0\times10^{-5}$ and $3.0\times10^{-4}$.} 
  \label{fig:field_loop}
\end{figure}

The field loop propagates in a periodic box, so should return to its starting location undisturbed. Fig. \ref{fig:field_loop} displays the magnetic field lines after zero, one, and two grid crossings. The lines are slightly distorted as they propagate across the grid, but the general shape and structure of the field loop is preserved. Further, the total magnetic energy on the grid drops by $< 13$\%. More importantly, we see no significant increase in $V^z$; after two grid crossings, its fractional error is $\delta V^z/V^z \le 10^{-7}$, and the $L$-2 norm error of divergence, $\vert \vert \nabla \cdotp \mathbf{B} \vert \vert_2$, remains below $6.40 \times 10^{-16}$.

\subsubsection{Orszag-Tang}
\label{sec:orszag}

\begin{figure}
\includegraphics[width=0.235\linewidth]{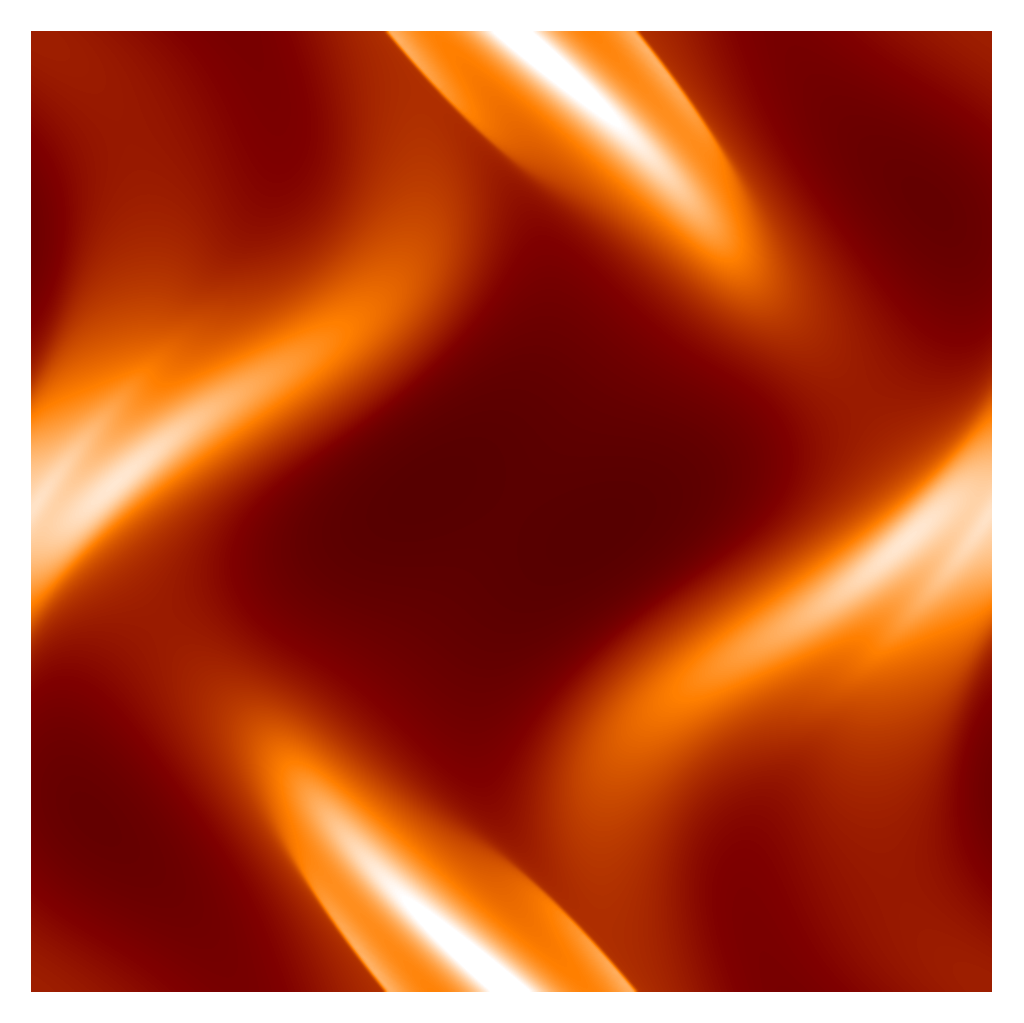}
\includegraphics[width=0.235\linewidth]{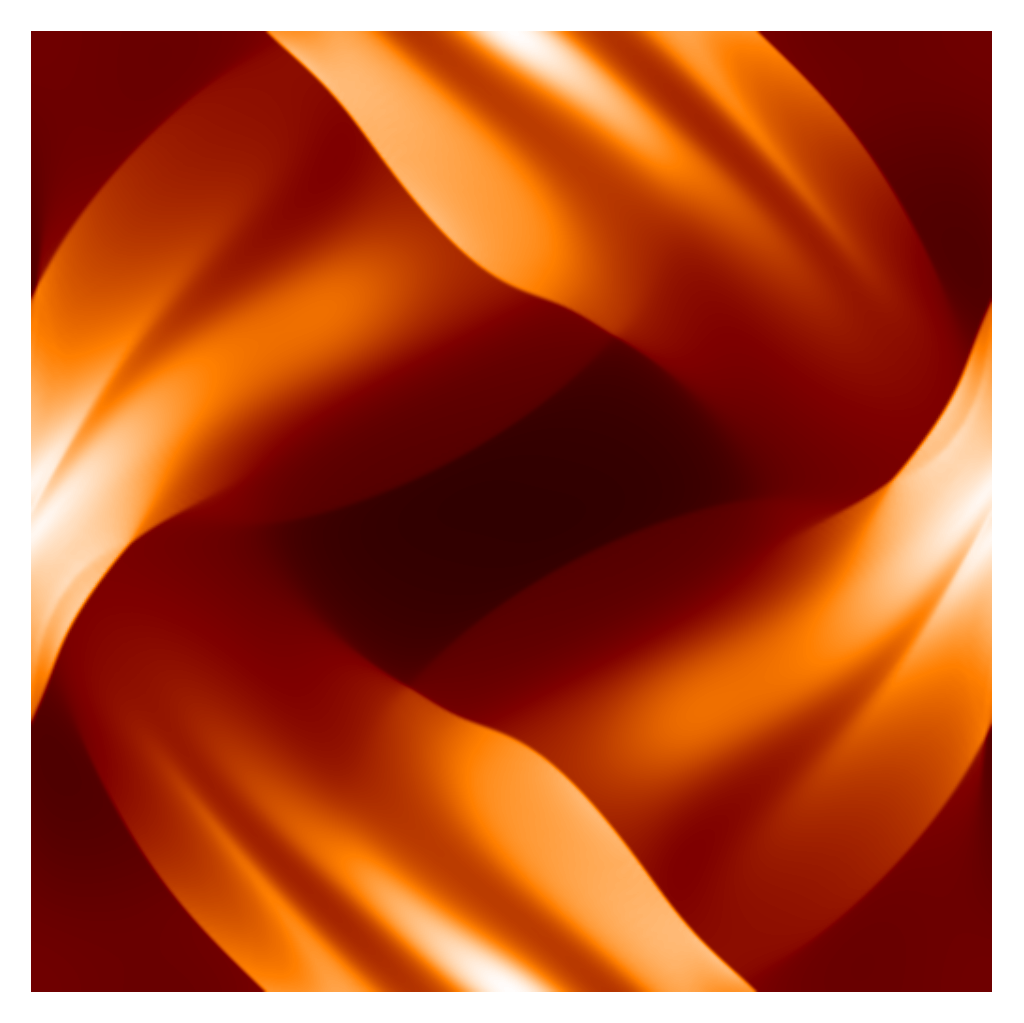}
\includegraphics[width=0.235\linewidth]{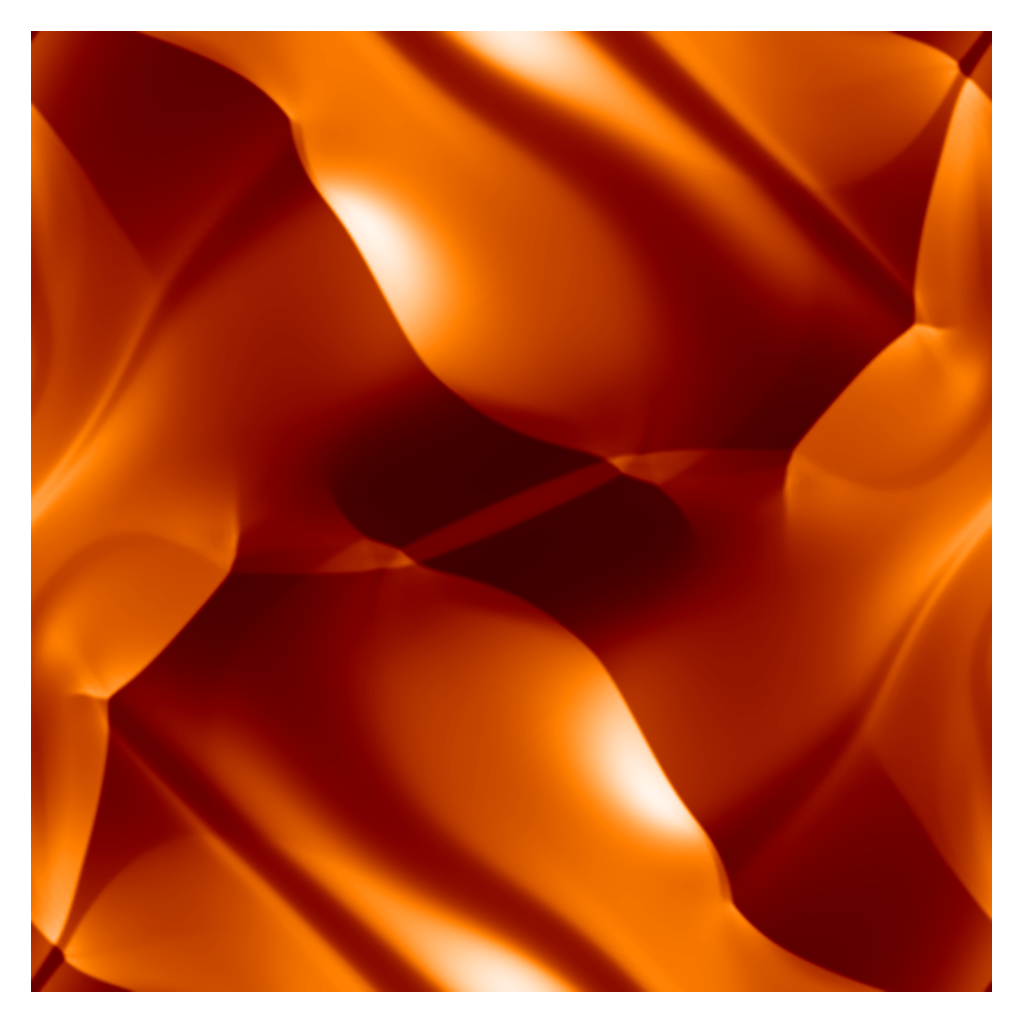}
\includegraphics[width=0.275\linewidth]{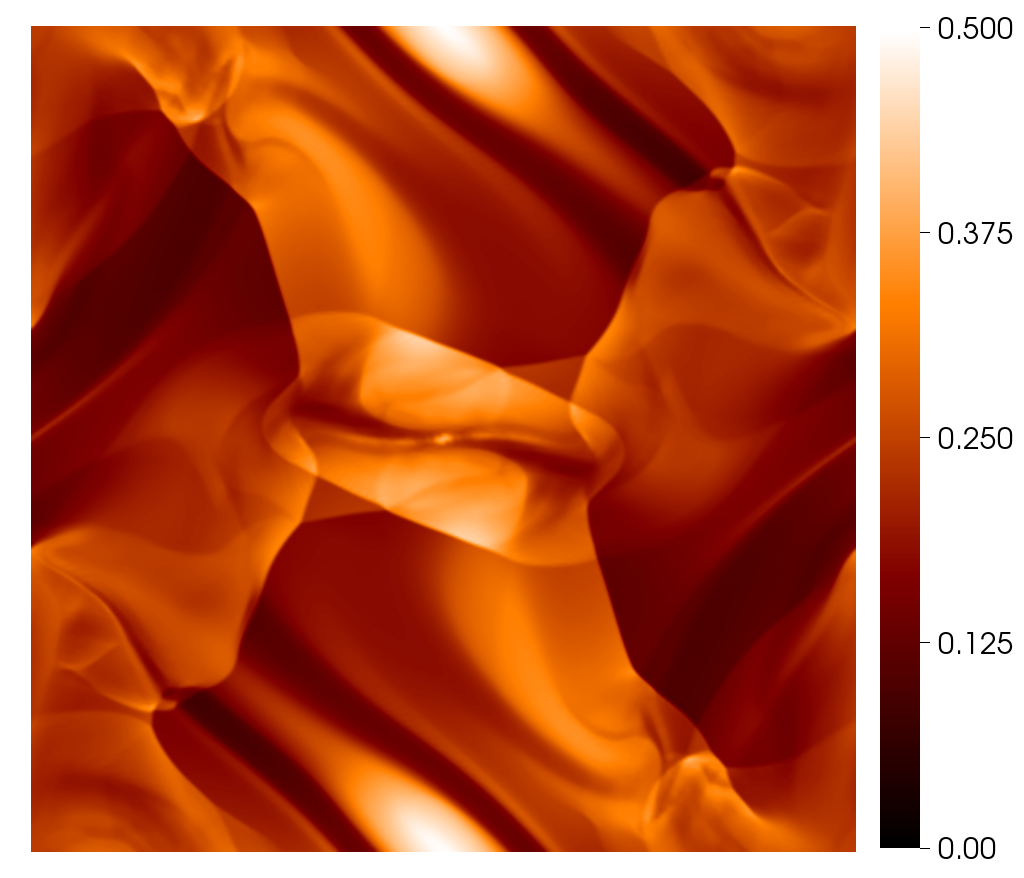}
\\
\includegraphics[width=0.235\linewidth]{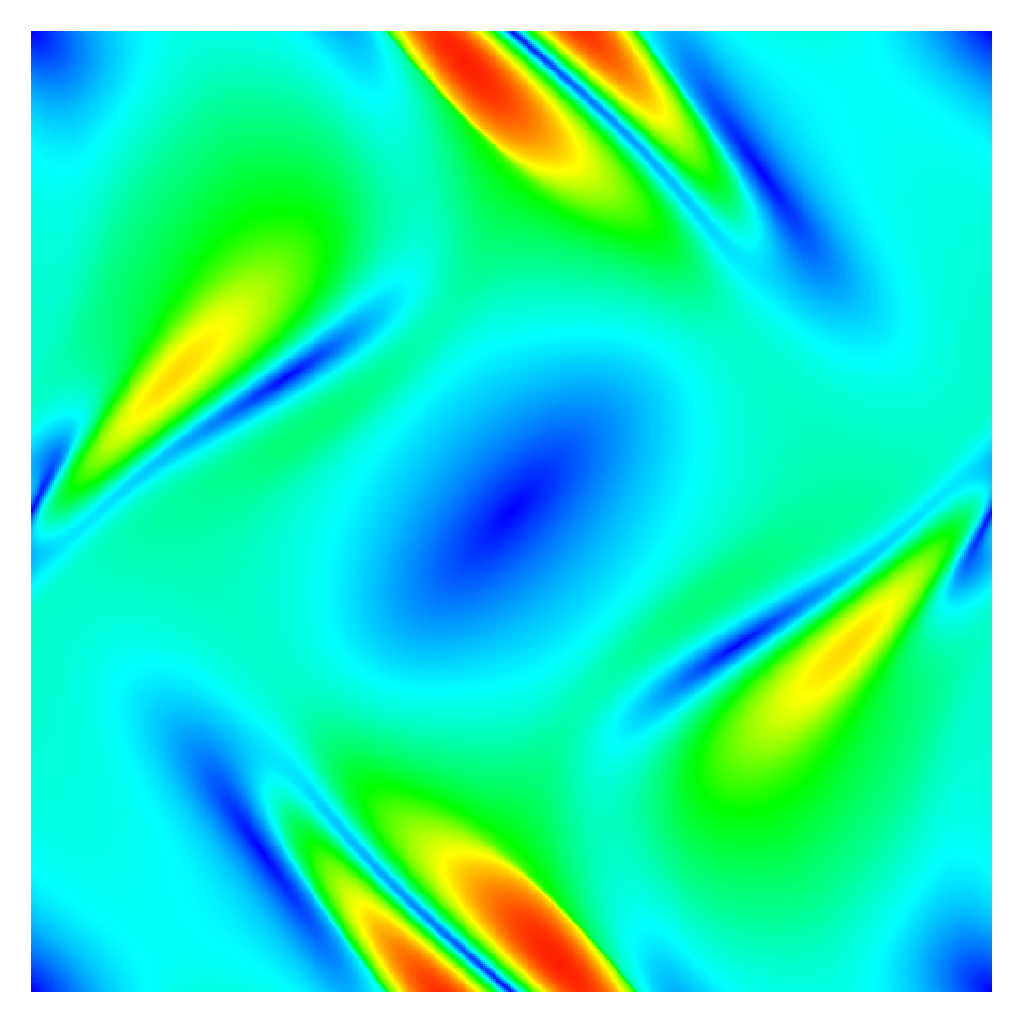}
\includegraphics[width=0.235\linewidth]{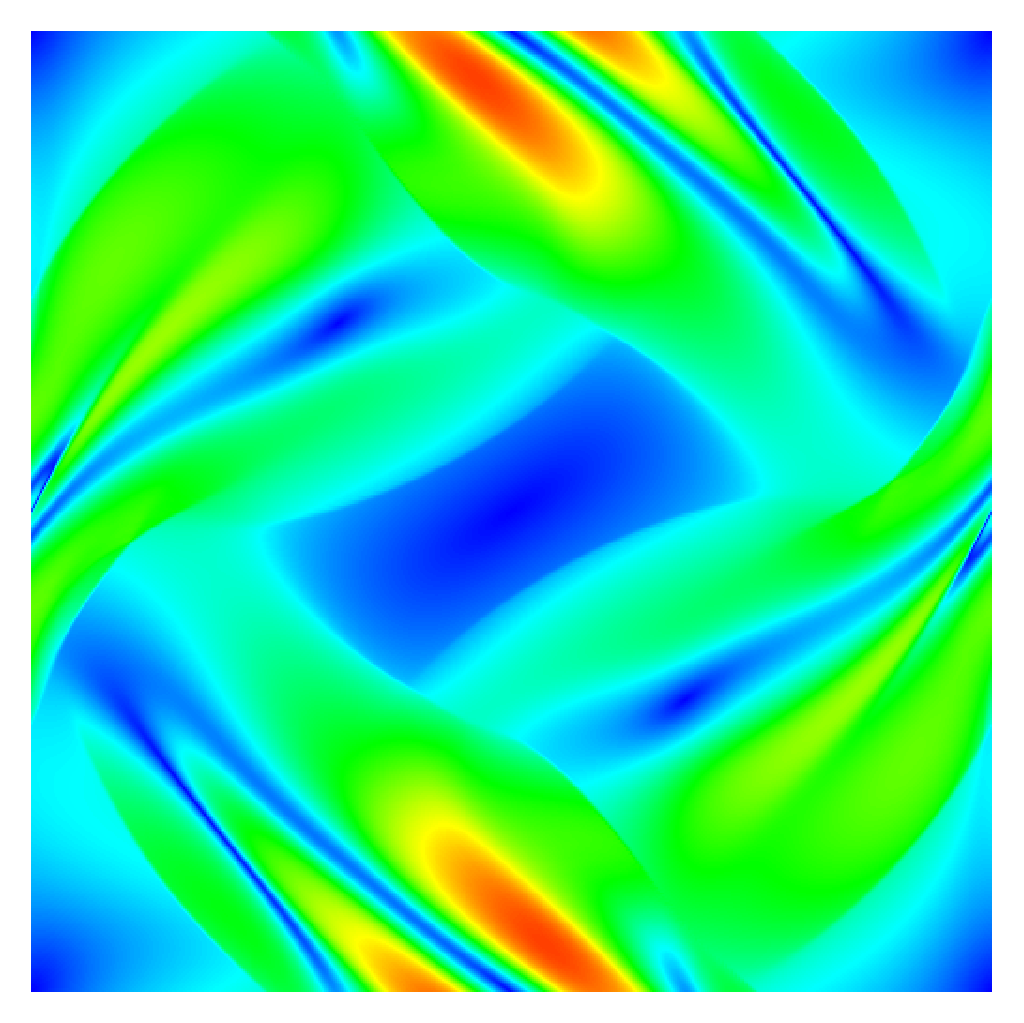}
\includegraphics[width=0.235\linewidth]{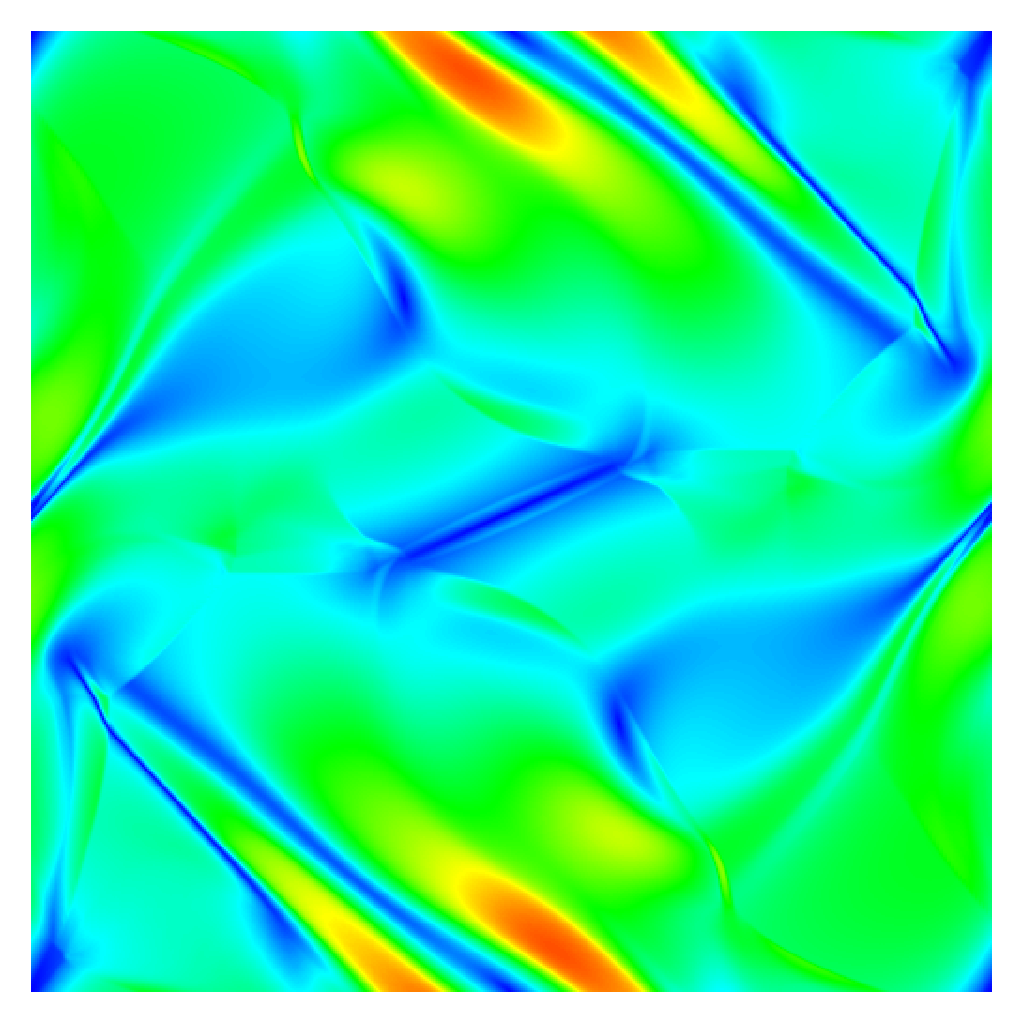}
\includegraphics[width=0.275\linewidth]{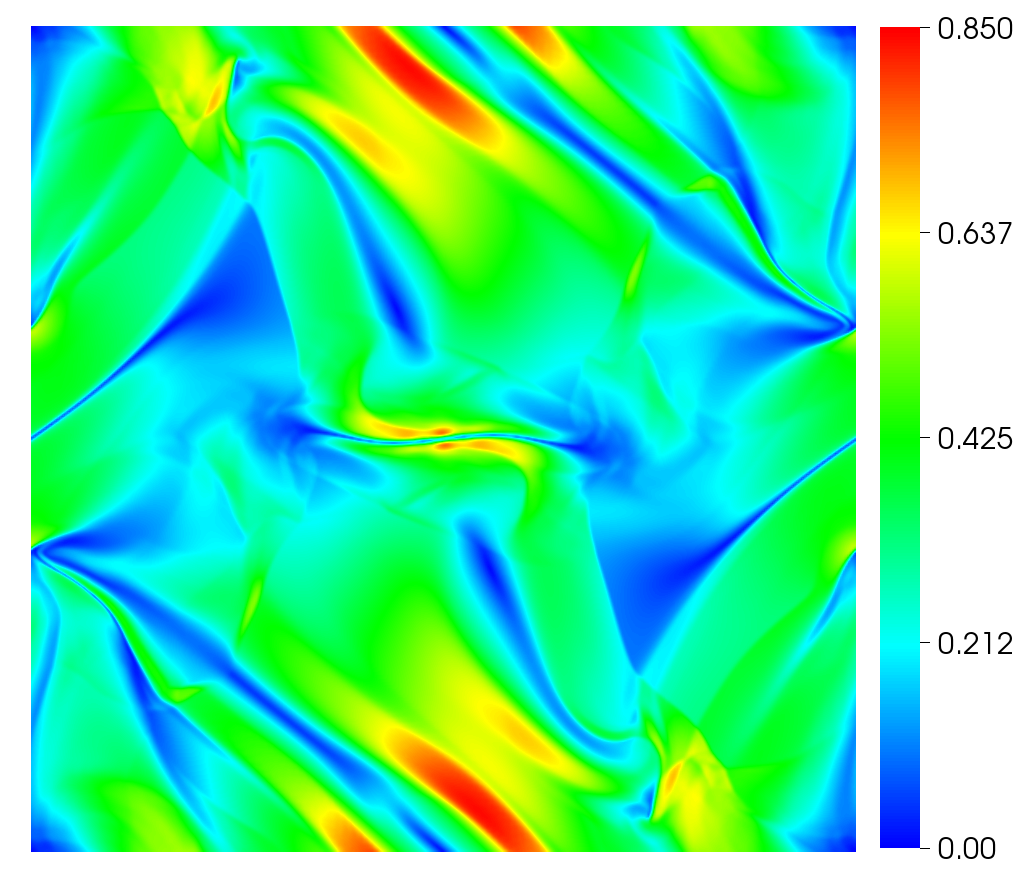}
\caption{Evolution of the mass density $\rho$ ({\it top}) and magnetic field magnitude $\vert \mathbf{B} \vert$ ({\it bottom}) at times $t = 0.15$, 0.25, 0.35, and 0.5 for the Orszag-Tang problem (resolution $512 \times 512$) using the VP scheme.}
  \label{figure:ot_evolution}
\end{figure}

The Orszag-Tang problem \citep{Orszag79} is another MHD vortex problem that has been adopted as one of the standard test problems for MHD. Unlike the smooth MHD vortex above, the Orszag-Tang problem produces turbulence and shocks. To initialize it, we set the mass density to $\rho=\Gamma^2/4 \pi$, the gas pressure to $P=\Gamma/4 \pi$, and the adiabatic index to $\Gamma=5/3$. The initial velocity and magnetic fields are functions of the position, which creates the vortex: $\mathbf{v}=\{-\sin(2 \pi y\}, \sin(2 \pi x))$, and $\mathbf{B}=\{-\sin(2 \pi y), \sin(4 \pi x)\}$.  The evolution of the mass density and magnetic field using the VP scheme at $512 \times 512$ resolution is shown in Fig. \ref{figure:ot_evolution}. These results are consistent with previously published solutions \cite[e.g.][]{Dai98,Ryu98,Londrillo00,Toth00,Mocz14} and have very low divergence error, with the $L$-2 norm error, $\vert \vert \nabla \cdotp \mathbf{B} \vert \vert_2$, never exceeding $3.11 \times 10^{-20}$. Notably, the solution remains symmetric, which is significant given the face-centering of our conserved magnetic field components and edge centering of our vector potential components. 

\subsubsection{Rotor}
\label{sec:rotor}

The next test looks at torsional \alf waves given off by a rotating cylinder (the rotor) of fluid threaded by a magnetic field. The magnetic field transfers angular momentum from the rotor to the background fluid. As the field wraps around the rotor, it also distorts its shape. 

\begin{figure}
\includegraphics[width=0.283\linewidth]{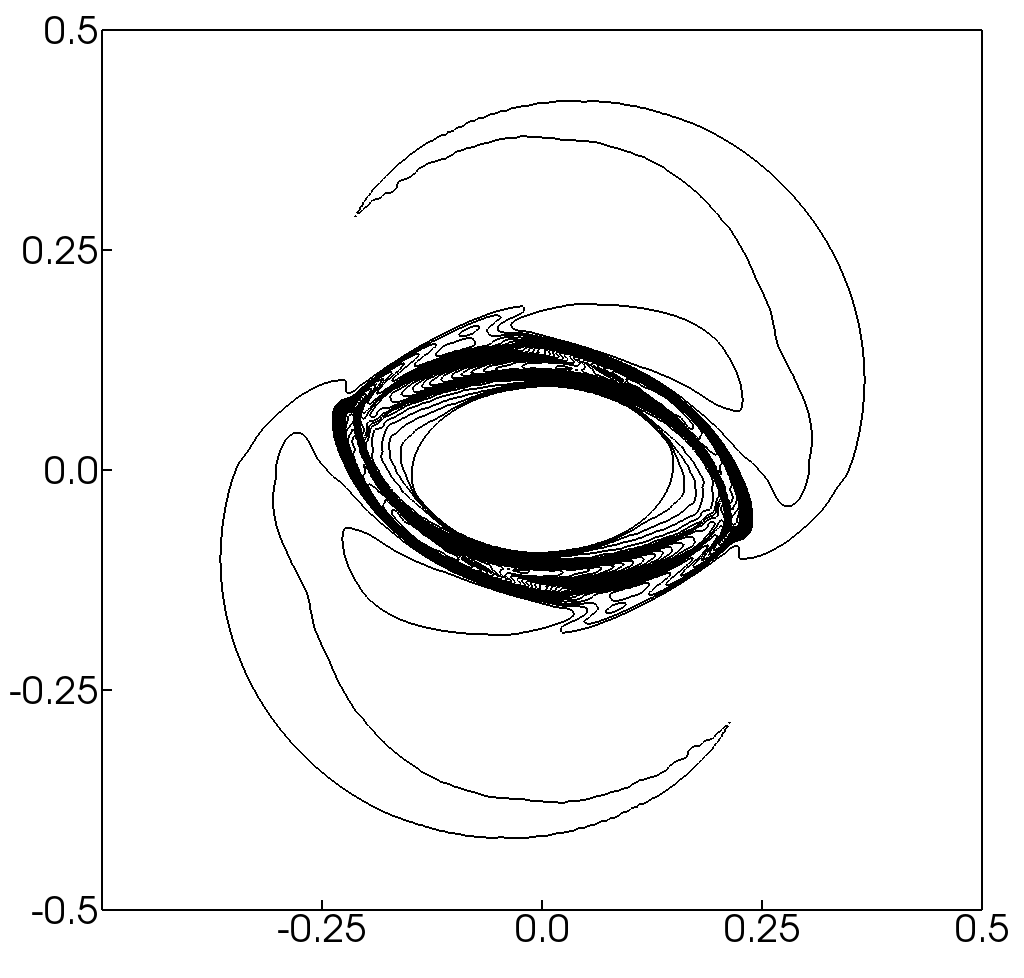}
\includegraphics[width=0.283\linewidth]{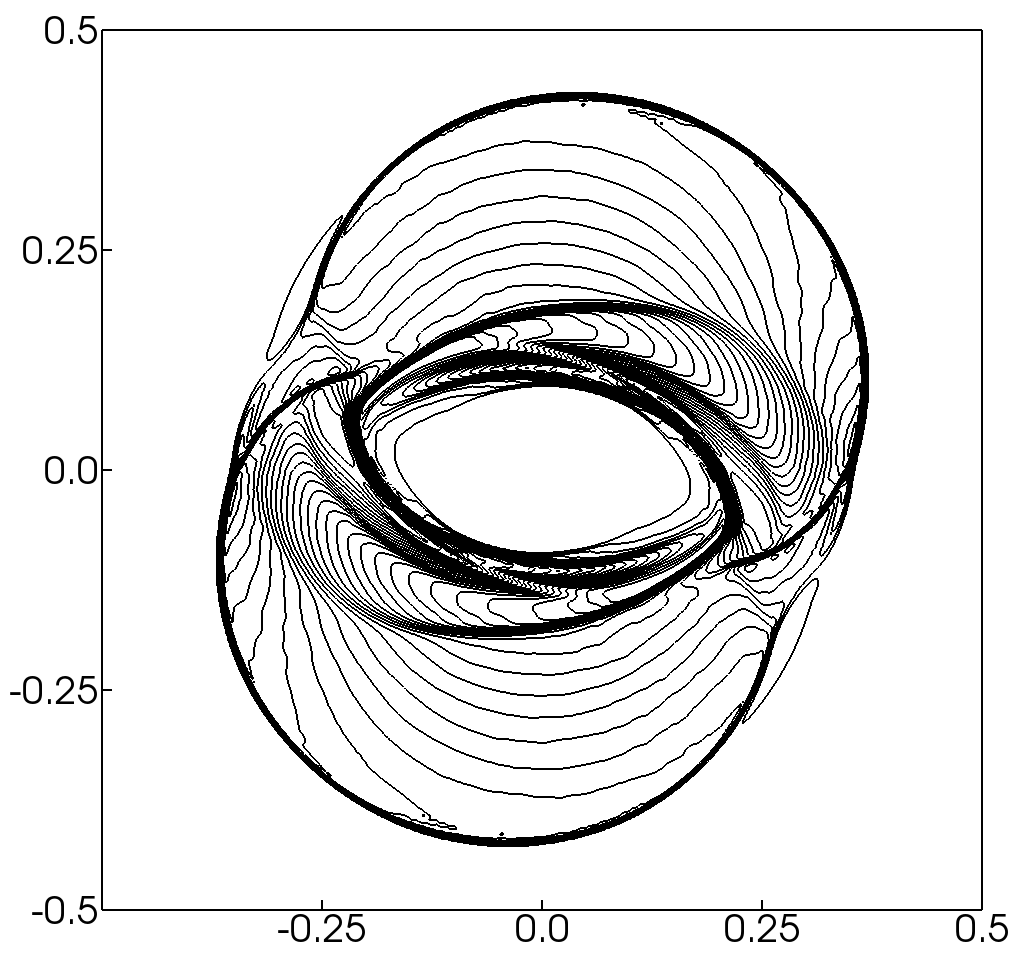} \\
\includegraphics[width=0.283\linewidth]{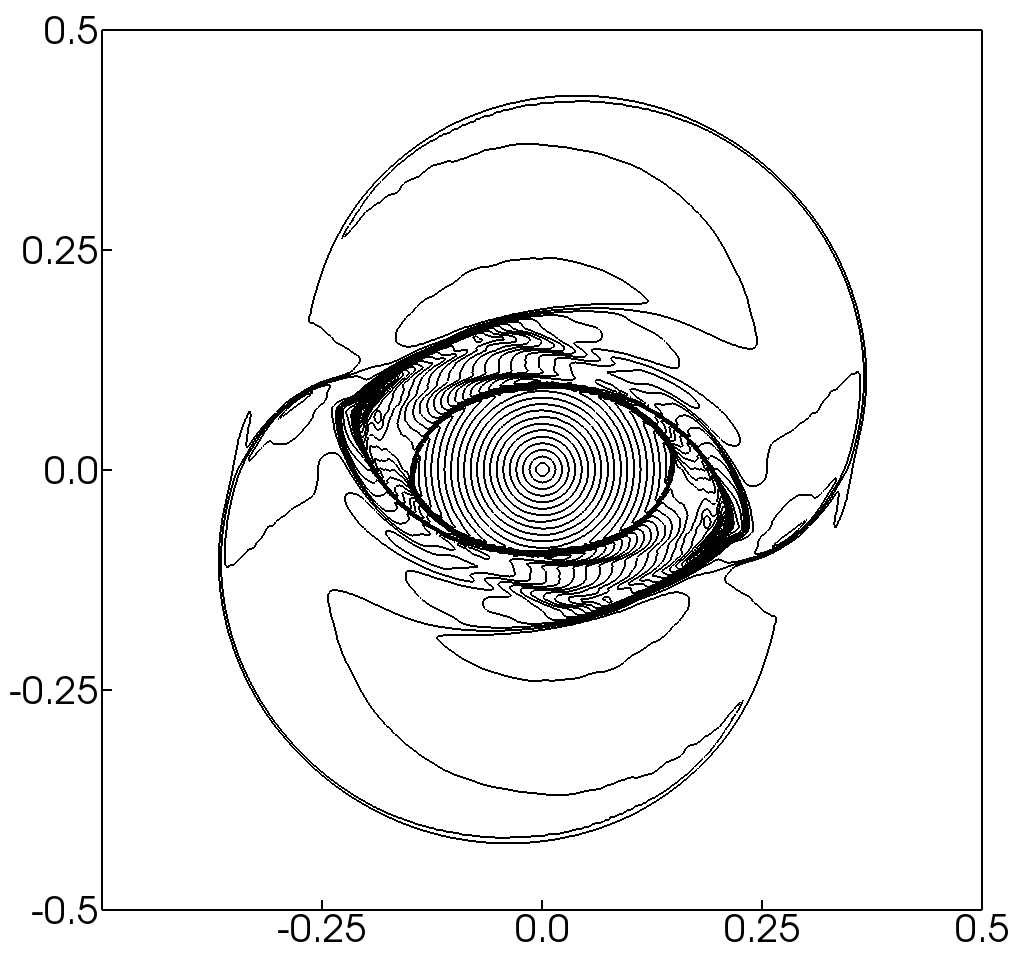}
\includegraphics[width=0.283\linewidth]{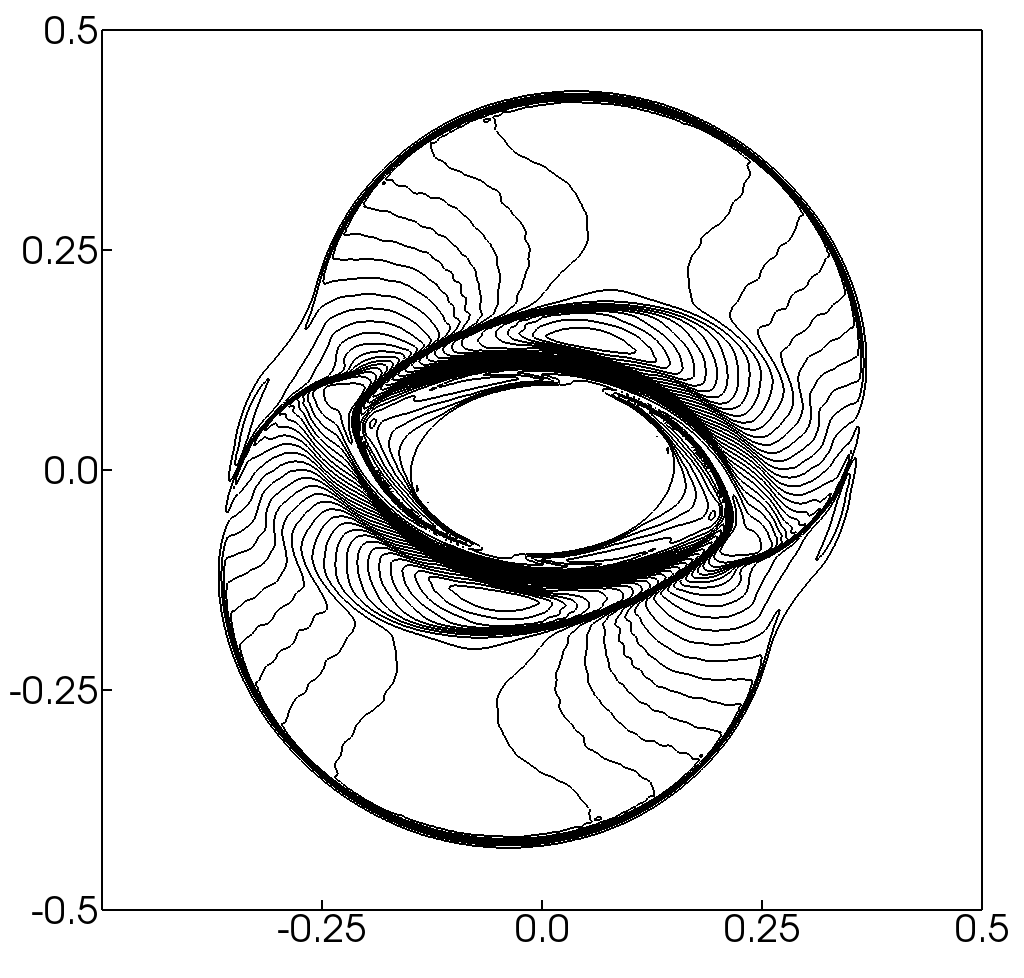}
\caption{Thirty evenly spaced contours of the mass density $\rho$ ({\it top, left}), pressure $P$ ({\it top, right}), Mach number ({\it bottom, left}), and magnetic pressure $P_B$ ({\it bottom, right}) for the rotor problem (resolution $400 \times 400$) using the VP scheme.}
  \label{fig:rotor}
\end{figure}

This problem is set up on a two-dimensional unit square, resolved with $400 \times 400$ zones. A static  ($\mathbf{v} = \mathbf{0}$) background is set with $\rho = 1$, $P = 1$, and $\Gamma = 1.4$. The rotor is initialized in the center of the grid with a density of $\rho = 10$ inside $r < 0.1$, tapering off to the background value over a distance of $\Delta r = 0.015$. Similarly, the rotor has a uniform angular frequency of 20 inside $r < 0.1$, tapering off to zero over the same $\Delta r = 0.015$. A uniform magnetic field in the $x$-direction of magnitude $B^x = 5/\sqrt{4\pi}$ threads the entire domain. This is initialized from a vector potential of the form ${\mathcal A}_z = B^x y$. The problem is run to a time $t = 0.15$. Figure \ref{fig:rotor} shows contours of the final density, gas pressure, Mach number, and magnetic pressure. The results look quite comparable to previously published results \cite{Balsara99,Toth00}, and again the $L$-2 norm error of divergence, $\vert \vert \nabla \cdotp \mathbf{B} \vert \vert_2$, remains low ($<1.68 \times 10^{-19}$).

\subsubsection{Blast}
\label{sec:blast}

Our final test problem is the explosion of a centrally over pressurized region into a low pressure, low $\beta$ ambient medium. Like the Orszag-Tang problem, this test combines shocked flows and strong magnetic fields. Like the rotor problem, there is also a smooth flow region. Similar to those tests, the results are not particularly quantitative in their measure of the accuracy, but this test adds to our confidence in the robustness of the method. Variants on this problem have been presented by a number of authors \citep[e.g.,][]{Balsara99,Londrillo00,Gardiner05}.

\begin{figure}
\includegraphics[width=0.283\linewidth]{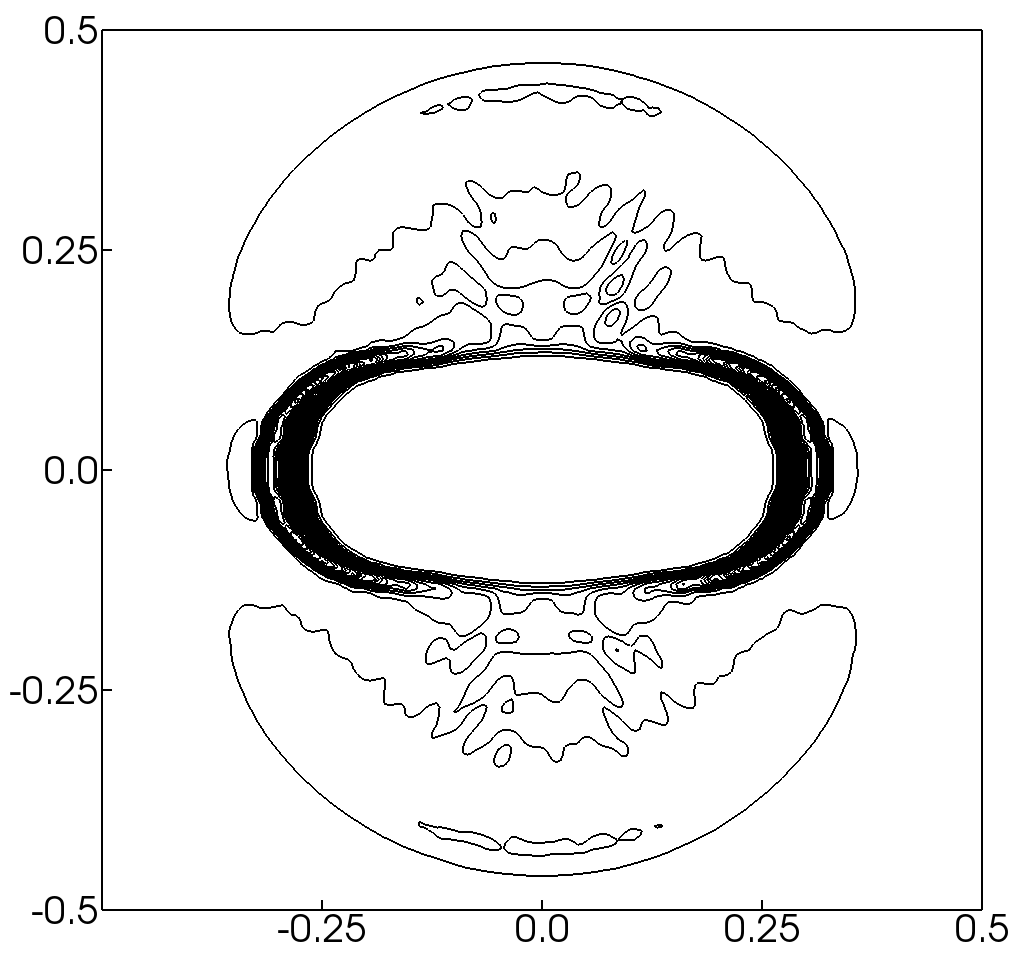}
\includegraphics[width=0.283\linewidth]{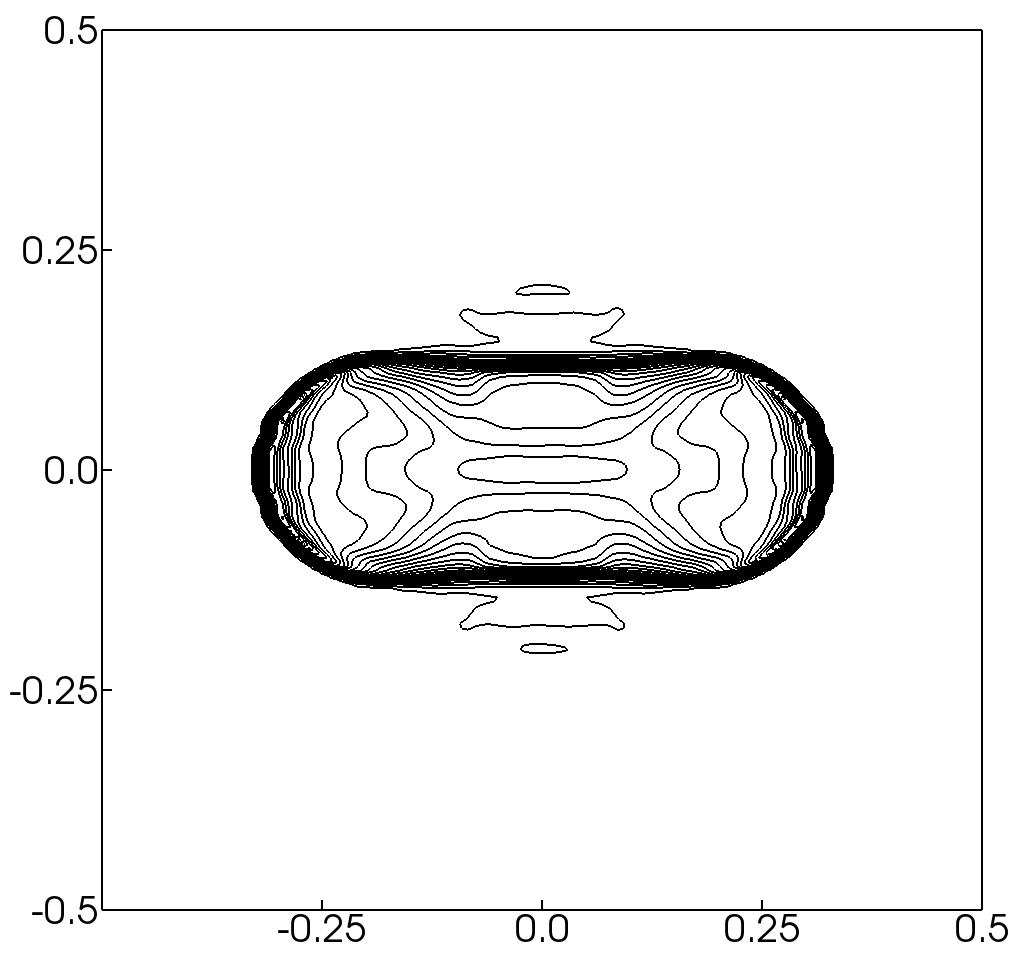} \\
\includegraphics[width=0.283\linewidth]{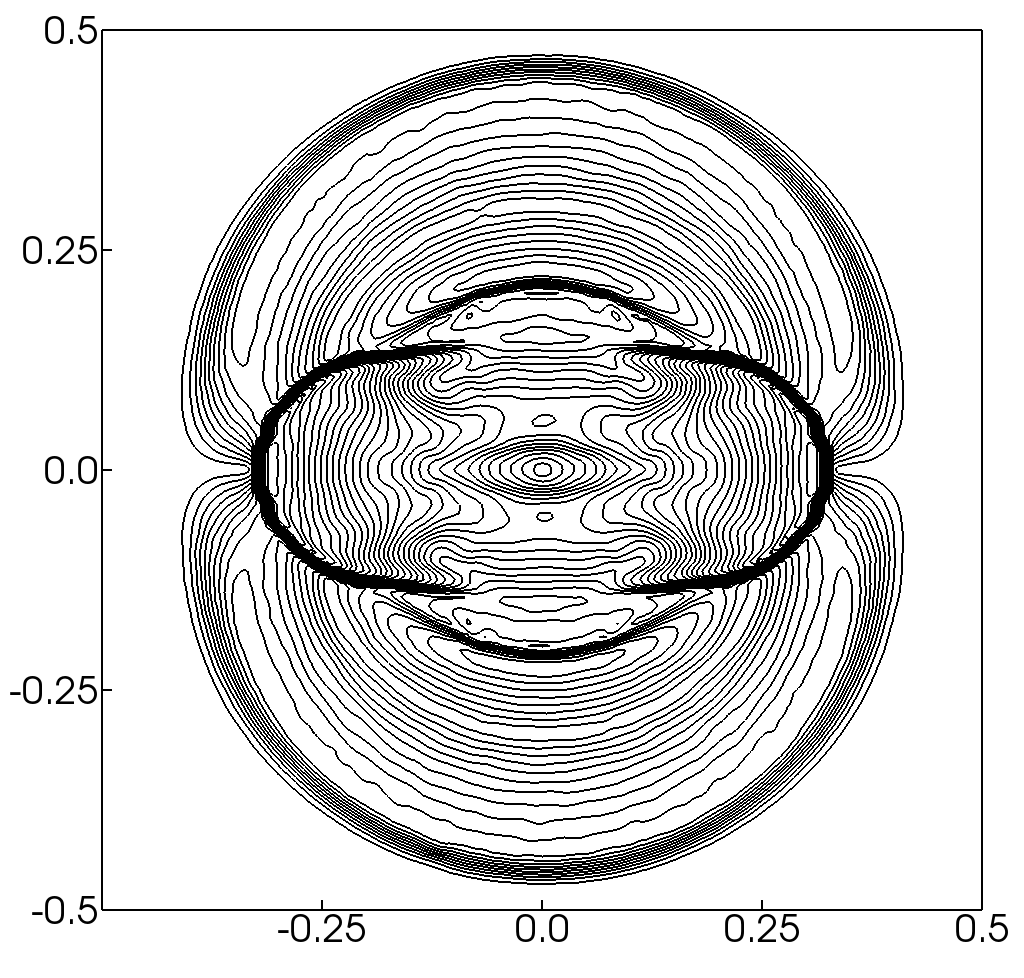}
\includegraphics[width=0.283\linewidth]{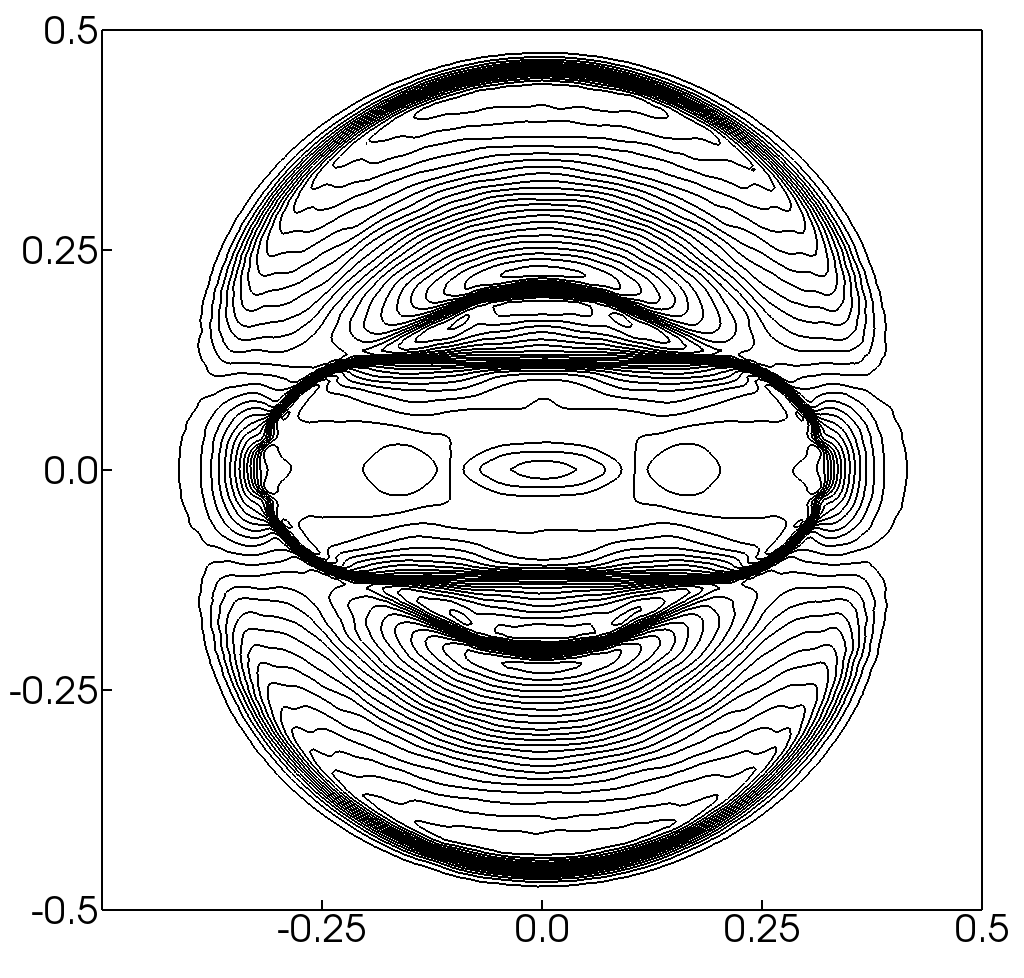}
\caption{Thirty evenly spaced contours of the mass density $\rho$ ({\it top, left}), pressure $P$ ({\it top, right}), velocity magnitude $\vert \mathbf{v} \vert$ ({\it bottom, left}), and magnetic pressure $P_B$ ({\it bottom, right}) for the blast problem (resolution $200 \times 200$) using the VP scheme.}
  \label{fig:blast}
\end{figure}

The computational domain extends over $-0.5 \le x \le 0.5$ and $-0.5 \le y \le 0.5$. A uniform, static ($\mathbf{v} = \mathbf{0}$) background of $\rho = 1$, $P = 0.1$, and $\Gamma = 1.4$ is set everywhere. A magnetic field of $\{B^x, B^y\} = \{100/\sqrt{4\pi}, 0\}$ threads the entire domain. As in the previous problem, this is initialized with a vector potential of the form ${\mathcal A}_z = B^x y$. This gives a background $\beta$ of $2.5 \times 10^{-4}$. Within a sphere of radius $r = 0.1$, centered at the origin, we set the pressure to 1000 to initiate the blast. The problem is run to a stop time of $t = 0.01$. Figure \ref{fig:blast} shows contours of the final density, gas pressure, velocity magnitude, and magnetic pressure. The results look quite comparable to previously published results \cite{Balsara99,Gardiner08}, and once again the $L$-2 norm error of divergence, $\vert \vert \nabla \cdotp \mathbf{B} \vert \vert_2$, remains low ($<3.10 \times 10^{-19}$).

\subsection{Code Tests with Mesh Refinement}

Now that we have demonstrated that our VP method performs well on uniform, static meshes, we move on to consider problems involving refined (static and active) and moving (uniform translation) meshes.

\subsubsection{MHD Vortex Revisited}

First, we repeat the MHD vortex problem with mesh motion. We set the problem up exactly as in Section \ref{sec:vortex}, except we add a uniform grid velocity equal to the propagation speed of the vortex, $\{V^x_g, V^y_g\} = \{1, 1\}$. This causes the vortex to remain stationary within the simulation domain, allowing us to quantitatively confirm that our convergence order is preserved whenever the grid velocity in non-zero. This is confirmed in Table \ref{tab:vortex2}, where we report the $L$-1 errors for this test. These are comparable to the errors reported in Table \ref{tab:vortex} for the static mesh, and more importantly, show the same second-order convergence. Also important, the $L$-2 norm error of divergence, $\vert \vert \nabla \cdotp \mathbf{B} \vert \vert_2$, is still below roundoff ($3.41 \times 10^{-18}$ for the $N=50$ test).

\begin{table}
\caption{$L$-1 Norm Errors for MHD Vortex with Mesh Motion \label{tab:vortex2}}
\centering
\begin{tabular}{ccccccc}
\hline\hline
$N$ & $\vert \vert E(D) \vert \vert_1$ & $\vert \vert E(E) \vert \vert_1$ & $\vert \vert E(s_x) \vert \vert_1$ & $\vert \vert E(s_y) \vert \vert_1$ & $\vert \vert E(B^x) \vert \vert_1$ & $\vert \vert E(B^y) \vert \vert_1$ \\
\hline
50 & $1.02 \times 10^{-4}$ & $1.26 \times 10^{-3}$ & $8.00 \times 10^{-4}$ & $7.96 \times 10^{-4}$ & $1.57 \times 10^{-3}$ & $1.35 \times 10^{-3}$  \\
100 & $2.67 \times 10^{-5}$ & $3.27 \times 10^{-4}$ & $2.07 \times 10^{-4}$ & $2.08 \times 10^{-4}$ & $4.00 \times 10^{-4}$ & $3.43 \times 10^{-4}$  \\
200 & $6.74 \times 10^{-6}$ & $8.23 \times 10^{-5}$ & $5.23 \times 10^{-5}$ & $5.25 \times 10^{-5}$ & $1.01 \times 10^{-4}$ & $8.64 \times 10^{-5}$ \\
\hline
\end{tabular}
\end{table}

\subsubsection{Field-loop Advection Revisited}

\begin{figure}
\includegraphics[width=0.33\linewidth]{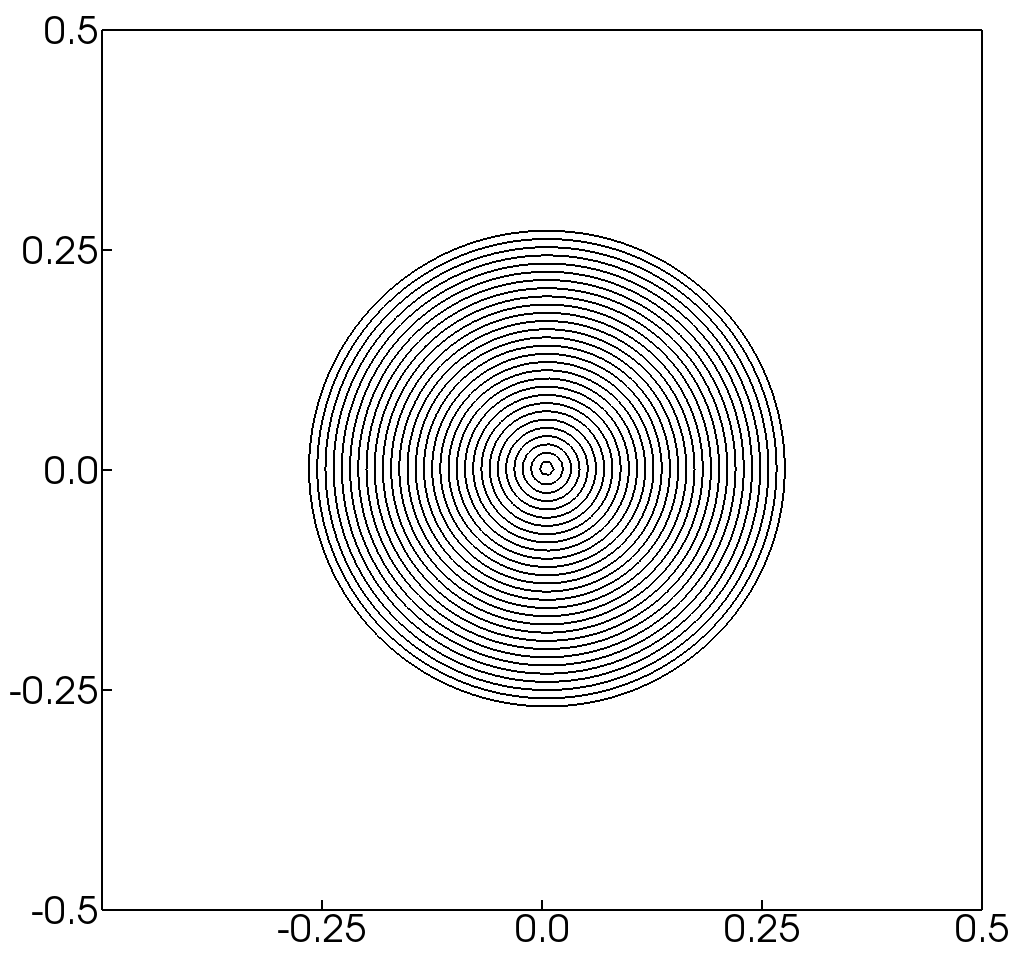}
\includegraphics[width=0.33\linewidth]{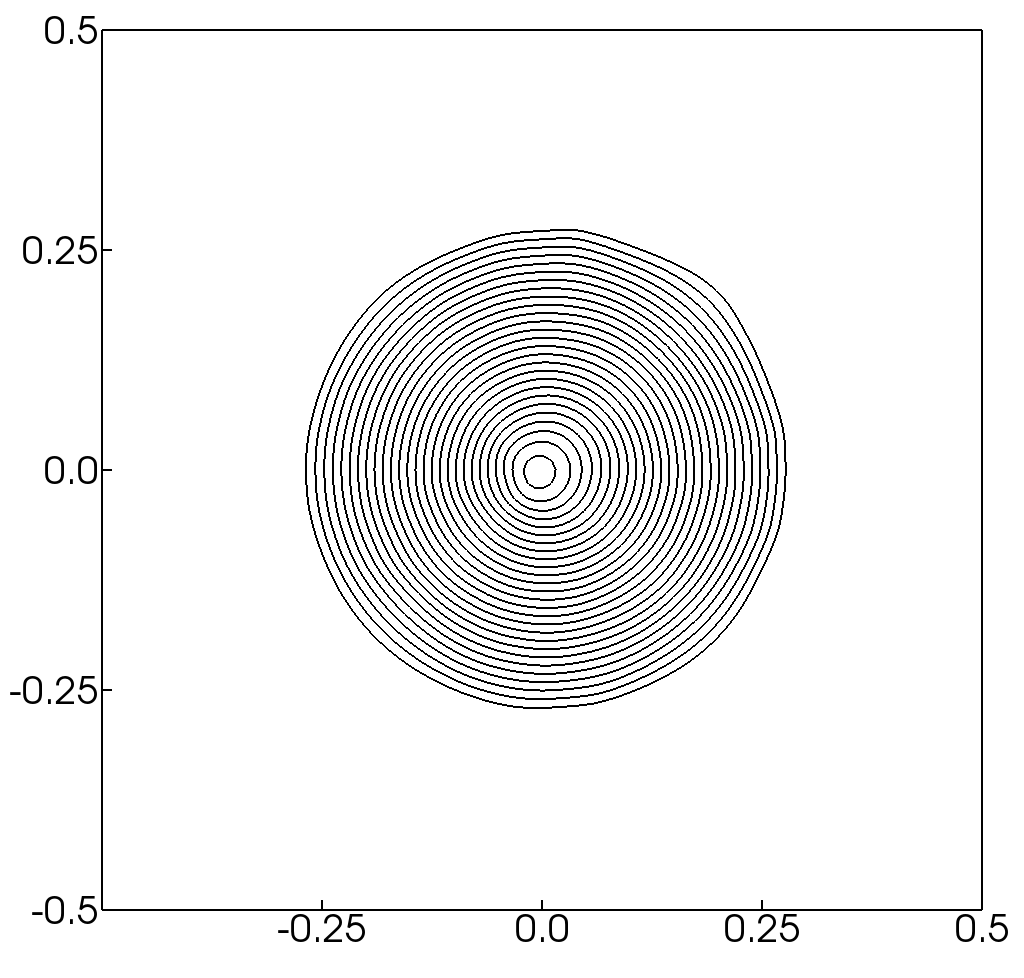}
\includegraphics[width=0.33\linewidth]{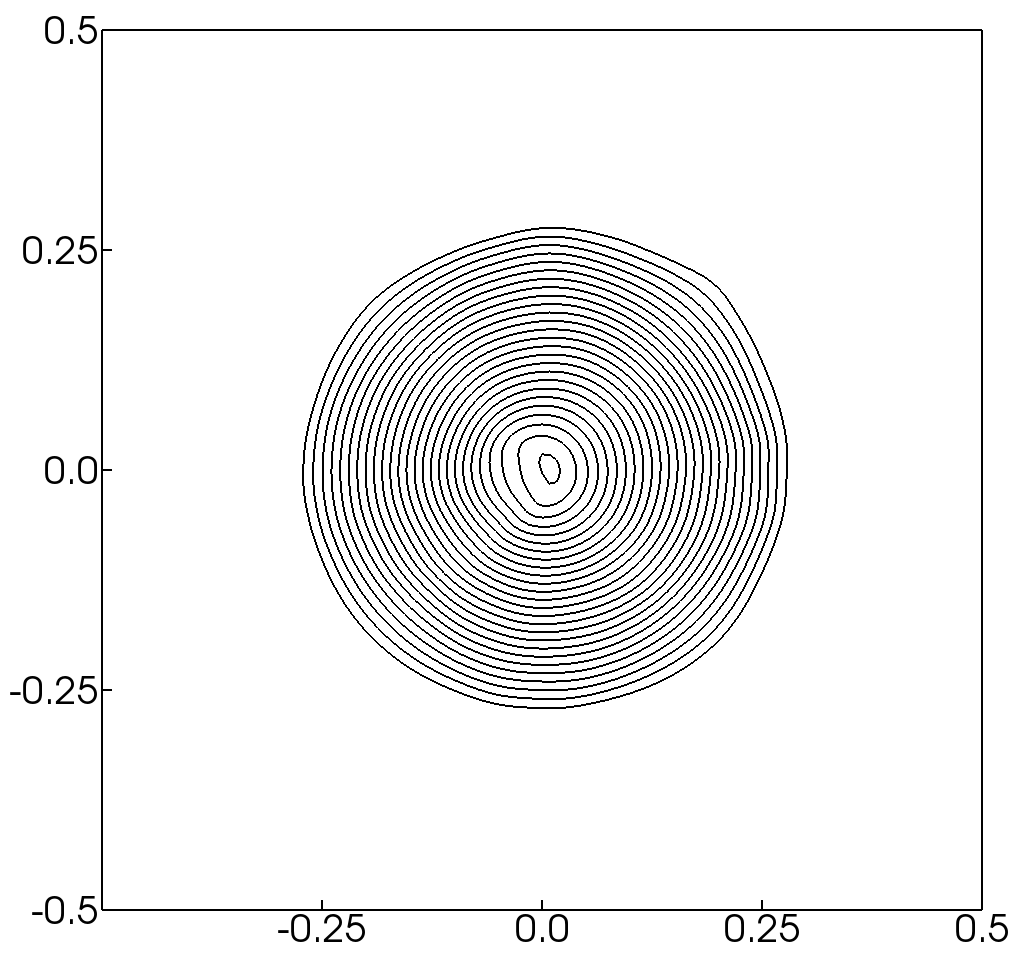}
\caption{Same as Fig. \ref{fig:field_loop} except using a $64\times 32$ base mesh with two levels of adaptive refinement (peak resolution of $256\times128$)}.
  \label{fig:field_loop_amr}
\end{figure}

Next we redo the field-loop advection test using independently both adaptive and moving mesh refinement. In Fig. \ref{fig:field_loop_amr}, we reproduce our results from Fig. \ref{fig:field_loop}, but instead of using a single-level $256\times 128$ mesh, we use a $64\times32$ base mesh with two levels of adaptive refinement. Hence, both simulations have the same peak resolution. We begin with the entire grid at the highest level of refinement to ensure proper initialization, but then allow the grid to de-refine and refine as the simulation evolves and the field loop moves. The refinement variable in this case is the magnetic pressure, with refinement set whenever $P_B > 10^{-9}$ and de-refinement whenever $P_B < 10^{-10}$. We use the strict refinement setting (Sec. \ref{sec:amr}) and test for refinement every tenth cycle. The point of this test is to demonstrate that we can achieve similar results on an adaptive, multi-layer mesh as we do on a single, static mesh of equivalent peak resolution. We note specifically that the drop in total magnetic energy is now 17\% and the error in $V^z$ is comparable with and without refinement.  

Even more so than adaptive mesh refinement, the field loop advection test lends itself to the application of moving mesh. We, therefore, set up another case where we introduce a grid velocity equal to the fixed-frame advective velocity of the field loop, i.e. $V^x_g = 0.2/\sqrt{6}$ and $V^y_g = 0.1/\sqrt{6}$. In this way, we keep the loop centered on the {\em moving} grid. We, again, let the simulation evolve for what would be two crossing periods. We find that the results are quite comparable to the fixed mesh results in Sec. \ref{sec:field_loop} with the same level of magnetic energy decay, $V^z$ errors, and $L$-2 norm error of divergence.

\subsubsection{Orszag-Tang Revisited}

\begin{figure}
\includegraphics[width=0.5\linewidth]{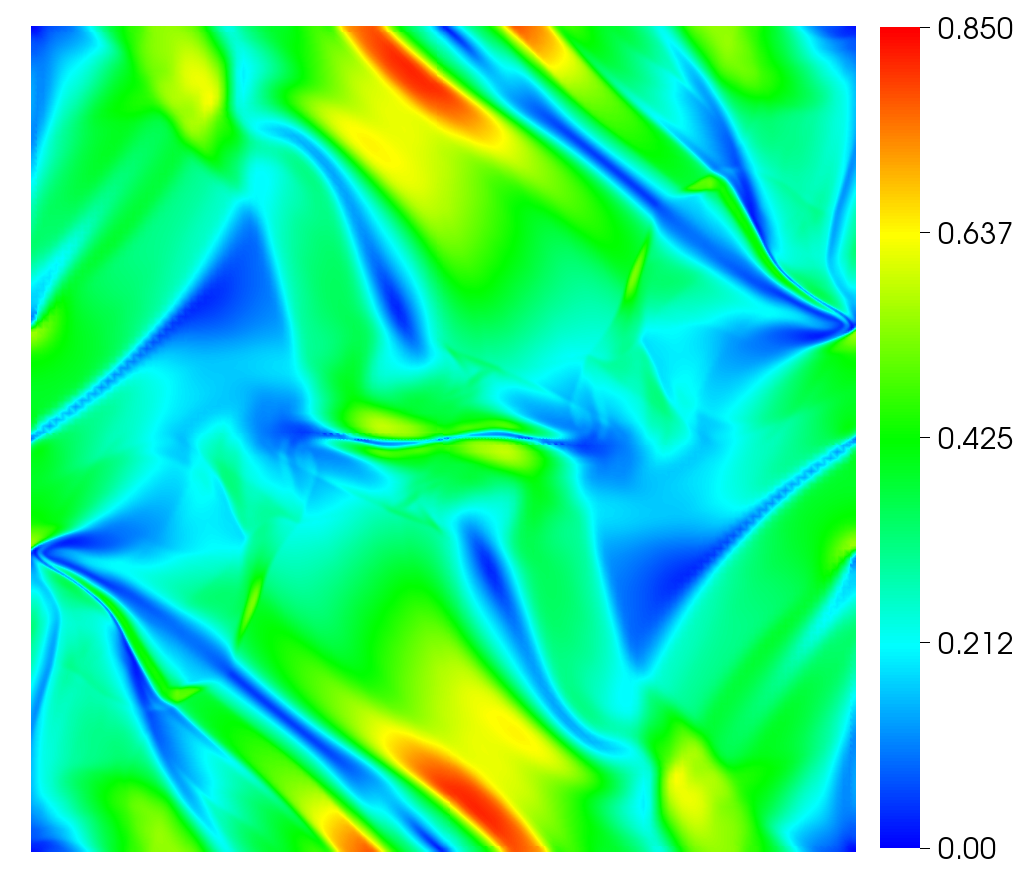}
\includegraphics[width=0.5\linewidth]{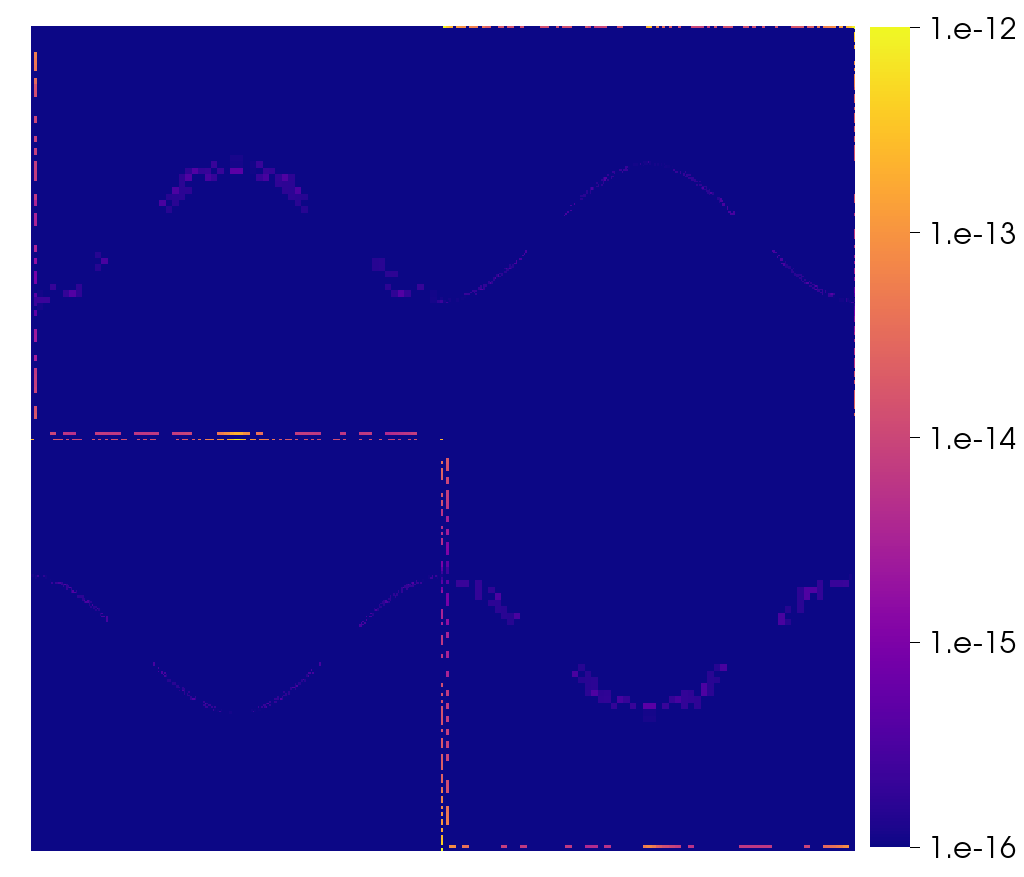}
\caption{Plots of the magnetic field magnitude $\vert \mathbf{B} \vert$ (left panel) and normalized divergence error (right panel) at $t=0.5$ for the Orszag-Tang problem using the VP evolution scheme on a mesh with $128 \times 128$ base resolution and 2 levels of static refinement in the lower-left and upper-right quadrants. No mesh imprinting is seen in the fluid variables and we maintain a very reasonable normalized divergence error.}
  \label{figure:ot_refined}
\end{figure}

The next test we try with mesh refinement is the Orszag-Tang problem from Sec. \ref{sec:orszag}. Here we set up static refinement over two quadrants of the grid, leaving the other two quadrants at a lower resolution ($128\times 128$ base resolution with two levels of refinement added in the lower left and upper right quadrants). The goal is to confirm that no artificial features or mesh imprinting are seen whenever a very dynamical, magnetized fluid moves back and forth across refinement boundaries. Fig. \ref{figure:ot_refined}a shows that the final solution shows the same characteristic features as our single layer mesh case (Fig. \ref{figure:ot}) though not as sharp in the unrefined quadrants. Crucially, we find no evidence of mesh imprinting, while maintaining a divergence error very close to machine precision ($\vert \vert \nabla \cdotp \mathbf{B} \vert \vert_2 < 1.10 \times 10^{-16}$; Fig. \ref{figure:ot_refined}b).

\subsubsection{Magnetized Black Hole Torus}

Next we consider the astrophysically more interesting problem of a magnetized torus of gas orbiting a rotating black hole. Although there is no known analytic solution for this problem, it is sufficiently well documented in the literature \citep[e.g.][]{Gammie03,Shiokawa12,White16,Porth18} to serve as a useful test of the code. It also represents one class of problems for which the new VP method is intended to be used. The particular equilibrium configuration we start from is the constant specific angular momentum torus solution presented in \cite{Fishbone76}. Specifically, the spin of the black hole is $a/M = 0.9375$, the inner radius of the torus is set to $r_{\rm in} = 6 GM/c^2$,
the density maximum is located at $r_{\rm max}=12 GM/c^2$, and we use an ideal gas EOS with an adiabatic
index of $\Gamma=4/3$. To the hydrodynamic solution we add an initially weak poloidal magnetic
field loop defined by the vector potential
\begin{equation}
\mathcal{A}_{\phi} \propto {\rm max} (\rho/\rho_{\rm max} - 0.2, 0) ~,
\end{equation}
which produces poloidal field loops coincident with the isodensity contours of the torus. The field strength is set such that
$P_{\rm max}/P_{B,{\rm max}}=100$, where the global maxima of the gas
$P_{\rm max}$ and magnetic $P_{B,{\rm max}}$ pressures do not
necessarily coincide. 
In order to spur growth of the magneto-rotational instability inside the torus, the gas pressure is perturbed as
\begin{equation}
P^* = P(1+X_p) ~,
\end{equation}
where $X_p$ is a uniformly distributed random variable between $-0.02$ and $0.02$.  

In the background region not specified by the torus solution, we set a floor model with $\rho_{\mathrm{fl}} = 10^{-5}r^{-3/2}$, $e_{\mathrm{fl}} =10^{-7} r^{-5/2}$, and $V^i = 0$. Obviously once the evolution begins, the background is no longer static. At first it falls toward the hole, but ultimately most of the background region is filled with a magnetized ``corona'' and a low-density, high-magnetization outflow launched by pressure forces and the opening of magnetic field lines. 

\begin{figure}
\includegraphics[width=0.5\linewidth]{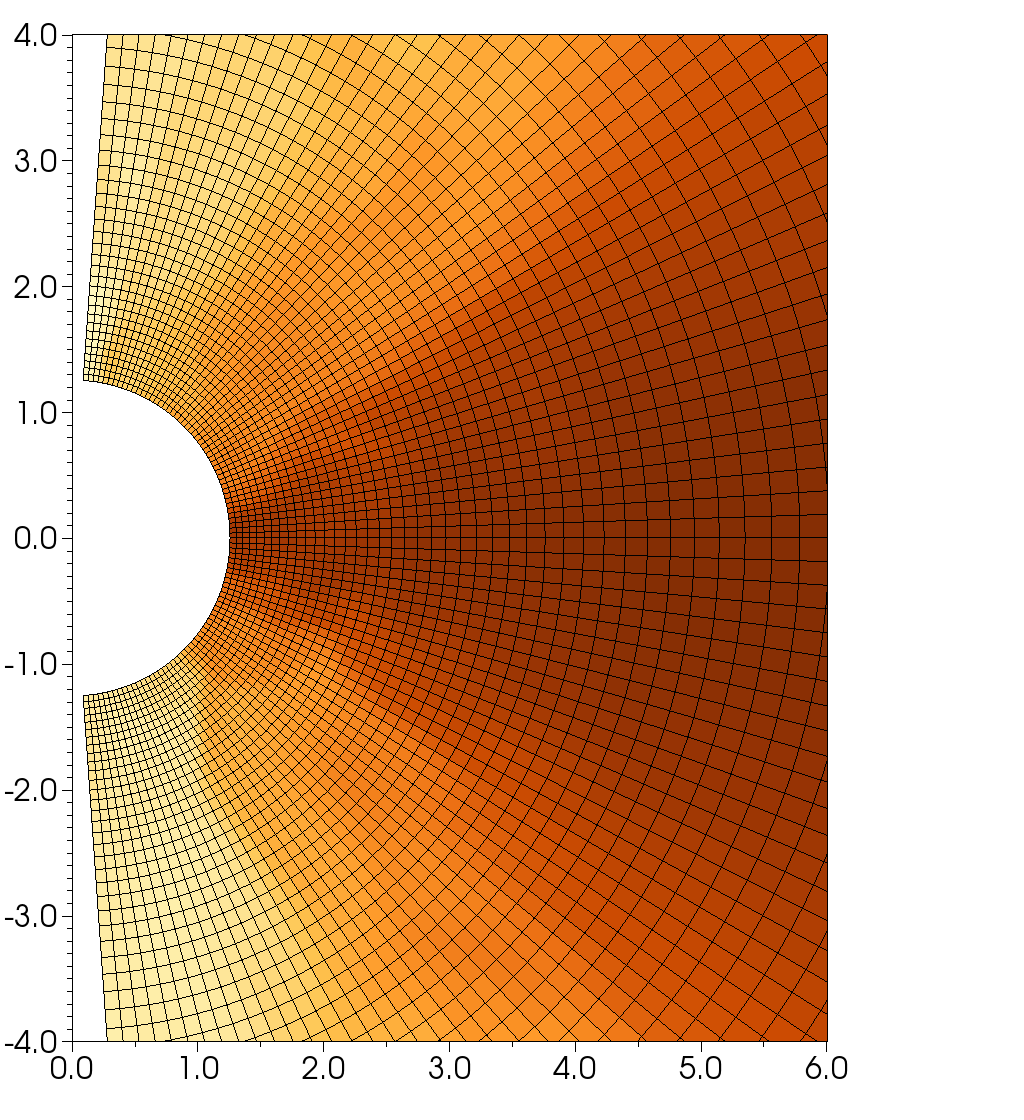}
\includegraphics[width=0.5\linewidth]{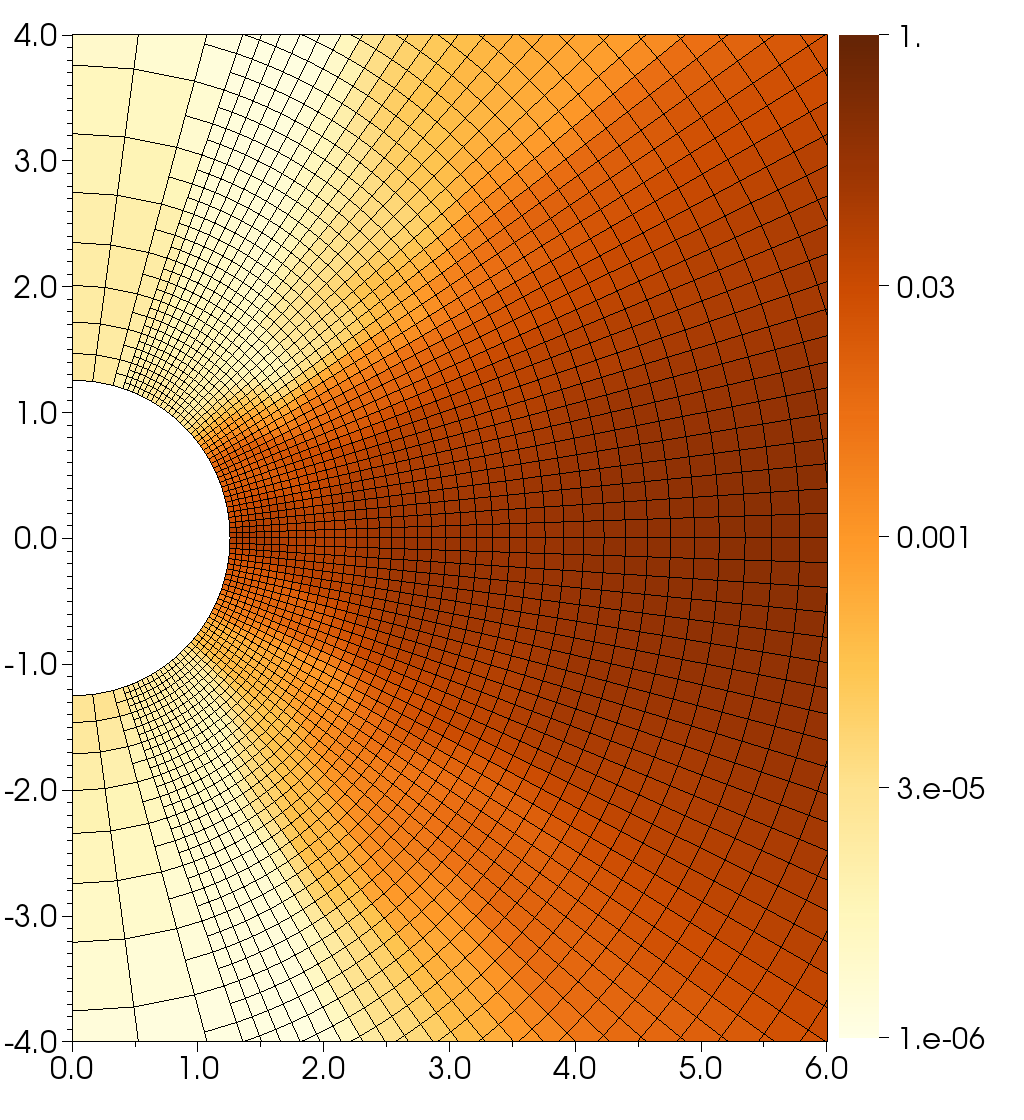}
\caption{Plots of the mass density $\rho$ at $t=10000 GM/c^3$ for the magnetized black hole torus problem using the CT evolution scheme on a single-layer mesh (left panel) and the VP scheme on a mesh with 2 levels of static refinement.}
\label{fig:BH_mesh}
\end{figure}

Part of our motivation for running this particular test is that we already have results available from the CT version of {\c} with which to compare. As mentioned throughout this paper, one of the advantages of the VP method over CT is that we are able to utilize mesh refinement. For black hole accretion problems done on spherical polar grids, one of the difficulties one faces is the very small zone sizes near the polar axis. On a single layer mesh, the only way to avoid this problem are to cut out a small region around the pole, as in the left panel of Fig. \ref{fig:BH_mesh}. However, with (static) refinement, one can extend the base level grid all the way to the pole without suffering prohibitive timestep constraints because the zones do not necessarily need to be small, while the desired level of refinement can be added away from the pole, as in the right panel of Fig. \ref{fig:BH_mesh}. Here, both simulations are run with comparable peak resolutions, yet the VP one runs with almost an order of magnitude larger timestep, while also including the full range of $\theta$. 

Both simulations use a logarithmic radial coordinate of the form $x_1 = 1 + \ln(r/r_\mathrm{BH})$ with radial boundaries at $r_\mathrm{min} = 0.98 r_\mathrm{BH}$ and $r_\mathrm{max} = 354 GM/c^2$, where $r_\mathrm{BH}$ is the radius of the black hole event horizon, and a concentrated latitude coordinate of the form $\theta = x_2 + 0.25\sin(2x_2)$ with the $x_1$ and $x_2$ coordinates being uniformly distributed. For the single-layer ($96 \times 96$) mesh (CT) simulation, the grid covers the angular range $0.02\pi \le \theta \le 0.98\pi$, while the base-level ($24 \times 24$) mesh for VP covers the full range $0 \le \theta \le \pi$. Two levels of refinement are added on top of this base mesh in the region $r_\mathrm{min} \le r \le 50 GM/c^2$ and $0.02\pi \le \theta \le 0.98\pi$. Both simulations cover the full range of $0 \le \phi \le 2\pi$, with periodic boundary conditions at $\phi = \{0, 2\pi\}$. Outflow boundary conditions are applied on the inner and outer radial boundaries, as well as the cutout region in the CT simulation. Special reflecting boundaries are used along the pole in the VP case.

\begin{figure}
\includegraphics[width=\linewidth]{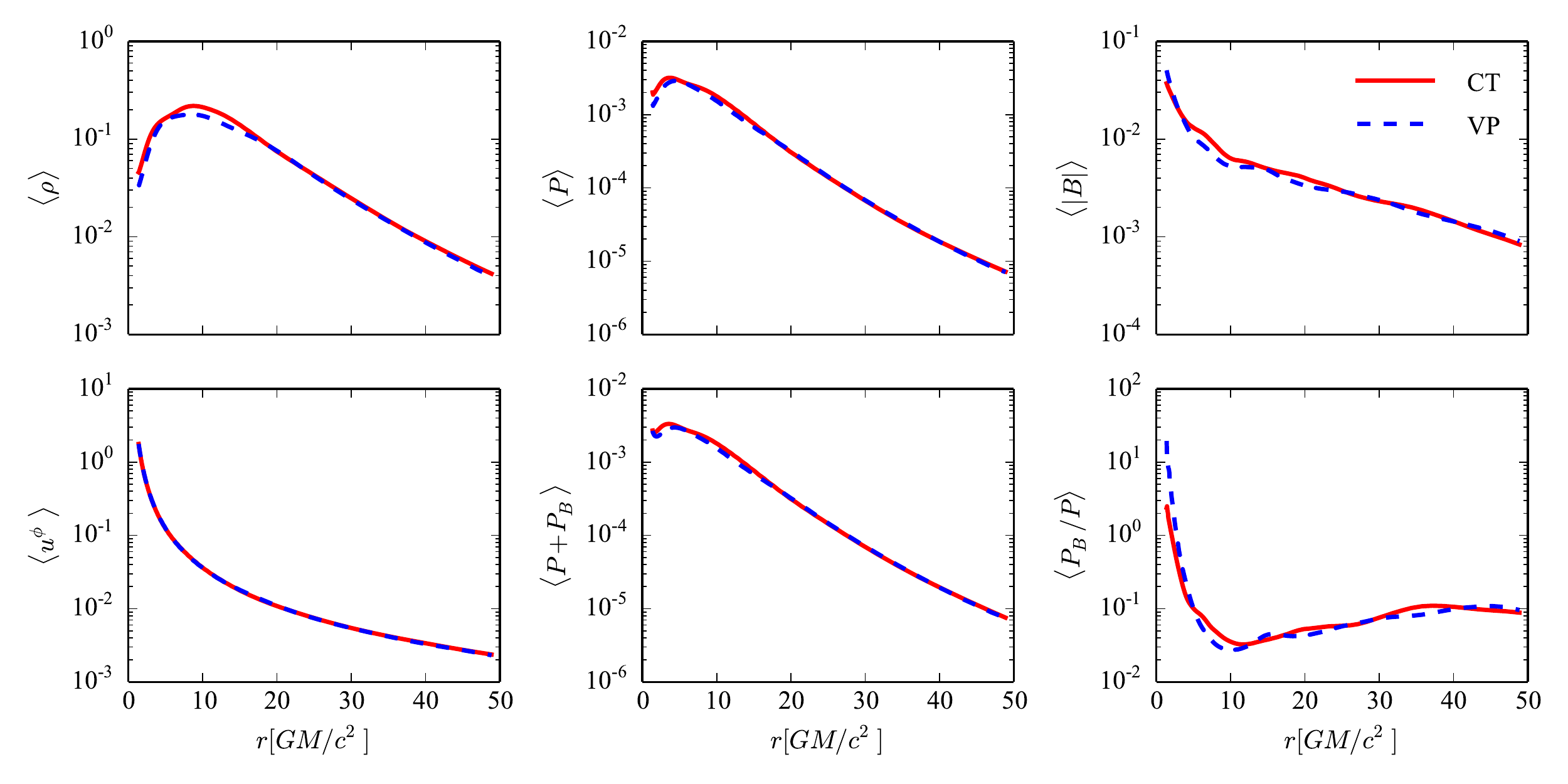}
\caption{Time-averaged profiles of select disk quantities. Data have been averaged over $\theta$ and $\phi$, as well as over the time-interval $5000 GM/c^3 \le t \le 10000 GM/c^3$.}
  \label{fig:BH_profiles}
\end{figure}

Fig. \ref{fig:BH_profiles} gives a more quantitative comparison between the two simulations. It shows time-averaged radial profiles of various disk quantities for both the CT and VP simulations. Spatial averages are calculated as 
\begin{equation}
\langle q(r) \rangle = \frac{\int_0^{2\pi}\int_{\theta{\rm min}}^{\theta{\rm max}} q(r,\theta,\phi,t) \sqrt{-g} d\theta d\phi}{\int_0^{2\pi}\int_{\theta{\rm min}}^{\theta{\rm max}} \sqrt{-g} d\theta d\phi} ~,
\end{equation}
which are then time averaged over approximately 100 dumps in the interval $5000 \le t \le 10000 GM/c^3$. Both simulations show nearly identical profiles, which also compare favorably to the collection of results published in \cite{Porth18}, confirming that the new VP method works quite well for this target problem. 

\section{Discussion \& Conclusions}
\label{sec:conclusions}

In this paper, we have reported our implementation of the vector potential method for divergence-free evolution of magnetic fields in both Newtonian and relativistic MHD. The key to this approach is to evolve the magnetic vector potential ($\vec{\mathcal{A}}$) directly, rather than the magnetic field ($\mathbf{B}$). The magnetic field components can then be recovered as needed from $\mathbf{B} = \nabla \times \vec{\mathcal{A}}$, a procedure that is guaranteed to satisfy $\nabla \cdot \mathbf{B} = 0$ to machine precision. The advantage in our case is that this approach makes it much easier to handle adaptive and moving mesh refinement (and hopefully adaptive order refinement in the future), while still preserving a divergence-free field.

We have shown through a series of multi-dimensional test problems that the method performs as anticipated. These tests were chosen to facilitate comparison with previously published results and to provide quantitative measures of the code's accuracy and convergence. We found the code maintained stable, accurate solutions in a number of challenging setups, performing as well or better than other published codes on fixed meshes. Further, the method showed no sign of any of the worrisome pathologies that were feared for vector potential evolution \cite{Evans88} nor the critical failures on select tests identified in other MHD approaches \cite{Toth00,Gardiner05,Gardiner08,Beckwith11}. Most importantly, the method was able to do all of this on adaptively refined and uniformly translating meshes, with no signs of mesh imprinting or other issues.

\section*{Acknowledgements}
We thank the anonymous referees for their helpful suggestions in improving this manuscript. This work used the Extreme Science and Engineering Discovery Environment (XSEDE), which is supported by National Science Foundation grant number ACI-1053575. P.C.F. acknowledges support from National Science Foundation grant AST-1616185. Work by P.A. was performed in part under the auspices of the U. S. Department of Energy by Lawrence Livermore National Laboratory under contract DE-AC52-07NA27344. P.L.S. acknowledges support from the College of Charleston Undergraduate Research and Creative Activities Board, through SURF grant SU2018-24.

\section*{References}
\bibliographystyle{elsarticle-num}

\begin{thebibliography}{10}
\expandafter\ifx\csname url\endcsname\relax
  \def\url#1{\texttt{#1}}\fi
\expandafter\ifx\csname urlprefix\endcsname\relax\def\urlprefix{URL }\fi
\expandafter\ifx\csname href\endcsname\relax
  \def\href#1#2{#2} \def\path#1{#1}\fi

\bibitem{Rossmanith13}
J.~A. {Rossmanith}, {High-Order Discontinuous Galerkin Finite Element Methods
  with Globally Divergence-Free Constrained Transport for Ideal MHD}, ArXiv
  e-prints\href {http://arxiv.org/abs/1310.4251} {\path{arXiv:1310.4251}}.

\bibitem{Brackbill80}
J.~U. {Brackbill}, D.~C. {Barnes}, {The effect of nonzero product of magnetic
  gradient and B on the numerical solution of the magnetohydrodynamic
  equations}, Journal of Computational Physics 35 (1980) 426--430.
\newblock \href {http://dx.doi.org/10.1016/0021-9991(80)90079-0}
  {\path{doi:10.1016/0021-9991(80)90079-0}}.

\bibitem{Balsara99}
D.~S. {Balsara}, D.~S. {Spicer}, {A Staggered Mesh Algorithm Using High Order
  Godunov Fluxes to Ensure Solenoidal Magnetic Fields in Magnetohydrodynamic
  Simulations}, Journal of Computational Physics 149 (1999) 270--292.
\newblock \href {http://dx.doi.org/10.1006/jcph.1998.6153}
  {\path{doi:10.1006/jcph.1998.6153}}.

\bibitem{Evans88}
C.~R. {Evans}, J.~F. {Hawley}, {Simulation of magnetohydrodynamic flows - A
  constrained transport method}, Astrophysical Journal 332 (1988) 659--677.
\newblock \href {http://dx.doi.org/10.1086/166684} {\path{doi:10.1086/166684}}.

\bibitem{Dai98}
W.~{Dai}, P.~R. {Woodward}, {On the Divergence-free Condition and Conservation
  Laws in Numerical Simulations for Supersonic Magnetohydrodynamical Flows},
  Astrophysical Journal 494 (1998) 317--335.
\newblock \href {http://dx.doi.org/10.1086/305176} {\path{doi:10.1086/305176}}.

\bibitem{Ryu98}
D.~{Ryu}, F.~{Miniati}, T.~W. {Jones}, A.~{Frank}, {A Divergence-free Upwind
  Code for Multidimensional Magnetohydrodynamic Flows}, Astrophysical Journal
  509 (1998) 244--255.
\newblock \href {http://arxiv.org/abs/astro-ph/9807228}
  {\path{arXiv:astro-ph/9807228}}, \href {http://dx.doi.org/10.1086/306481}
  {\path{doi:10.1086/306481}}.

\bibitem{Toth00}
G.~{T{\'o}th}, {The ${\nabla} \cdot B=0$ Constraint in Shock-Capturing
  Magnetohydrodynamics Codes}, Journal of Computational Physics 161 (2000)
  605--652.
\newblock \href {http://dx.doi.org/10.1006/jcph.2000.6519}
  {\path{doi:10.1006/jcph.2000.6519}}.

\bibitem{Powell99}
K.~G. {Powell}, P.~L. {Roe}, T.~J. {Linde}, T.~I. {Gombosi}, D.~L. {De Zeeuw},
  {A Solution-Adaptive Upwind Scheme for Ideal Magnetohydrodynamics}, Journal
  of Computational Physics 154 (1999) 284--309.
\newblock \href {http://dx.doi.org/10.1006/jcph.1999.6299}
  {\path{doi:10.1006/jcph.1999.6299}}.

\bibitem{Dedner02}
A.~{Dedner}, F.~{Kemm}, D.~{Kr{\"o}ner}, C.-D. {Munz}, T.~{Schnitzer},
  M.~{Wesenberg}, {Hyperbolic Divergence Cleaning for the MHD Equations},
  Journal of Computational Physics 175 (2002) 645--673.
\newblock \href {http://dx.doi.org/10.1006/jcph.2001.6961}
  {\path{doi:10.1006/jcph.2001.6961}}.

\bibitem{Mocz14}
P.~{Mocz}, M.~{Vogelsberger}, D.~{Sijacki}, R.~{Pakmor}, L.~{Hernquist}, {A
  discontinuous Galerkin method for solving the fluid and magnetohydrodynamic
  equations in astrophysical simulations}, Monthly Notices of the Royal
  Astronomical Society 437 (2014) 397--414.
\newblock \href {http://arxiv.org/abs/1305.5536} {\path{arXiv:1305.5536}},
  \href {http://dx.doi.org/10.1093/mnras/stt1890}
  {\path{doi:10.1093/mnras/stt1890}}.

\bibitem{Balsara15}
D.~S. {Balsara}, M.~{Dumbser}, {Divergence-free MHD on unstructured meshes
  using high order finite volume schemes based on multidimensional Riemann
  solvers}, Journal of Computational Physics 299 (2015) 687--715.
\newblock \href {http://dx.doi.org/10.1016/j.jcp.2015.07.012}
  {\path{doi:10.1016/j.jcp.2015.07.012}}.

\bibitem{Balsara10}
D.~S. {Balsara}, {Multidimensional HLLE Riemann solver: Application to Euler
  and magnetohydrodynamic flows}, Journal of Computational Physics 229 (2010)
  1970--1993.
\newblock \href {http://arxiv.org/abs/0911.1613} {\path{arXiv:0911.1613}},
  \href {http://dx.doi.org/10.1016/j.jcp.2009.11.018}
  {\path{doi:10.1016/j.jcp.2009.11.018}}.

\bibitem{Balsara12}
D.~S. {Balsara}, {A two-dimensional HLLC Riemann solver for conservation laws:
  Application to Euler and magnetohydrodynamic flows}, Journal of Computational
  Physics 231 (2012) 7476--7503.
\newblock \href {http://arxiv.org/abs/1110.0750} {\path{arXiv:1110.0750}},
  \href {http://dx.doi.org/10.1016/j.jcp.2011.12.025}
  {\path{doi:10.1016/j.jcp.2011.12.025}}.

\bibitem{Brandenburg02}
A.~{Brandenburg}, W.~{Dobler}, {Hydromagnetic turbulence in computer
  simulations}, Computer Physics Communications 147 (2002) 471--475.
\newblock \href {http://arxiv.org/abs/astro-ph/0111569}
  {\path{arXiv:astro-ph/0111569}}, \href
  {http://dx.doi.org/10.1016/S0010-4655(02)00334-X}
  {\path{doi:10.1016/S0010-4655(02)00334-X}}.

\bibitem{Etienne10}
Z.~B. {Etienne}, Y.~T. {Liu}, S.~L. {Shapiro}, {Relativistic
  magnetohydrodynamics in dynamical spacetimes: A new adaptive mesh refinement
  implementation}, Physical Review 82~(8) (2010) 084031.
\newblock \href {http://arxiv.org/abs/1007.2848} {\path{arXiv:1007.2848}},
  \href {http://dx.doi.org/10.1103/PhysRevD.82.084031}
  {\path{doi:10.1103/PhysRevD.82.084031}}.

\bibitem{Choptuik86}
M.~W. {Choptuik}, {a Study of Numerical Techniques for Radiative Problems in
  General Relativity.}, Ph.D. thesis, THE UNIVERSITY OF BRITISH COLUMBIA
  (CANADA). (1986).

\bibitem{Anninos05}
P.~{Anninos}, P.~C. {Fragile}, J.~D. {Salmonson}, {Cosmos++: Relativistic
  Magnetohydrodynamics on Unstructured Grids with Local Adaptive Refinement},
  Astrophysical Journal 635 (2005) 723--740.
\newblock \href {http://arxiv.org/abs/arXiv:astro-ph/0509254}
  {\path{arXiv:arXiv:astro-ph/0509254}}, \href
  {http://dx.doi.org/10.1086/497294} {\path{doi:10.1086/497294}}.

\bibitem{Anninos12}
P.~{Anninos}, P.~C. {Fragile}, J.~{Wilson}, S.~D. {Murray}, {Three-dimensional
  Moving-mesh Simulations of Galactic Center Cloud G2}, Astrophysical Journal
  759 (2012) 132.
\newblock \href {http://arxiv.org/abs/1209.1638} {\path{arXiv:1209.1638}},
  \href {http://dx.doi.org/10.1088/0004-637X/759/2/132}
  {\path{doi:10.1088/0004-637X/759/2/132}}.

\bibitem{Anninos17}
P.~{Anninos}, C.~{Bryant}, P.~C. {Fragile}, A.~M. {Holgado}, C.~{Lau},
  D.~{Nemergut}, {CosmosDG: An hp-adaptive Discontinuous Galerkin Code for
  Hyper-resolved Relativistic MHD}, Astrophysical Journal Supplement Series 231
  (2017) 17.
\newblock \href {http://arxiv.org/abs/1706.09939} {\path{arXiv:1706.09939}},
  \href {http://dx.doi.org/10.3847/1538-4365/aa7ff5}
  {\path{doi:10.3847/1538-4365/aa7ff5}}.

\bibitem{Anninos18}
P.~{Anninos}, P.~C. {Fragile}, S.~S. {Olivier}, R.~{Hoffman}, B.~{Mishra},
  K.~{Camarda}, {Relativistic Tidal Disruption and Nuclear Ignition of White
  Dwarf Stars by Intermediate-mass Black Holes}, Astrophysical Journal 865
  (2018) 3.
\newblock \href {http://arxiv.org/abs/1808.05664} {\path{arXiv:1808.05664}},
  \href {http://dx.doi.org/10.3847/1538-4357/aadad9}
  {\path{doi:10.3847/1538-4357/aadad9}}.

\bibitem{Anninos03b}
P.~{Anninos}, P.~C. {Fragile}, S.~D. {Murray}, {Cosmos: A
  Radiation-Chemo-Hydrodynamics Code for Astrophysical Problems}, Astrophysical
  Journal Supplement Series 147 (2003) 177--186.
\newblock \href {http://arxiv.org/abs/astro-ph/0303209}
  {\path{arXiv:astro-ph/0303209}}, \href {http://dx.doi.org/10.1086/375184}
  {\path{doi:10.1086/375184}}.

\bibitem{Font03}
J.~A. {Font}, {Numerical Hydrodynamics in General Relativity}, Living Reviews
  in Relativity 6 (2003) 4.
\newblock \href {http://dx.doi.org/10.12942/lrr-2003-4}
  {\path{doi:10.12942/lrr-2003-4}}.

\bibitem{Spiteri02}
R.~J. {Spiteri}, S.~J. {Ruuth}, {A new class of optimal high-order
  strong-stability-preserving time discretization methods}, SIAM Journal on
  Numerical Analysis 40 (2002) 469--491.

\bibitem{Noble06}
S.~C. {Noble}, C.~F. {Gammie}, J.~C. {McKinney}, L.~{Del Zanna}, {Primitive
  Variable Solvers for Conservative General Relativistic Magnetohydrodynamics},
  Astrophysical Journal 641 (2006) 626--637.
\newblock \href {http://arxiv.org/abs/astro-ph/0512420}
  {\path{arXiv:astro-ph/0512420}}, \href {http://dx.doi.org/10.1086/500349}
  {\path{doi:10.1086/500349}}.

\bibitem{Budd09}
C.~J. {Budd}, J.~F. {Williams}, {Moving Mesh Generation Using the Parabolic
  Monge?Amp\`ere Equation}, SIAM Journal on Scientific Computing 31 (2009)
  3438--3465.
\newblock \href {http://dx.doi.org/10.1137/080716773}
  {\path{doi:10.1137/080716773}}.

\bibitem{Cao02}
W.~{Cao}, W.~{Huang}, R.~D. {Russell}, {A Moving Mesh Method Based on the
  Geometric Conservation Law}, SIAM Journal on Scientific Computing 24 (2002)
  118--142.
\newblock \href {http://dx.doi.org/10.1137/S1064827501384925}
  {\path{doi:10.1137/S1064827501384925}}.

\bibitem{Huang01}
W.~{Huang}, {Practical Aspects of Formulation and Solution of Moving Mesh
  Partial Differential Equations}, Journal of Computational Physics 171 (2001)
  753--775.
\newblock \href {http://dx.doi.org/10.1006/jcph.2001.6809}
  {\path{doi:10.1006/jcph.2001.6809}}.

\bibitem{Chacon11}
L.~{Chac{\'o}n}, G.~L. {Delzanno}, J.~M. {Finn}, {Robust, multidimensional
  mesh-motion based on Monge-Kantorovich equidistribution}, Journal of
  Computational Physics 230 (2011) 87--103.
\newblock \href {http://dx.doi.org/10.1016/j.jcp.2010.09.013}
  {\path{doi:10.1016/j.jcp.2010.09.013}}.

\bibitem{Sulman11}
M.~{Sulman}, J.~F. {Williams}, R.~D. {Russell}, {Optimal mass transport for
  higher dimensional adaptive grid generation}, Journal of Computational
  Physics 230 (2011) 3302--3330.
\newblock \href {http://dx.doi.org/10.1016/j.jcp.2011.01.025}
  {\path{doi:10.1016/j.jcp.2011.01.025}}.

\bibitem{Fragile05a}
P.~C. {Fragile}, P.~{Anninos}, K.~{Gustafson}, S.~D. {Murray},
  {Magnetohydrodynamic Simulations of Shock Interactions with Radiative
  Clouds}, Astrophysical Journal 619 (2005) 327--339.
\newblock \href {http://arxiv.org/abs/astro-ph/0410285}
  {\path{arXiv:astro-ph/0410285}}, \href {http://dx.doi.org/10.1086/426313}
  {\path{doi:10.1086/426313}}.

\bibitem{Fragile07}
P.~C. {Fragile}, O.~M. {Blaes}, P.~{Anninos}, J.~D. {Salmonson}, {Global
  General Relativistic Magnetohydrodynamic Simulation of a Tilted Black Hole
  Accretion Disk}, Astrophysical Journal 668 (2007) 417--429.
\newblock \href {http://arxiv.org/abs/0706.4303} {\path{arXiv:0706.4303}},
  \href {http://dx.doi.org/10.1086/521092} {\path{doi:10.1086/521092}}.

\bibitem{Fragile12}
P.~C. {Fragile}, A.~{Gillespie}, T.~{Monahan}, M.~{Rodriguez}, P.~{Anninos},
  {Numerical Simulations of Optically Thick Accretion onto a Black Hole. I.
  Spherical Case}, Astrophysical Journal Supplement Series 201 (2012) 9.
\newblock \href {http://arxiv.org/abs/1204.5538} {\path{arXiv:1204.5538}},
  \href {http://dx.doi.org/10.1088/0067-0049/201/2/9}
  {\path{doi:10.1088/0067-0049/201/2/9}}.

\bibitem{DelZanna03}
L.~{Del Zanna}, N.~{Bucciantini}, P.~{Londrillo}, {An efficient shock-capturing
  central-type scheme for multidimensional relativistic flows. II.
  Magnetohydrodynamics}, Astronomy and Astrophysics 400 (2003) 397--413.
\newblock \href {http://arxiv.org/abs/astro-ph/0210618}
  {\path{arXiv:astro-ph/0210618}}, \href
  {http://dx.doi.org/10.1051/0004-6361:20021641}
  {\path{doi:10.1051/0004-6361:20021641}}.

\bibitem{Etienne12}
Z.~B. {Etienne}, V.~{Paschalidis}, Y.~T. {Liu}, S.~L. {Shapiro}, {Relativistic
  magnetohydrodynamics in dynamical spacetimes: Improved electromagnetic gauge
  condition for adaptive mesh refinement grids}, Physical Review 85~(2) (2012)
  024013.
\newblock \href {http://arxiv.org/abs/1110.4633} {\path{arXiv:1110.4633}},
  \href {http://dx.doi.org/10.1103/PhysRevD.85.024013}
  {\path{doi:10.1103/PhysRevD.85.024013}}.

\bibitem{Gardiner08}
T.~A. {Gardiner}, J.~M. {Stone}, {An unsplit Godunov method for ideal MHD via
  constrained transport in three dimensions}, Journal of Computational Physics
  227 (2008) 4123--4141.
\newblock \href {http://arxiv.org/abs/0712.2634} {\path{arXiv:0712.2634}},
  \href {http://dx.doi.org/10.1016/j.jcp.2007.12.017}
  {\path{doi:10.1016/j.jcp.2007.12.017}}.

\bibitem{Goldstein78}
M.~L. {Goldstein}, {An instability of finite amplitude circularly polarized
  Alfven waves}, Astrophysical Journal 219 (1978) 700--704.
\newblock \href {http://dx.doi.org/10.1086/155829} {\path{doi:10.1086/155829}}.

\bibitem{DelZanna01}
L.~{Del Zanna}, M.~{Velli}, P.~{Londrillo}, {Parametric decay of circularly
  polarized Alfv{\'e}n waves: Multidimensional simulations in periodic and open
  domains}, Astronomy \& Astrophysics 367 (2001) 705--718.
\newblock \href {http://dx.doi.org/10.1051/0004-6361:20000455}
  {\path{doi:10.1051/0004-6361:20000455}}.

\bibitem{Balsara04}
D.~S. {Balsara}, {Second-Order-accurate Schemes for Magnetohydrodynamics with
  Divergence-free Reconstruction}, Astrophysical Journal Supplement Series 151
  (2004) 149--184.
\newblock \href {http://arxiv.org/abs/astro-ph/0308249}
  {\path{arXiv:astro-ph/0308249}}, \href {http://dx.doi.org/10.1086/381377}
  {\path{doi:10.1086/381377}}.

\bibitem{Hawley95a}
J.~F. {Hawley}, J.~M. {Stone}, {MOCCT: A numerical technique for astrophysical
  MHD}, Computer Physics Communications 89 (1995) 127--148.
\newblock \href {http://dx.doi.org/10.1016/0010-4655(95)00190-Q}
  {\path{doi:10.1016/0010-4655(95)00190-Q}}.

\bibitem{Beckwith11}
K.~{Beckwith}, J.~M. {Stone}, {A Second-order Godunov Method for
  Multi-dimensional Relativistic Magnetohydrodynamics}, Astrophysical Journal
  Supplement Series 193 (2011) 6.
\newblock \href {http://arxiv.org/abs/1101.3573} {\path{arXiv:1101.3573}},
  \href {http://dx.doi.org/10.1088/0067-0049/193/1/6}
  {\path{doi:10.1088/0067-0049/193/1/6}}.

\bibitem{Gardiner05}
T.~A. {Gardiner}, J.~M. {Stone}, {An unsplit Godunov method for ideal MHD via
  constrained transport}, Journal of Computational Physics 205 (2005) 509--539.
\newblock \href {http://arxiv.org/abs/astro-ph/0501557}
  {\path{arXiv:astro-ph/0501557}}, \href
  {http://dx.doi.org/10.1016/j.jcp.2004.11.016}
  {\path{doi:10.1016/j.jcp.2004.11.016}}.

\bibitem{Orszag79}
S.~A. {Orszag}, C.-M. {Tang}, {Small-scale structure of two-dimensional
  magnetohydrodynamic turbulence}, Journal of Fluid Mechanics 90 (1979)
  129--143.
\newblock \href {http://dx.doi.org/10.1017/S002211207900210X}
  {\path{doi:10.1017/S002211207900210X}}.

\bibitem{Londrillo00}
P.~{Londrillo}, L.~{Del Zanna}, {High-Order Upwind Schemes for Multidimensional
  Magnetohydrodynamics}, Astrophysical Journal 530 (2000) 508--524.
\newblock \href {http://arxiv.org/abs/astro-ph/9910086}
  {\path{arXiv:astro-ph/9910086}}, \href {http://dx.doi.org/10.1086/308344}
  {\path{doi:10.1086/308344}}.

\bibitem{Gammie03}
C.~F. {Gammie}, J.~C. {McKinney}, G.~{T{\'o}th}, {HARM: A Numerical Scheme for
  General Relativistic Magnetohydrodynamics}, Astrophysical Journal 589 (2003)
  444--457.
\newblock \href {http://arxiv.org/abs/astro-ph/0301509}
  {\path{arXiv:astro-ph/0301509}}, \href {http://dx.doi.org/10.1086/374594}
  {\path{doi:10.1086/374594}}.

\bibitem{Shiokawa12}
H.~{Shiokawa}, J.~C. {Dolence}, C.~F. {Gammie}, S.~C. {Noble}, {Global General
  Relativistic Magnetohydrodynamic Simulations of Black Hole Accretion Flows: A
  Convergence Study}, Astrophysical Journal 744 (2012) 187.
\newblock \href {http://arxiv.org/abs/1111.0396} {\path{arXiv:1111.0396}},
  \href {http://dx.doi.org/10.1088/0004-637X/744/2/187}
  {\path{doi:10.1088/0004-637X/744/2/187}}.

\bibitem{White16}
C.~J. {White}, J.~M. {Stone}, C.~F. {Gammie}, {An Extension of the Athena++
  Code Framework for GRMHD Based on Advanced Riemann Solvers and Staggered-mesh
  Constrained Transport}, Astrophysical Journal Supplement Series 225 (2016)
  22.
\newblock \href {http://arxiv.org/abs/1511.00943} {\path{arXiv:1511.00943}},
  \href {http://dx.doi.org/10.3847/0067-0049/225/2/22}
  {\path{doi:10.3847/0067-0049/225/2/22}}.

\bibitem{Porth18}
O.~{Porth}, C.~{Gammie}, Y.~{Mizuno}, L.~{Rezzolla}, H.~{Olivares},
  S.~{Markoff}, B.~{Ryan}, J.~{McKinney}, R.~{Narayan}, S.~{Noble}, C.~J.
  {White}, J.~{Stone}, F.~{Foucart}, P.~C. {Fragile}, Z.~{Etienne}, M.~{Liska},
  G.~S. {Ryan}, L.~{Del Zanna}, K.~{Chatterjee}, M.~{Bugli}, C.~K. {Chan},
  Z.~{Younsi}, {The Event Horizon GRMHD Code Comparison Project}, Astrophysical
  Journal.

\bibitem{Fishbone76}
L.~G. {Fishbone}, V.~{Moncrief}, {Relativistic fluid disks in orbit around Kerr
  black holes}, Astrophysical Journal 207 (1976) 962--976.
\newblock \href {http://dx.doi.org/10.1086/154565} {\path{doi:10.1086/154565}}.

\end{thebibliography}

\end{document}